\PassOptionsToPackage{pdfpagelabels=false}{hyperref}
\documentclass[useAMS,usenatbib]{mnras}
\usepackage[pdftex]{graphicx}
\usepackage{newtxtext,newtxmath}
\usepackage[T1]{fontenc}
\usepackage{xspace}

\newcommand{\msun}{\,M$_{\odot}$\xspace}

\title[The TNG50 CGM: MgII Emission]{The cold circumgalactic medium in emission: MgII halos in TNG50}
\author[D. Nelson et al.]{Dylan Nelson$^{1}$\thanks{E-mail: dnelson@uni-heidelberg.de},
Chris Byrohl$^{2}$,
Celine Peroux$^{3,4}$,
Kate H. R. Rubin$^{5}$,
Joseph N. Burchett$^{6}$
\\\\
$^{1}$Universit\"{a}t Heidelberg, Zentrum f\"{u}r Astronomie, Institut f\"{u}r theoretische Astrophysik, Albert-Ueberle-Str. 2, 69120 Heidelberg, Germany\\
$^{2}$Max-Planck-Institut f\"{u}r Astrophysik, Karl-Schwarzschild-Str. 1, 85741 Garching, Germany\\
$^{3}$European Southern Observatory, Karl-Schwarzschildstrasse 2, D-85748 Garching bei M{\"u}nchen, Germany\\
$^{4}$Aix Marseille Universit\'e, CNRS, LAM (Laboratoire d'Astrophysique de Marseille) UMR 7326, 13388, Marseille, France\\
$^{5}$San Diego State University, Department of Astronomy, San Diego, CA 92182, USA\\
$^{6}$Department of Astronomy, New Mexico State University, Las Cruces, NM 88003, USA\\
}

\begin{document}

\maketitle

\begin{abstract}
We outline theoretical predictions for extended emission from MgII, tracing cool $\sim 10^4$\,K gas in the circumgalactic medium (CGM) of star-forming galaxies in the high-resolution TNG50 cosmological magnetohydrodynamical simulation. We synthesize surface brightness maps of this strong rest-frame ultraviolet metal emission doublet ($\lambda\lambda2796, 2803$), adopting the assumption that the resonant scattering of MgII can be neglected and connecting to recent and upcoming observations with the Keck/KCWI, VLT/MUSE, and BlueMUSE optical integral field unit spectrographs. Studying galaxies with stellar masses $7.5 < \log{(M_\star/M_\odot)} < 11$ at redshifts $z=0.3, 0.7, 1$ and $2$ we find that extended MgII halos in emission, similar to their Ly$\alpha$ counterparts, are ubiquitous across the galaxy population. Median surface brightness profiles exceed $10^{-19}$ erg s$^{-1}$ cm$^{-2}$ arcsec$^{-2}$ in the central $\sim\,$10s of kpc, and total halo MgII luminosity increases with mass for star-forming galaxies, reaching \mbox{$10^{40}$ erg s$^{-1}$} for $M_\star \sim 10^{9.5}$\msun. MgII halo sizes increase from a few kpc to $\gtrsim 20$\,kpc at the highest masses, and sizes are larger for halos in denser environments. MgII halos are highly structured, clumpy, and asymmetric, with isophotal axis ratio increasing with galaxy mass. Similarly, the amount and distribution of MgII emission depends on the star formation activity of the central galaxy. Kinematically, inflowing versus outflowing gas dominates the MgII luminosity at high and low galaxy masses, respectively, although the majority of MgII halo emission at $z \sim 0.7$ traces near-equilibrium fountain flows and gas with non-negligible rotational support, rather than rapidly outflowing galactic winds.
\end{abstract}



\section{Introduction}

The promise of the circumgalactic medium is that it offers a new window into the formation and evolution of galaxies \citep{tumlinson17} as well as the cosmic baryon cycle \citep{peroux20a}. Despite significant progress in recent years, two main challenges remain: (i) observations of this low-density, diffuse baryonic reservoir are demanding, as are their interpretation; (ii) hydrodynamical simulations of the CGM are confronted by its complex multi-physics, multi-phase, and multi-scale nature. As a result, making robust and quantitative connections between theory and data in this regime remains challenging.

For several decades studies of the intergalactic and circumgalactic media have relied heavily upon detection and characterization in absorption \citep{boksenberg78}. Most typically, this method constrains the column densities and covering fractions of intervening gas along one random sightline per halo, typically towards a background quasar \citep{tripp00,chen01}. The statistical characterization of circumgalactic gas enabled by large sky surveys, connecting of order millions of absorption systems to similar numbers of foreground galaxies, demonstrates the power of this technique \citep{bouche06,lan14,anand21}. However, configurations which enable more than a single pencil-beam view of the CGM of a given halo are rare \citep{rauch01,lopez18,peroux18}. 

On the other hand, circumgalactic metals -- tracing the heavy element enrichment of the Universe -- are notoriously difficult to detect in emission, primarily due to their low surface brightness \citep{cantalupo19}. Despite this drawback, emission offers a more direct constraint than absorption on density and other gas physical properties \citep{corlies16}. Furthermore, spatially resolved emission offers the unique possibility to `map' or `directly image' the CGM for a more holistic view of circumgalactic gas \citep{rupke19}. The obstacles in measurement and interpretation are different than for absorption techniques, but no less challenging \citep{zhang18,lokhorst19}. Ultimately our goal is to infer the true, three-dimensional distribution, physical properties, and kinematics of circumgalactic gas and metals -- across its diversity of phases and observables -- from spatially resolved emission spectroscopy.

For instance, the hotter phases of galactic outflows and the CGM are observed in emission for the nearest local galaxies \citep{anderson15,hayes16,lijt17}, including our own Milky Way \citep{predehl20}. Extended emission from colder, ionized phases has been detected in a number of ultraluminous systems \citep{rupke19}, and is increasing accessible at high redshift through lines such as CII \citep{ginolfi20} and Lyman-$\alpha$ \citep{leclercq17}.

A particularly useful target is singly ionized magnesium, MgII, which has one electron in the ground state (Mg$^{+}$). Absorption into this ionization state is followed with $\sim 100$\% probability by spontaneous decay back to the ground state. This produces a doublet of transitions at $\lambda \simeq \{2796.352, 2803.532\}$\AA, corresponding to a velocity separation of $\Delta v \simeq 770$ km\,s$^{-1}$ \citep[][see their Fig. 1 for the MgII energy level diagram]{prochaska11a}. The ionization potential to produce MgII is 7.646 eV, while the second ionization potential (to produce Mg$^{++}$) is 15.035 eV. The doublet nature of this transition makes the line easily identifiable, and its wavelength makes it accessible from the ground for $z \gtrsim 0.2$. It is also bright, due to its atomic transition characteristics, as well as the relatively high abundance of magnesium as a metal species. 

Like the Lyman-$\alpha$ line of hydrogen, MgII is resonant. The lack of fine-structure splitting enables emitted photons to be efficiently and repeatedly reabsorbed\footnote{`Scattering' meaning excitation from the ground state to the first excited state, followed by de-excitation back to the ground state. (As opposed to fluorescence, meaning de-excitation to a lower excited level).}, a trapping process which hinders escape and confuses interpretation of the intrinsic origin of MgII emission \citep{prochaska11a}. In the classic view of a galactic-scale outflow as an expanding shell, this produces a P-Cygni like profile for each component of the MgII doublet, with blue-shifted absorption together with red-shifted emission.\footnote{Emission infill, where the absorption features are partially suppressed due to emission at the same wavelength, further complicates interpretation. At reasonable velocities, the 2796 absorption feature is infilled only by that component, while the 2803 absorption feature can be covered by emission from both components to either side.} The absorption profile arises from MgII moving towards the observer as it absorbs photons in its rest-frame, in which case the typically saturated shape (and total equivalent width) of the absorption profile mainly encodes information about the velocity distribution of the gas, rather than column density (i.e. the total number of MgII ions), and typical blueshifts are of order a few hundred km\,s$^{-1}$. There is no constraint as to where the gas is located along the sightline. On the other hand, the emission peak arises from the receding component of the shell (outflow), whereby photons emitted in the rest-frame, having scattered to escape towards the observer, are already redshifted and so pass through any intervening gas. Typical redshifts are closer to the systemic redshift of the galaxy.

In tandem with MgII absorption, the MgII emission line was also identified in down-the-barrel spectra of $0.7 \lesssim z \lesssim 2$ star forming galaxies \citep{weiner09,rubin10,erb12}. This was followed by the first studies of spatially extended MgII emission \citep{rubin11,martin13} using Keck/LRIS, which hinted towards circumgalactic and/or outflowing gas extending beyond the stellar continuum, possibly illuminated due to resonantly scattered photons. The handful of galaxies observed to host extended MgII emission halos are luminous and highly star-forming, while more typical (main sequence) galaxies at these redshifts may not show such bright metal-line emission \citep[e.g. see the five non-detections in the narrow-band imaging search of][]{rickards19}.

Only in the past several years have two crucial instruments come online: the VLT/MUSE and Keck/KCWI integral field unit (IFU) spectrographs. There has been significant investigation of Lyman-$\alpha$ emission with MUSE at $z \gtrsim 3$, including unbiased studies based on the UDF \citep{leclercq17,leclercq20}, quasar-host targeted samples \citep{borisova16,battaia19,farina19,mackenzie21}, and tentative detections of the large-scale cosmic web \citep{witstok19,bacon21}. This mirrors Ly$\alpha$ surveys and halo studies down to $z \sim 2$ with KCWI \citep{cai19,martin19_kcwi,osullivan20}. Recently MgII halos have also been seen with KCWI \citep{burchett20}, as have OII and MgII halos in emission \citep{rupke19}. The MUSE UDF enables statistical study of FeII* halos across the galaxy population \citep{finley17a,finley17b}, while MgII halos in emission have been characterized in the MEGAFLOW \citep{zabl21} and MAGG \citep{dutta20} surveys. The MUSE spectrograph, with a lower wavelength cutoff of 480\,nm, can observe MgII at $z \gtrsim 0.7$ \citep{feltre18}, whereas KCWI and the future BlueMUSE instrument \citep{richard19} can both observe MgII at $z \gtrsim 0.3$.

Simulations have long provided predictions for emission from the gravitationally bound gaseous halos around larger groups and clusters \citep{kravtsov02,yoshikawa03,fang05}. Predictions for UV and x-ray line emission across the galaxy population have been made from older cosmological volumes \citep{bertone10a,bertone10b,bertone12,vdv13}, while detectable emission from the CGM in particular lines has also been forecast \citep{frank12,corlies16,sravan16}. In the past several years cosmological zoom simulations have studied prospects for detecting the CGM in emission \citep{augustin19,corlies20}, including fully radiation-hydrodynamical calculations focused on the Ly$\alpha$ line \citep{mitchell21} as well as nebular emission \citep{katz19}. Recently, \cite{byrohl20} presented a new method to treat the resonant scattering of Ly$\alpha$, applied this to TNG50 and made a comparison to MUSE. Specifically, none of these works have considered MgII emission. And no theoretical studies of ultraviolet emission from the CGM have yet been based on cosmological simulations with demonstrably realistic galaxy populations such as IllustrisTNG.

In this work we develop a relatively simple model for MgII emission, neglecting resonant scattering effects, and apply it to the TNG50 cosmological magnetohydrodynamical simulation. This enables us to outline predictions for extended MgII halos observable in emission for $0.3 < z < 2$ and for a statistical ensemble of thousands of galaxies with $10^{7.5} < M_\star < 10^{11}$\msun. We note that the importance of resonant scattering in producing observable MgII halos in emission is currently unknown, and will be investigated in future work. Comparison with the quantitative profiles predicted from TNG50 in the absence of this process can thus constrain not only the distribution and thermal state of cool circumgalactic gas around these galaxies, but also the role of scattering.

The paper is organized as follows. Section \ref{sec_methods} introduces the simulations and our modeling methods. We explore the results in Section \ref{sec_results}, covering MgII halo surface brightness profiles in Section \ref{subsec_sb}, halo size in Section \ref{subsec_size}, halo shape in Section \ref{subsec_shape}, and the relationship with gas kinematics, inflows, and outflows in Section \ref{subsec_kinematics}. We discuss our findings and future directions in Section \ref{sec_discussion}, including justification for why neglecting resonant scattering may be a reasonable assumption. We summarize our main results in Section \ref{sec_conclusions}.


\section{Methods} \label{sec_methods}

\subsection{The TNG Simulations} \label{sec_sims}

The IllustrisTNG project\footnote{\url{http://www.tng-project.org}} \citep{pillepich18b, nelson18a, naiman18, marinacci18, springel18} is a series of three large cosmological volumes, simulated with gravo-magnetohydrodynamics (MHD) and incorporating a comprehensive model for galaxy formation physics \citep{weinberger17,pillepich18a}. All aspects of the model, including parameter values and the simulation code, are described in these two methods papers and remain unchanged for our production simulations, and we give here only a brief overview.

The TNG project includes three distinct simulation volumes: TNG50, TNG100, and TNG300. Here we exclusively use the high-resolution TNG50 simulation \citep{pillepich19,nelson19b} which includes 2$\times$2160$^3$ resolution elements in a $\sim$\,50 Mpc (comoving) box. The baryon mass resolution is $8.5 \times 10^4$\msun, the collisionless softening is 0.3 kpc at $z$\,=\,0, and the minimum gas softening is 74 comoving parsecs. For reference, the average size of star-forming gas cells is 130 physical parsecs at $z=0.7$. We adopt a cosmology consistent with  \cite{planck2015_xiii}; $\Omega_{\Lambda,0}=0.6911$, $\Omega_{m,0}=0.3089$, $\Omega_{b,0}=0.0486$, $\sigma_8=0.8159$, $n_s=0.9667$ and $h=0.6774$. Dark matter halos and subhalos (galaxies) are identified with the \textsc{Subfind} algorithm \citep{spr01}.

TNG uses the \textsc{Arepo} code \citep{spr10} which solves for the coupled evolution under (self-)gravity and ideal, continuum MHD \citep{pakmor11,pakmor13}. Gravity uses a Tree-PM approach, while the fluid dynamics employ a Godunov type finite-volume scheme where an unstructured, moving, Voronoi tessellation provides the spatial discretization. The simulations include a physical model for the most important processes relevant for the formation and evolution of galaxies. Specifically: (i) gas radiative processes, including primordial/metal-line cooling and heating from the background radiation field, (ii) star formation in the dense ISM, (iii) stellar population evolution and chemical enrichment following supernovae Ia, II, as well as AGB stars, with individual accounting for the nine elements H, He, C, N, O, Ne, Mg, Si, and Fe, (iv) supernova driven galactic-scale outflows \citep[see][for details]{pillepich18a}, (v) the formation, coalescence, and growth of supermassive blackholes, (vi) local radiation effects from actively accreting black holes, including self-shielding for dense gas, (vii) and blackhole feedback in a thermal `quasar' state at high accretion rates, versus a kinetic `wind' state at low accretion rates \citep{weinberger17}.

The TNG model has been shown to produce galaxy and circumgalactic medium properties in broad agreement with available observational constraints. 
In particular, the properties of the CGM have been compared in terms of their OVI column densities \citep{nelson18b}, gas contents \citep{pillepich18a,terrazas20}, x-ray properties \citep{barnes18,truong20}, outflows and dynamics \citep{nelson19b}, MgII covering fractions around $z \sim 0.5$ luminous red galaxies \citep{nelson20}, and $3 < z < 5$ Lyman-$\alpha$ halos in emission \citep{byrohl20}. The overall realism of these simulations gives us baseline confidence in the outcomes of the TNG model in new and previously unexplored regimes, such as emission from cool gas in the CGM.

\subsection{MgII Line Emissivities and Surface Brightness Maps} \label{sec_emission}

We compute line volume emissivities for UV, optical, and x-ray lines of interest using \textsc{Cloudy} \citep[][v17.00]{ferland17} including collisional and photo-ionization processes and assuming ionization equilibrium in the presence of a UV + X-ray background \citep[the 2011 update of][]{fg09}. All processes are always included regardless of which dominates in any particular regime.\footnote{This UVB includes frequencies from 0.5 to $\sim$4000 Rydberg, and is the same as used in the TNG simulations themselves, in order to be self-consistent. The exact CLOUDY version and configuration does, however, differ slightly between our current runs and the cooling tables of TNG.} We use \textsc{Cloudy} in single-zone mode and iterate to equilibrium, accounting for a frequency dependent shielding from the background radiation field (UVB) at high densities \citep[following][]{bird14,rahmati13}. As gas cells in the simulation are single-zone with no internal structure, it would be inconsistent to assume a multi-zone or more complex geometry in the photoionization calculation. We run \textsc{Cloudy} in the `constant temperature' mode, with no induced processes \citep[following][]{wiersma09} and assuming the solar abundances of \cite{grevesse10}.

The emissivity $\epsilon_{\rm V}$(n$_{\rm H}$,T,z,Z) is tabulated in units of \mbox{[erg cm$^{-3}$ s$^{-1}$]} for each emission line of interest over a 4D grid in (n$_{\rm H}$,\,T,\,Z,\,z), hydrogen number density, temperature, metallicity, and redshift with the following parameter space: $-7.0 < \log(\rm{n}_{\rm H} [\rm{cm}^{-3}]) < 4.0$ with $\Delta$n$_{\rm H}$ = 0.1, $3.0 < \log(\rm{T} [\rm{K}]) < 9.0$ and $\Delta$T = 0.05, $-3.0 < \log(\rm{Z} [\rm{Z}_{\rm sun}]) < 1.0$ and $\Delta$Z = 0.4, and $0 < \rm{z} < 8$ with $\Delta$z = 0.5 (a total of 2.5 million grid points). Dense, star-forming gas cells in the TNG model adopt a two-phase pressurization model \citep{spr03}, and we place these cells at their cold-phase temperature of 1000\,K, rather than adopting an effective weighted temperature, or neglecting them entirely, as the cold phase dominates by mass ($> 90$\%).

To derive the emission for each gas cell in the simulation we then interpolate this table as a function of these four parameters. We then multiply this emissivity by the cell volume and gas-phase species fraction relative to solar to obtain per cell line luminosities [erg s$^{-1}$]. The Mg abundance is taken directly from the TNG simulation which follows its production and mixing \citep{pillepich18a}, such that we do not need to assume solar abundances.

We develop a simple model to account for the effects of dust depletion, whereby some fraction of Mg will be locked into solid dust grains and thus removed from the gas-phase. To do so we take the observed dependence of the dust-to-metal (DTM) ratio on metallicity \citep{peroux20a} as measured for individual DLAs \citep{decia18}, using Si as a proxy for Mg \citep{decia16}. We fit this data to define a Mg-like DTM as a function of ISM metallicity [Fe/H], which we then apply on a gas cell-by-cell basis. For reference, the remaining gas-phase Mg fractions are 0.92, 0.52, and 0.29 at $\log{(Z/Z_\odot)} = -2, -1, 0$, respectively. Beyond dust grain depletion, we do not account for dust in any way, i.e. we neglect it as an additional source of opacity for photons.\footnote{Dust is important during resonant Ly$\alpha$ line transfer, for instance, as a resonant photon trapped in a dusty medium will be strongly extincted relative to a non-resonant photon, due to the much larger path length traveled.}

The luminosity distance $d_{\rm L}$ at a given redshift is used to convert luminosity into a photon flux as $F = L / (4\pi d_{\rm L}^2) \cdot \lambda(1+z) / (hc)$ where $\lambda$ is the rest vacuum wavelength of the emission line. This photon flux is gridded to produce a flux map which is normalized by the solid angle subtended by each pixel $\Omega_{\rm px}$ given the angular diameter distance, resulting in an observable surface brightness. All luminosities, profiles, and maps include gas cells gravitationally bound to the central halo, with no additional line-of-sight cuts imposed nor projection effects considered.

\subsection{MgII Emission Model Assumptions}

Our modeling of MgII emission makes a number of simplifying assumptions. First, $C=1$ (i.e. no sub-resolution clumping factor). Any physical mechanism producing metal emission in gas structures with very small physical scales ($\lesssim $100 pc) would not be resolved in the TNG50 simulation, and not accounted for herein. Emission from highly ionized HII regions close proximity to the strong radiation fields of young stellar populations is thus neglected \citep[see][]{byler17}. This intrinsic ISM emission from MgII could be non-negligible particularly in lower mass galaxies \citep{chisholm20}, and will propagate and scatter in the CGM with some non-zero escape fraction \citep{henry18}, potentially boosting extended emission.

Second, emitted line photons are assumed to propagate through optically thin media, i.e. there is no intervening absorption, either intergalactic or atmospheric, and no scattering. As a result $v_{\rm emit}=0$ i.e. peculiar gas motions are not relevant. Specifically, we neglect the resonant nature of MgII, which results in a complex multi-scattering process similar to the Lyman-$\alpha$ line of hydrogen \citep{prochaska11a}. The optical depths involved are, however, significantly lower: resonantly trapped MgII photons require $\mathcal{O}(1)$ scatterings to escape a CGM environment, whereas Lyman-$\alpha$ photons may require many orders of magnitude more. In the future we will extend the Lyman-$\alpha$ Monte Carlo RT method of \cite{byrohl20} to resonant metal lines, including MgII and FeII.\footnote{FeII is a resonant transition similar to MgII, except that FeII photons can excite, and then escape via, non-resonant FeII$^\star$ emission.} In the current work, however, we focus on the predictions and outcomes of TNG50 intentionally assuming the optically thin case. 

We also clarify that we do not include the stellar continuum at the energies of the MgII transitions, which would source additional photons (like HII regions) available for subsequent scattering. The scattering of \textit{continuum} photons by an outflow is in fact the most common modeling assumption applied to model observed MgII emission halos \citep{rubin11,martin13,rickards19,burchett20}, although some observations favor HII region production as the dominant source of scattered MgII photons \citep{erb12}.

Third, we do not include local radiation sources beyond the UVB or AGN, which could change the ionization balance of Mg close to bright/star-forming galaxies. Past studies have generally concluded that local sources (as well as variation of the UVB) have a small impact on rest-frame UV metal emission lines, as collisional excitation dominates \citep{vdv13,sravan16}. We can also re-evaluate this issue with future methodological advances.

For MgII we always combine the two emission lines at $2796.352$\AA, $2803.532$\AA, treating them as a blend. Unless otherwise noted we always apply a Gaussian point spread function (PSF) with a FWHM of $0.7\ $arcsec (5.1 pkpc at $z=0.7$), which roughly corresponds to that of MUSE (non-AO) UDF data \citep{leclercq17,leclercq20}.


\section{Results} \label{sec_results}

\begin{figure*}
\centering
\includegraphics[angle=0,width=6.8in]{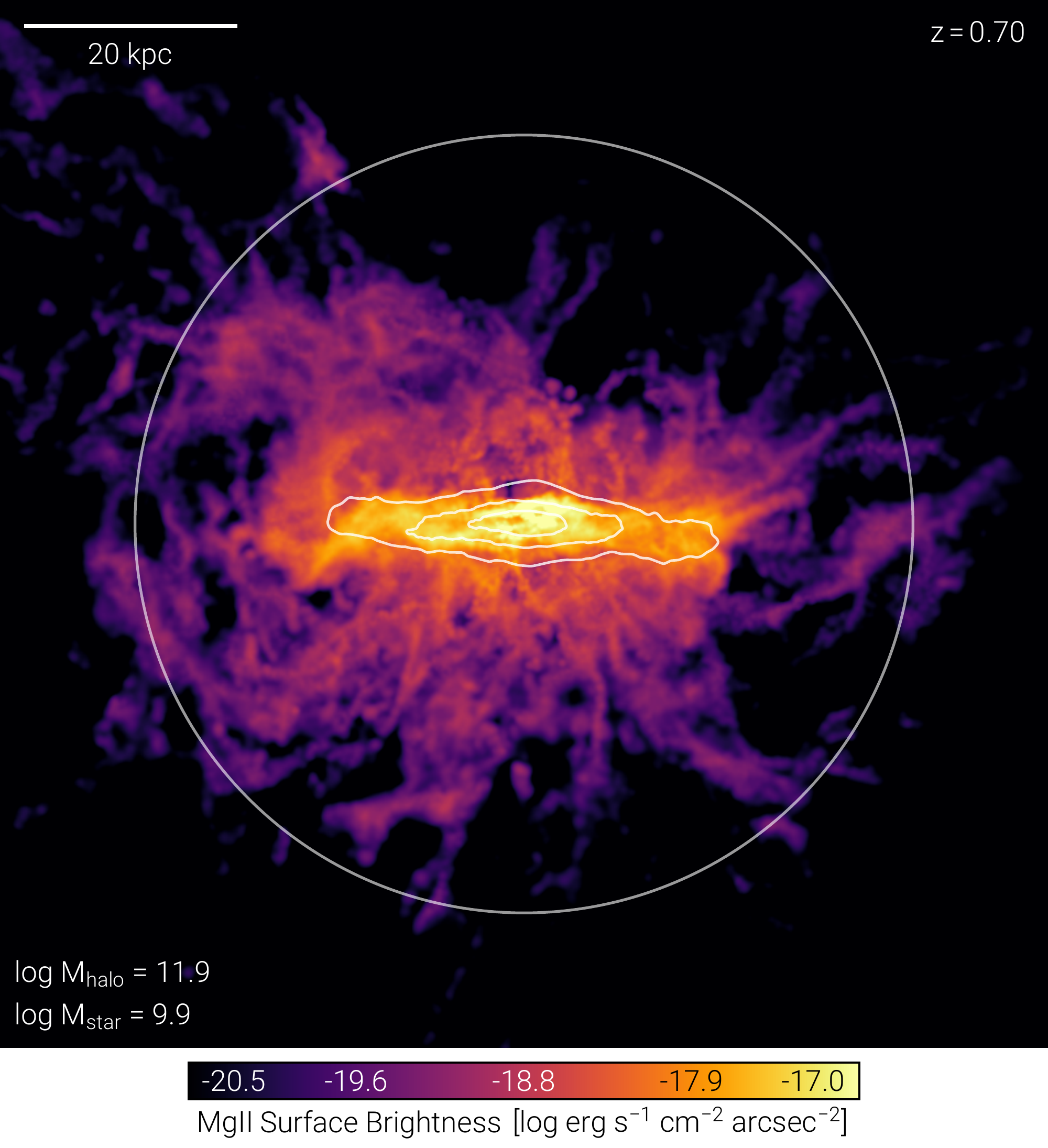}
\caption{ The MgII emission surface brightness map around an example star-forming galaxy (\href{https://www.tng-project.org/api/TNG50-1/snapshots/59/subhalos/396565/}{ID 396565}) at $z=0.7$ with intermediate stellar mass ($M_\star = 10^{10}$\msun) and star formation rate (SFR = 6.1 \msun\,yr$^{-1}$). Shown rotated edge-on, projected 100 pkpc across. The circle shows $5 r_{\rm 1/2,\star}$, while the three white contours enclose stellar mass surface densities of $\log{(\Sigma_\star / \rm{M}_\odot \,\rm{kpc}^{-2})} \in \{7.0, 7.5, 8.0\}$. We see MgII emission extending from the galaxy out into the circumgalactic medium at distances of $\sim 10-20$ kpc, with a complex morphology indicative of galactic fountain flows.
 \label{fig_vis_single}}
\end{figure*}

\begin{figure*}
\centering
\includegraphics[angle=0,width=6.8in]{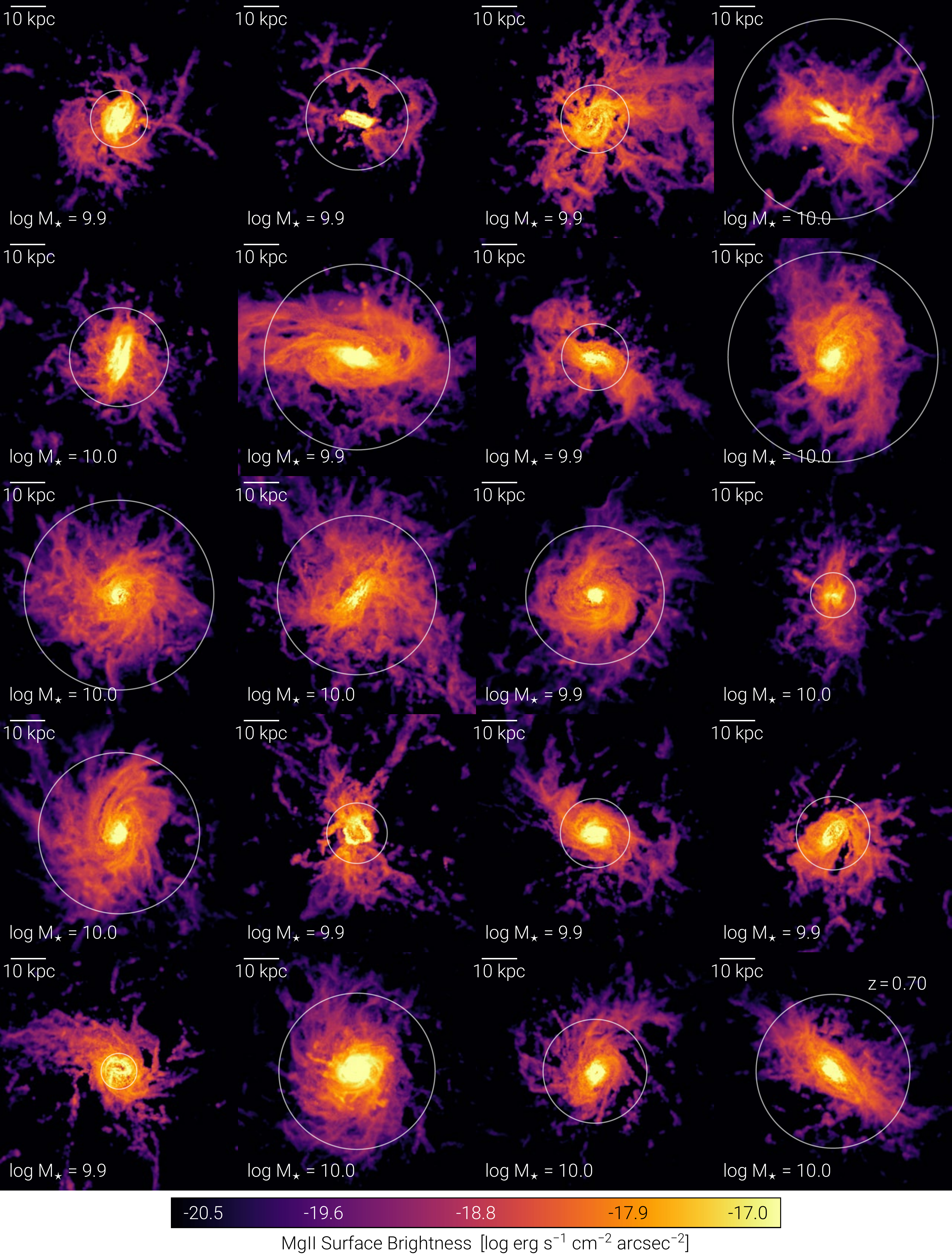}
\caption{ Gallery of circumgalactic MgII emission surface brightness maps around a set of twenty $M_\star \simeq 10^{10}$\msun galaxies. All are shown at $z=0.7$, and each image is 70 pkpc across, and with random orientations/inclinations. A diversity of disk-like, outflow, and inflow morphologies are visible in MgII emission. 
 \label{fig_vis_gallery}}
\end{figure*}

We begin with a visual inspection of the MgII emission from the extended gaseous halo surrounding a typical $z=0.7$ TNG50 galaxy residing on the star-forming main sequence. Figure \ref{fig_vis_single} shows the surface brightness map of emission from the MgII doublet for a galaxy with stellar mass $M_\star = 10^{9.9}$\msun and total halo mass $M_{\rm halo} = 10^{11.9}$\msun, with a star formation rate of $6.1$\msun\,yr$^{-1}$. The galaxy itself is a disk, as is characteristic at this mass and epoch, shown in an edge-on configuration. Contours of stellar mass surface density are shown in white to emphasize this orientation, and the relative directions of the major and minor axes, which extend along the horizontal and vertical directions of the image, respectively.

Tracing cool $\sim$\,10$^4$\,K gas, MgII emission is most luminous in the center of the halo, decreasing towards larger distances. The surface brightness exceeds $\gtrsim 10^{-17}$ erg s$^{-1}$ cm$^{-2}$ arcsec$^{-2}$ in the inner few kpc, $\gtrsim 10^{-18}$ erg s$^{-1}$ cm$^{-2}$ arcsec$^{-2}$ out to $\sim 10$ kpc, and $\gtrsim 10^{-19}$ erg s$^{-1}$ cm$^{-2}$ arcsec$^{-2}$ out to several tens of kpc. Beyond five times the stellar half mass radius ($r_{\rm 1/2,\star}$, marked with the white circle) the MgII emission continues to drop rapidly.

\begin{figure*}
\centering
\includegraphics[angle=0,width=3.4in]{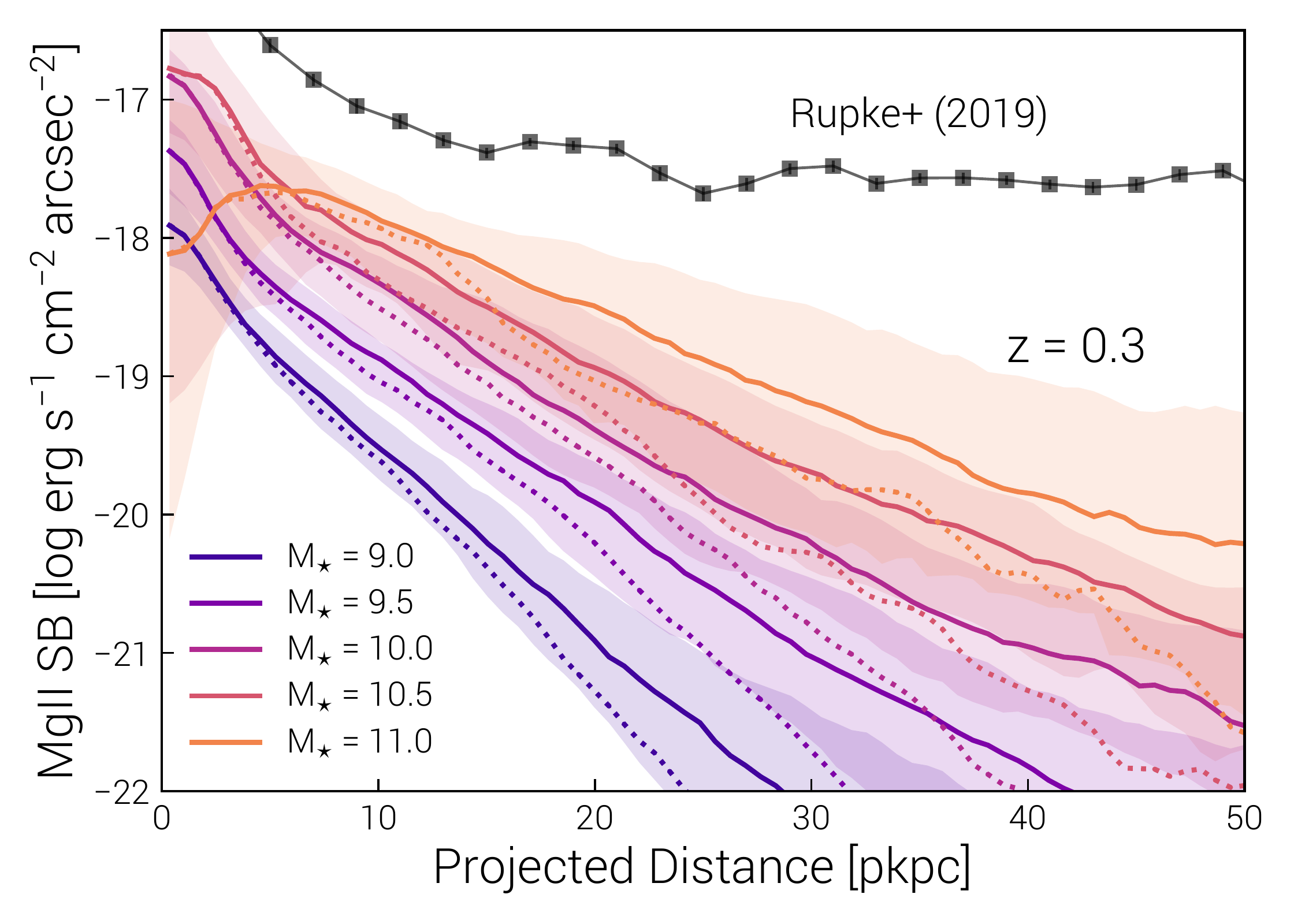}
\includegraphics[angle=0,width=3.4in]{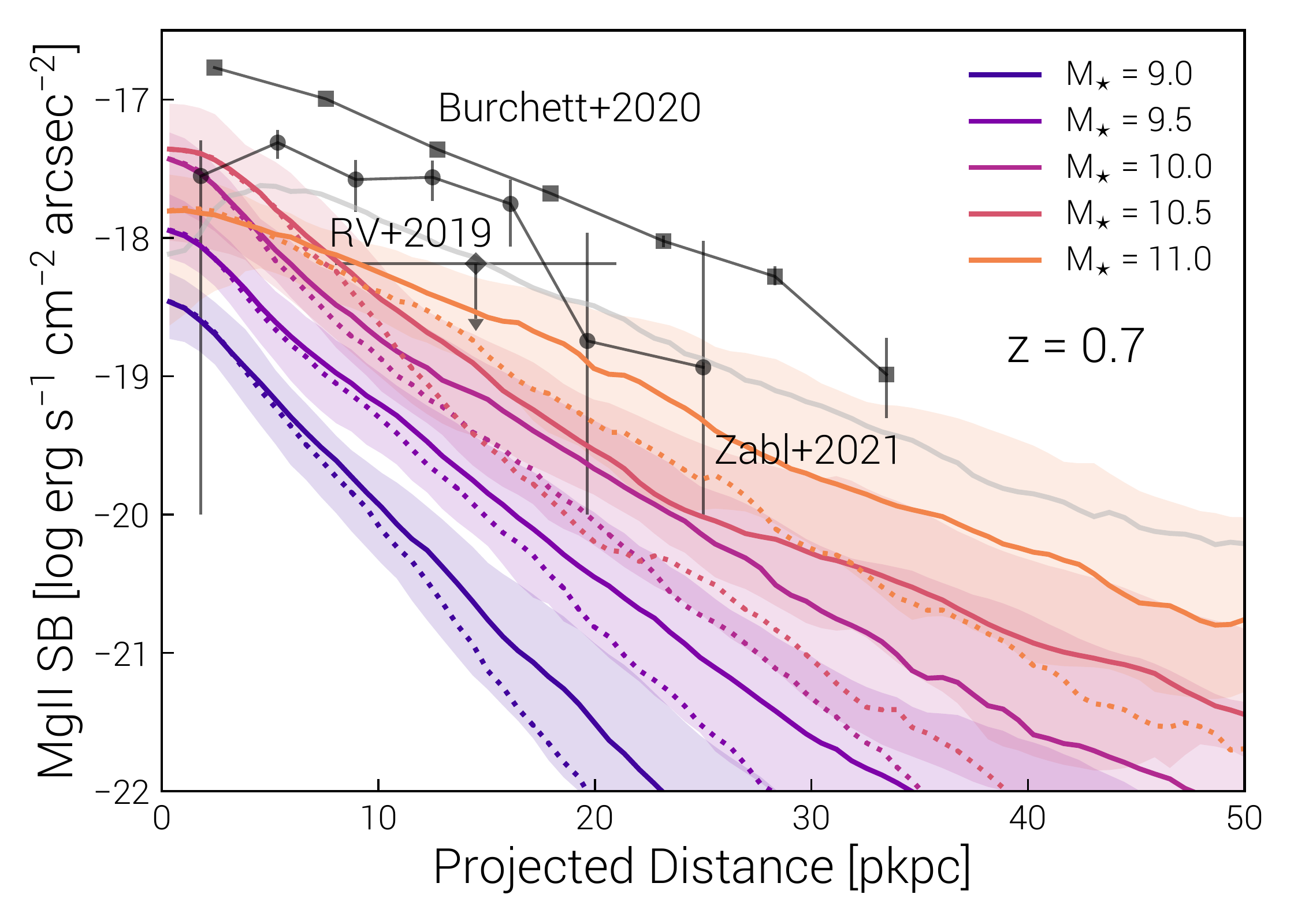}
\includegraphics[angle=0,width=3.4in]{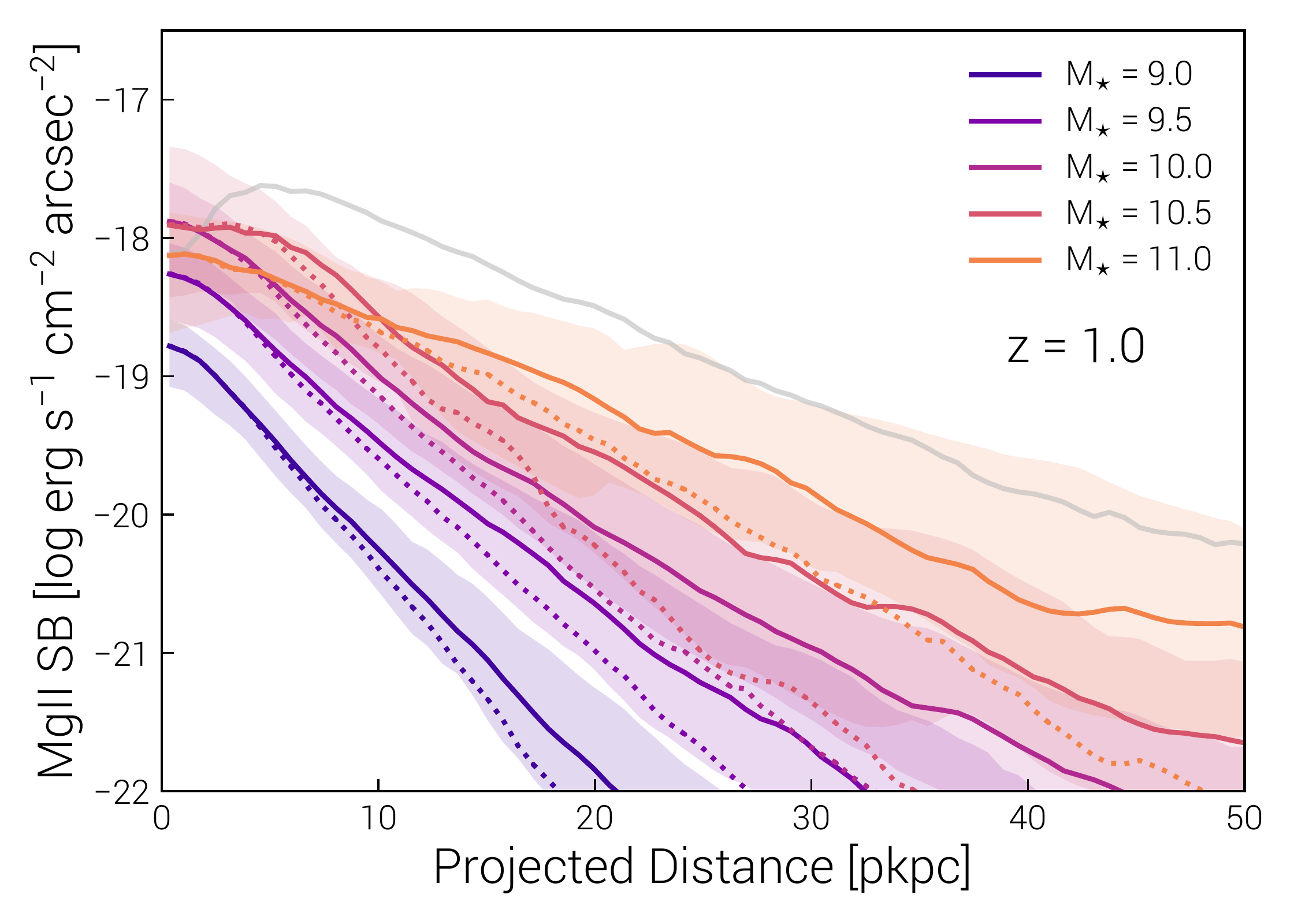}
\includegraphics[angle=0,width=3.4in]{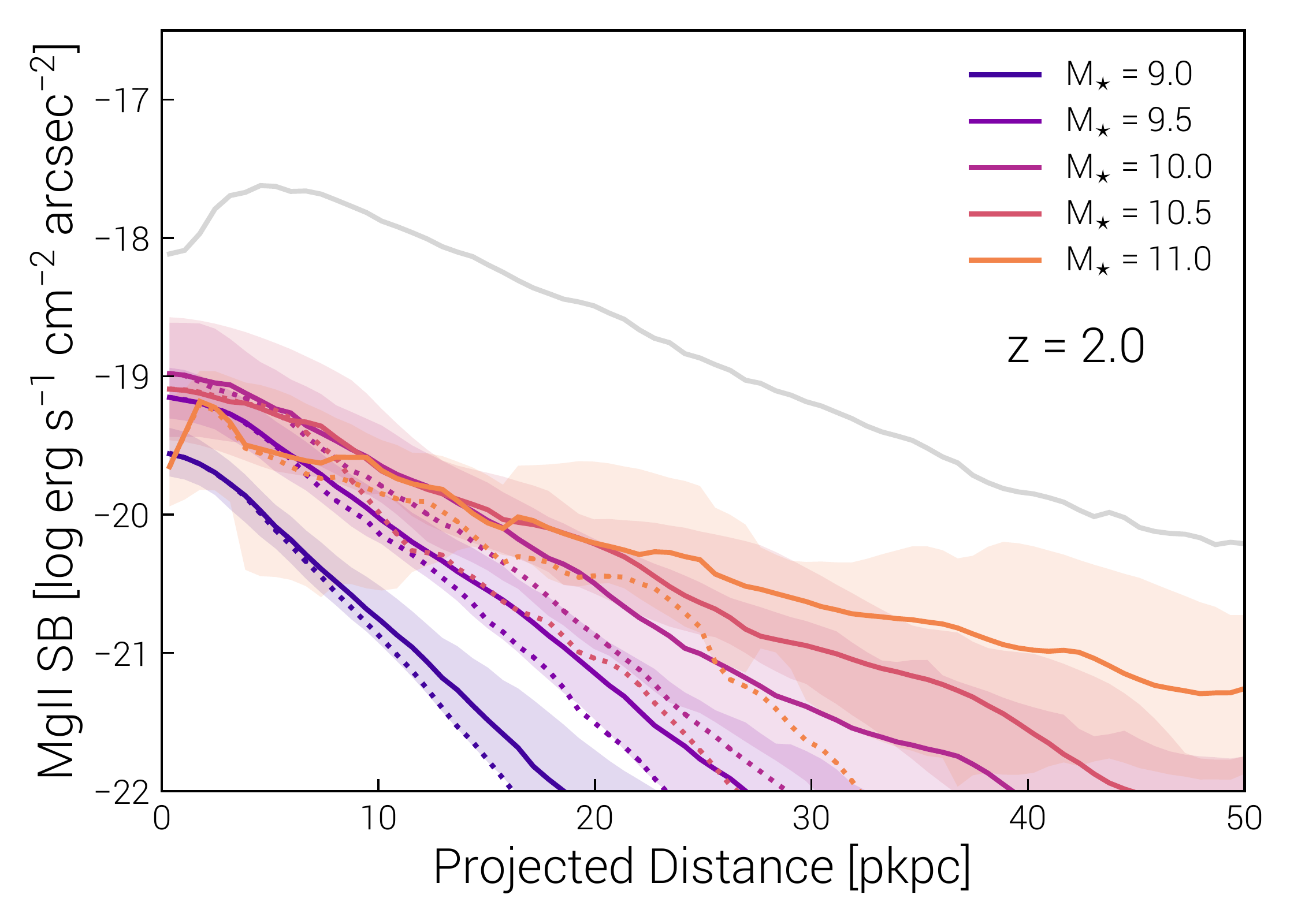}
\caption{ Stacked radial surface brightness profiles of MgII emission at $z= \{0.3, 0.7, 1, 2\}$ (four panels) in bins of stellar mass (line colors). Solid lines show the mean across individual halo profiles, while dotted lines show the median stack. Shaded bands indicate $\pm 1\sigma$ halo-to-halo variation. For the $z \geq 1$ panels, the light gray line shows the lowest redshift profile for the high mass bin, for comparison. Here we include convolution by a Gaussian PSF with FWHM of 0.7'' as appropriate for the MUSE UDF at $z=0.7$, which reduces the overall SB values and flattens profiles (see Appendix). However, we do not treat the resonant scattering of MgII, which would act to further flatten profiles and shift central emission towards halo outskirts. We include three recent observations of extended MgII emission from galactic halos, at their approximate stellar mass and redshifts as appropriate. These are: a $5\sigma$ upper limit \protect\citep{rickards19}, the main galaxy of \protect\cite{zabl21}, the TKRS4389 starburst system \protect\citep{burchett20}, and the Makani major merger superwind \protect\citep{rupke19}; the latter two were specifically targeted and, as expected given that they are most likely atypically bright examples, significantly above any of our median profiles.
 \label{fig_sbr_stacked}}
\end{figure*}

The MgII surface brightness distribution of this galaxy is not azimuthally isotropic, having a preference to be brighter along the major axis of the galaxy. Above and below the disk there are clear signatures of galactic fountain flows: outflows become inflows at $5-10$ kpc distances, as traced by MgII bright, loop-like structures in the gas. These are reminiscent of the observed morphology of extra-planar HI, and idealized models designed to capture such `small-scale fountain flows' across the disk-halo interface \protect\citep{fraternali06,fraternali08}. At larger distances $\gtrsim 20$ kpc there is evidence for gas accretion from larger scales, i.e. from the intergalactic medium, as indicated by infalling gas morphologies towards the upper left.

Galaxies with exactly the same stellar mass exhibit a diversity of MgII halo sizes, shapes, and morphologies. Figure \ref{fig_vis_gallery} presents a gallery of twenty examples, all $M_\star = 10^{10}$\msun systems at $z=0.7$. Here we show each galaxy as viewed from a random orientation, as would be the case in observations, in panels each 70 pkpc across. All twenty halos display extended MgII emission out to tens of kpc, i.e. reaching far beyond the luminous, stellar bodies of these galaxies. As we quantify below, TNG50 forecasts that extended halos of MgII emission are a ubiquitous feature of `normal' galaxies at this epoch. This finding is similar to the observational result that Lyman-alpha halos are a common feature for both normal star-forming and higher mass quasar host halos across $2 \lesssim z \lesssim 5$ \citep{leclercq17,borisova16,cai19}.

The most commonly visible characteristic at this mass scale is MgII emission tracing extended disk-like structures in the gas (cf. second halo, second row), as visible also in MgII absorption \citep{defelippis21,bouche21}. While most halos have a bright central core of MgII emission of just a few kpc in extent but exceeding $\rm{SB} > 10^{-17}$ erg s$^{-1}$ cm$^{-2}$ arcsec$^{-2}$, this is not always the case (cf. third halo, first row). Some halos exhibit clear signatures of fountain flows in MgII emitting gas, with infalling, head-tail morphology gas clumps at the edges of inner cavities blown by galactic-scale outflows \citep[cf. second halo, first row, and][]{nelson19b}. Some halos have filamentary structure extending into the outer halo (cf. second halo, fourth row). Several have a curious X-shaped, central structure in MgII surface brightness (cf. fourth halo, first row). All share one detail in common: MgII emission arising from a complex morphology of small-scale ($\sim$ kpc), inhomogenously distributed, clumpy gas.

\subsection{Surface Brightness Profiles} \label{subsec_sb}

Maps of spatially resolved MgII emission are naturally characterized, to zeroth order, by radial surface brightness profiles. Figure \ref{fig_sbr_stacked} shows the stacked MgII emission profiles from TNG50 halos. These are a strong function of galaxy stellar mass, where we include results from $M_\star = 10^9$\msun to $M_\star = 10^{11}$\msun in different line colors (solid denoting means, dotted medians), with shaded bands indicating $\pm 1\sigma$ halo-to-halo variation. The four different panels show independent results at the four redshifts of relevance for observational MgII halo detection: $z=0.3$ (KCWI and future BlueMUSE) as well as $z=0.7$, $z=1$, and $z=2$. To aid comparison between these different epochs, the lowest redshift profile for the highest mass bin is repeated in the three other panels as the faint gray line. In all cases we include convolution by the fiducial PSF with a FWHM of 0.7'', which reduces central luminosities and flattens profiles, making them more extended and shallower at large radii (see comparison Figure \ref{fig_appendix_psf}).

TNG50 predicts that the circumgalactic medium of galaxies produces a MgII surface brightness which is a steep function of galaxy mass and redshift, rising for more massive halos (at fixed redshift) and towards redshift zero (at fixed mass). At $z=0.7$ surface brightness values exceeding $10^{-18}$ erg s$^{-1}$ cm$^{-2}$ arcsec$^{-2}$ occur only in the centers of halos, at projected distances of $\lesssim$ 10 kpc, while values ten times fainter ($10^{-19}$ erg s$^{-1}$ cm$^{-2}$ arcsec$^{-2}$) reach out to $\sim 20$ kpc. These represent the rough limits of current observational sensitivity, at least for individual objects. This distance corresponds to only $r_{\rm vir}/10$ for $M_\star \simeq 10^{10}$\msun at $z=0.7$, i.e. such probes are currently confined to gas in the inner halo. Nonetheless we find a clear expectation that MgII halos, in the average, are smoothly declining and extend far beyond the luminous body of the galaxy itself.

Comparison of these predicted profiles with results from blind surveys where significant numbers of galaxies can be stacked, i.e. in the MUSE UDF data, will give us a first impression of the circumgalactic MgII expected from cosmological galaxy formation simulations. 

At present we include a number of recent detections of MgII halos. In particular, these are an upper limit from VLT/FORS2 narrow band imaging spanning $10 < \log{(M_\star / \rm{M}_\odot)} < 11$ is shown at $z=0.7$ \protect\citep{rickards19}. We also show the KCWI measurement of the surface brightness profile of TKRS4389, a $z = 0.7$ galaxy with $M_\star \simeq 10^{10}$\msun undergoing a starburst of \mbox{$\rm{SFR} \sim 50$\msun\,yr$^{-1}$}, specifically targeted as a known source of bright MgII emission \protect\citep{burchett20}. Finally, we include data for the Makani galaxy at $z=0.5$ with $M_\star \sim 10^{11}$\msun, which is an ultraluminous system driving an intense wind \protect\citep{rupke19}. Our results include all galaxies and are not restricted for instance to only star-forming systems, such that the highest mass bins will be dominated by passive galaxies \citep{donnari20a}.

The current detections of extended MgII emission are consistent with the idea that only very luminous galaxies can produce observable, resonantly scattered halos \citep{martin13}. Furthermore, the galaxies targeted in these observations are drawn from large parent samples. In particular, the TKRS4389 system of \cite{rubin11} and \cite{burchett20} is one of the brightest galaxies in the entire TKRS survey of GOODS-N. Similarly, galaxy 32016857 of \cite{martin13} is the only system from 145 available Keck/LRIS spectra with MgII coverage (drawn from DEEP2) which shows clear MgII emission more extended than the stellar continuum. The non-detections for the sample of five galaxies in \cite{rickards19}, which have more main-sequence star formation rates, further supports this picture. Modeling the scattering of MgII photons in the most (over)luminous TNG50 galaxies will thus be an interesting avenue for future work, even if they may not represent the typical behavior of the galaxy population as a whole. Scattered photons originating from either stellar continuum or dense HII regions would both represent an additional boost to the surface brightness profiles, which would presumably enable a subset of the highest star-forming TNG50 galaxies to reach levels as high as TKRS4389, for example.

To understand if our most MgII luminous halos arise due to mergers and increased star formation activity, we select galaxies similar to Makani, with $10.8 < \log(M_\star/\rm{M}_\odot) < 11.2$ and $\log(L_{\rm MgII}) > 41.2$, resulting in four systems. Two of these have had recent major mergers (stellar mass ratios $\gtrsim 1/2$) within the past few hundred Myr, while the other two have not. This somewhat ambiguous result clearly motivates a more detailed study of the relationship between the most (over)luminous MgII halos and merger history/dynamical state.

As the volume of TNG50 is relatively small, we also investigate whether the TNG300 simulation, with two hundred times larger volume, more fully populates the tails of the scatter of this $L_{\rm MgII}-M_\star$ relation. Despite the lower numerical resolution, which tends to hinder extreme starburst events, we find that TNG300 does in fact produce many galaxies with higher MgII luminosities than found in TNG50, at all masses. These remain outliers, i.e. above the median relation, but we note that the increased statistics produce tens of simulated galaxies consistent with each of the observed systems and their inferred uncertainties (not shown).

\begin{figure}
\centering
\includegraphics[angle=0,width=3.4in]{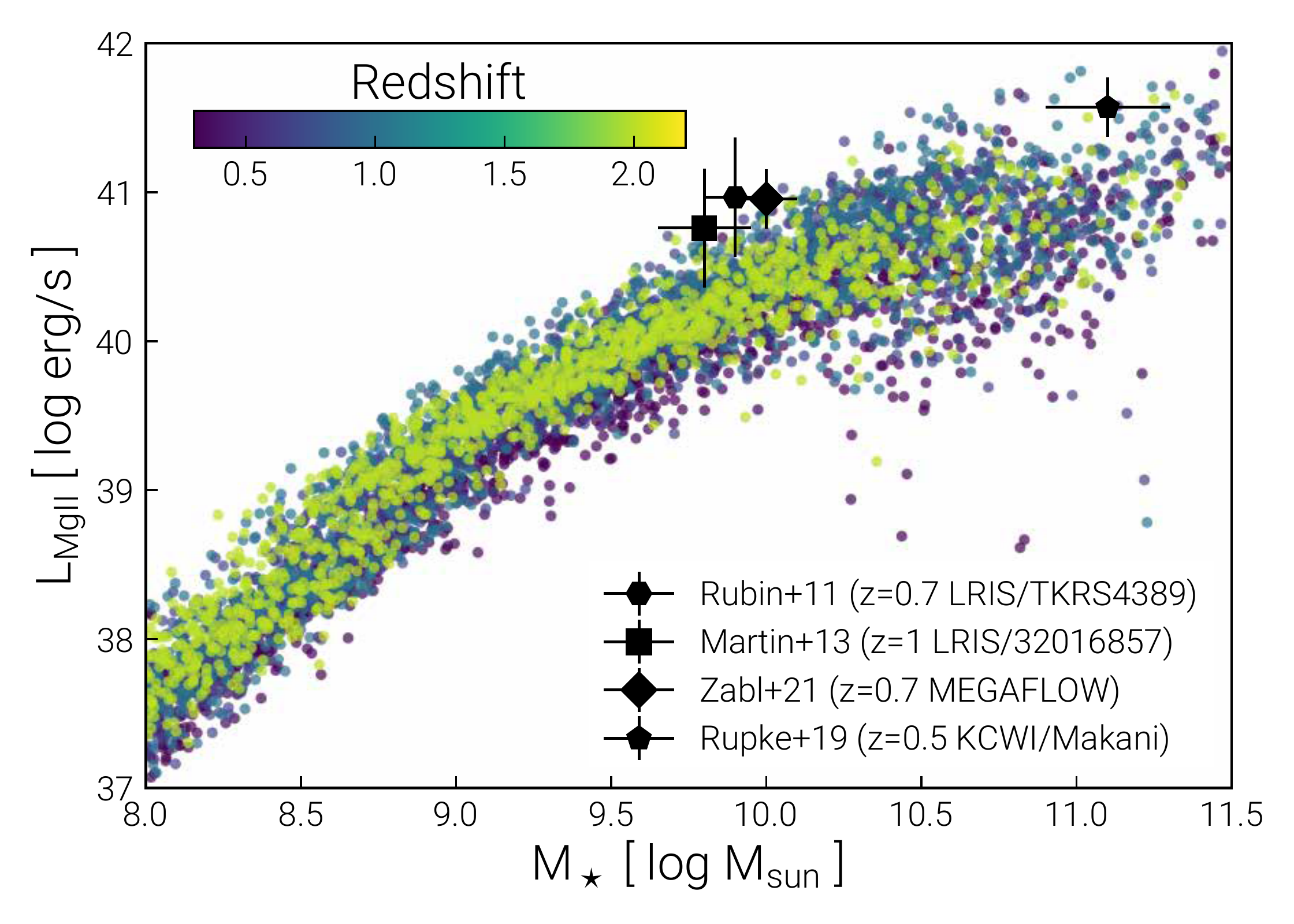}
\caption{ Total luminosity of MgII emission from the halo, as a function of galaxy stellar mass, at $z \in \{0.7, 1, 1.5, 2\}$, where individual colored points show TNG50 halos. At fixed stellar mass there is no significant redshift evolution of $L_{\rm MgII}$. We show face-value comparisons with the limited observational inferences \citep{zabl21}, $z=0.7$, SFMS, from the MEGAFLOW survey; \protect\cite{martin13}, $z=1$, high \mbox{$\rm{SFR}=60$\msun\,yr$^{-1}$}, from Keck/LRIS slit spectroscopy; \protect\cite{rubin11}, $z=0.7$, a similarly star-bursting \mbox{$\rm{SFR} \sim 50$\msun\,yr$^{-1}$} galaxy; \protect\cite{rupke19}, $z=0.5$, the massive Makani galaxy with $\gtrsim 100$ kpc extent emission in OII, observed with KCWI).
 \label{fig_lum_vs_mstar}}
\end{figure}

Integrating the total emission in these profiles across all gas in the halo we obtain the halo MgII luminosity, which we plot in Figure \ref{fig_lum_vs_mstar} as a function of galaxy stellar mass. Four different redshifts ($z=0.7, 1, 1.5, 2$) are overplotted with different color symbols -- we find no appreciable evolution of $L_{\rm MgII}$ with redshift at fixed stellar mass. MgII luminosity is, however, a strong function of mass, rising from an average of $\sim 10^{38}$ erg s$^{-1}$ for $M_\star = 10^8$\msun to $\sim 10^{39}$ erg s$^{-1}$ at $M_\star = 10^9$\msun and reaching $\sim 10^{40}$ erg s$^{-1}$ by $M_\star = 10^{10}$\msun. This roughly linear relation for galaxies on the star-forming main sequence begins to flatten towards even higher masses, as the cold gaseous reservoir of the central galaxy is disrupted and the hot gaseous halo becomes increasing inhospitable to cold MgII bearing gas. For example, \cite{nelson19b} shows that the trend of total MgII mass as a function of halo mass similarly begins to flatten above $M_{\rm halo} \gtrsim 10^{12}$\msun, in contrast to the total Mg mass and the total mass in metals, both of which continue to rise.

Figure \ref{fig_lum_vs_mstar} presents a theoretical measurement of $L_{\rm MgII}$, defined as the sum of the intrinsic luminosity of all gas cells gravitationally bound to the central galaxy of each halo. This excludes not only line-of-sight contributions beyond the halo virial radius, but also the contribution of any MgII emitting gas still bound to a satellite galaxy, which may become increasingly relevant at high mass. Nonetheless, we show a few `face-value' comparisons with observations of MgII halo luminosity \citep{rubin11,martin13,rupke19}. We note as before that these systems are some of the brightest known sources, often starbursts and/or starbursting major mergers, specifically targeted as such. In terms of total $L_{\rm MgII}$ we would then reasonably expect them to probe the high scatter end of the distribution at their stellar masses. We also include an early measurement of a main-sequence system from the MEGAFLOW survey which should be more characteristic of the average galaxy (\textcolor{blue}{Zahl et al. in prep}).

Note that of the $\sim 4,600$ galaxies included in this figure, only a handful $< 1$\% have low values of $L_{\rm MgII}$ which place them off the bottom of the panel. That is, essentially every central galaxy in this mass range at $z=0.7$ hosts a gaseous halo which has a non-negligible MgII luminosity. As we show below, this emission is also spatially extended, i.e. moreso than the stellar bodies of galaxies. Thus we conclude that extended MgII halos are, as now commonly appreciated to be true also for extended Ly$\alpha$ halos, a ubiquitous feature of essentially all galaxies. 

\subsection{Spatial Extent of MgII Halos} \label{subsec_size}

Galaxy (or halo) size, and galaxy (or halo) mass, are strongly correlated: more massive galaxies are larger, in terms of the extent of their stellar bodies, gaseous disks, and galactic atmospheres.

\begin{figure*}
\centering
\includegraphics[angle=0,width=6.2in]{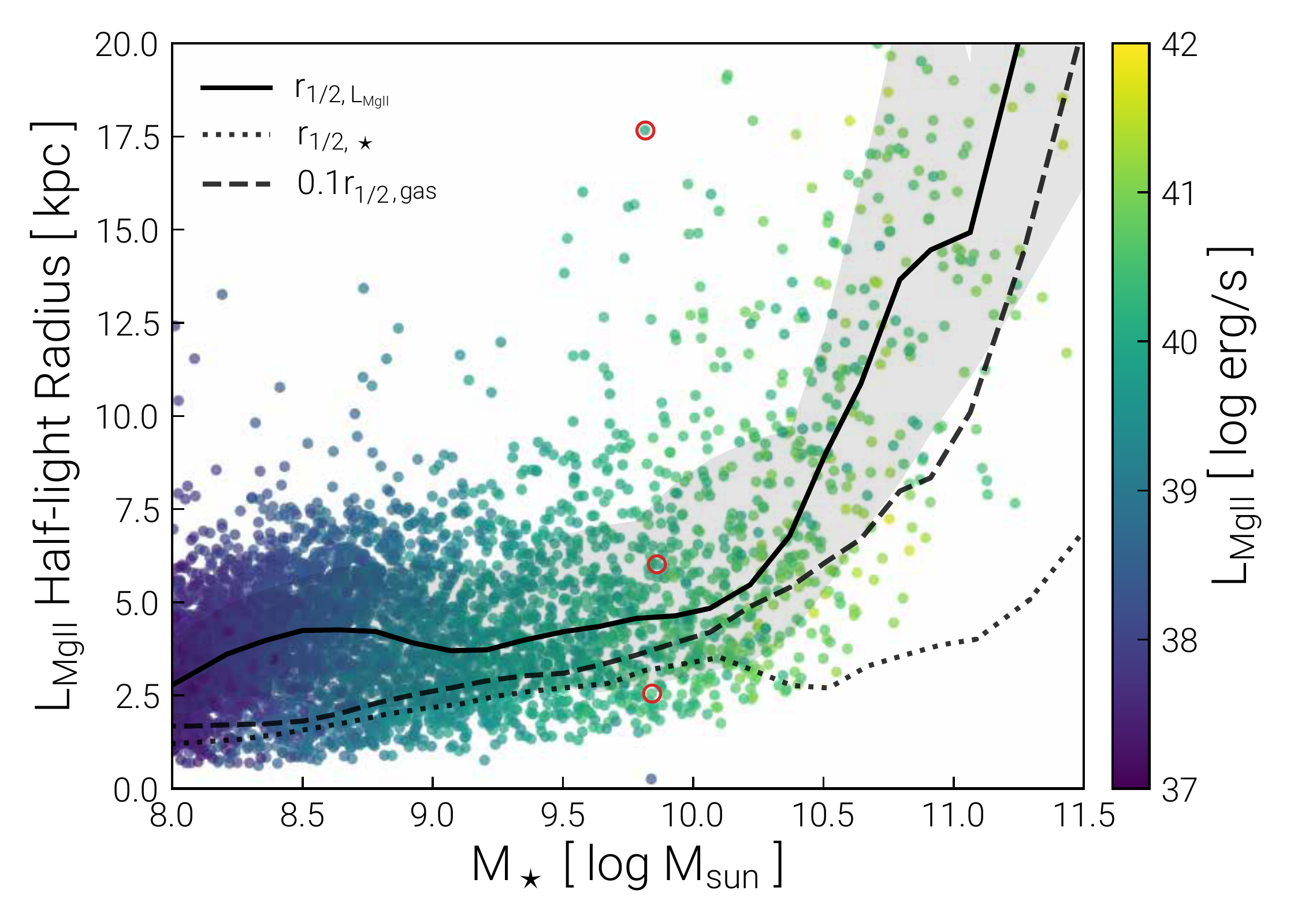}
\includegraphics[angle=0,width=2.25in]{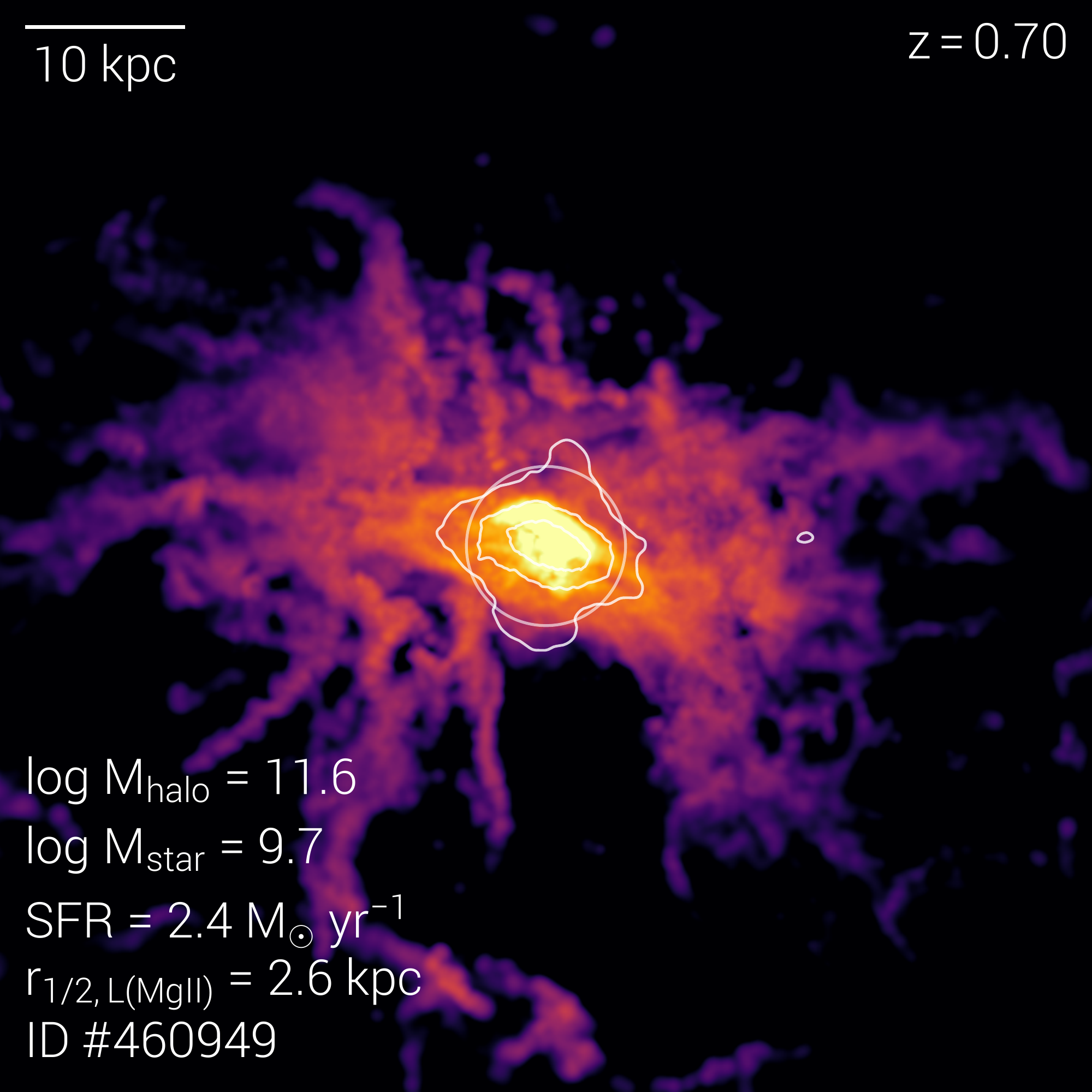}
\includegraphics[angle=0,width=2.25in]{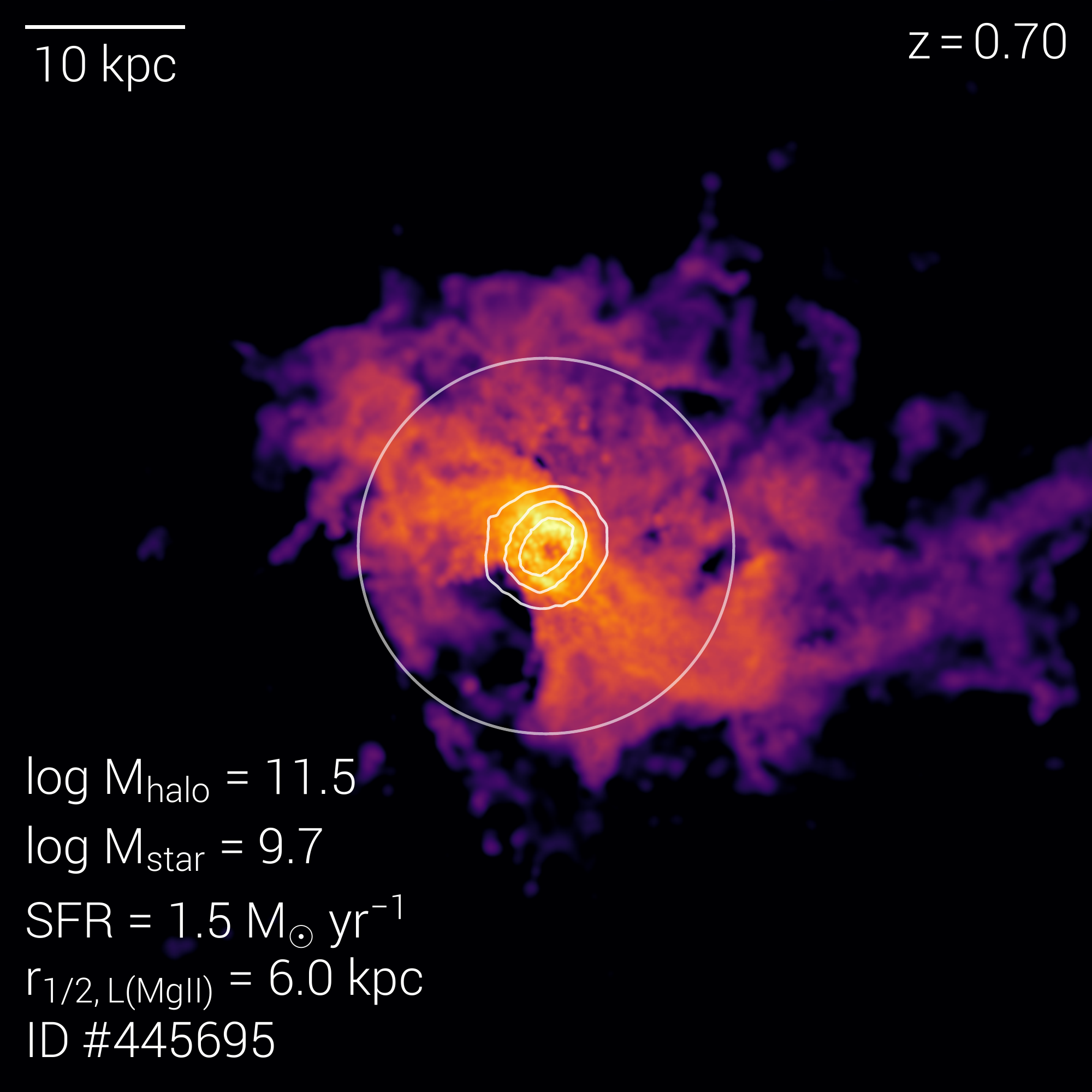}
\includegraphics[angle=0,width=2.25in]{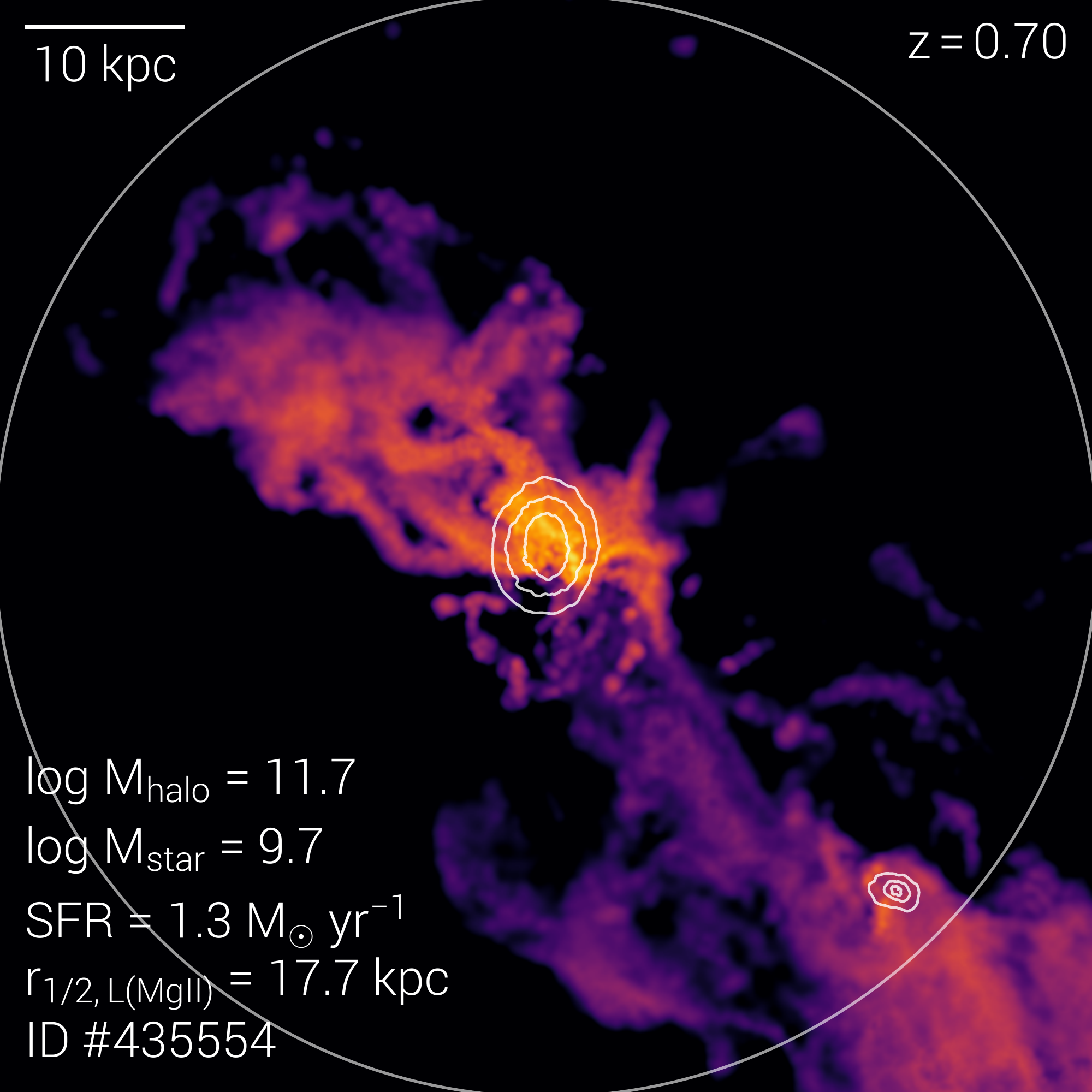}
\caption{ The half-light radius (extent) of the MgII halo in emission, as a function of galaxy stellar mass (top panel). Individual colored circles mark individual TNG50 halos, color indicating total MgII luminosity. The solid black line shows the median relation, and gray band the $\pm 1\sigma$ scatter, which is compared to the stellar sizes of galaxies (half mass radii, dotted line) and the size of the halo gas as a whole (gas half mass radii, scaled down by a factor of ten, for comparison). All are circular, rather than e.g. elliptical, radii. Overall $r_{\rm 1/2,L(MgII)}$ is a few kpc at $M_\star \lesssim 10^{10}$\msun, after which it begins to rise, following the characteristic trend of the total halo gas extent and in contrast to the shallower increase of the stellar effective radii. In red three particular halos are marked: with small, medium, and large MgII halos, respectively. The lower panels show images of these three systems, from left to right, where the white circles denote twice the MgII half light radii, and the white contours trace stellar mass surface density, with the same conventions and color ranges as previously. Note that we calculate half light radii including the impact of the PSF, but show here intrinsic, unsmoothed images. The corresponding smoothed images are shown for comparison in Figure \ref{fig_appendix_psf_images}.
 \label{fig_size}}
\end{figure*}

In Figure \ref{fig_size} we measure the spatial size of MgII halos in emission as a function of galaxy stellar mass (main panel). For sizes we adopt the circular half-light radii, computed in 3D for all gas in each subhalo. We show all galaxies in TNG50 at $z=0.7$ as individual circles, coloring by total MgII luminosity. The median relation of $r_{\rm 1/2,MgII} - M_\star$ is shown with the solid black line; the shaded gray band indicates the $\pm 1\sigma$ scatter. We find that the median MgII halo size monotonically increases with stellar mass, from a few kpc at $M_\star \lesssim 10^{10}$\msun to tens of kpc at $M_\star \gtrsim 10^{10.5}$\msun. This general shape is reminiscent of the $z=0$ galaxy stellar size-mass relation, which is nearly flat and only gradually increasing at low mass, until high-mass quenched galaxies begin to dominate with a much steeper relation at $M_\star \gtrsim 10^{10}$\msun. For comparison we also include the half-mass radii of the stellar component of galaxies (dotted line), and the half-mass radii of the entire gaseous component of the dark matter halo (dashed line) scaled to 10\% to fit on the same scale. At all masses MgII emission is more extended than the stellar continuum, and this is particularly the case at the high mass end.

In order to better understand the significant scatter in MgII half-light radii, we have marked three individual galaxies in the main panel (red circles), all of which have essentially the same stellar mass of $M_\star \sim 10^{9.8}$\msun. These three MgII halos are shown in the bottom three images, from smallest to largest (left to right). The halo mass, stellar mass, SFR, half-light radius, and ID of each are labeled. Despite having the same stellar mass, the first galaxy (left) has an extremely small MgII half-light radius due to a peak of central brightness, although lower SB features extend to similar distance as the second galaxy (middle). The more than twice as large $r_{\rm 1/2,L(MgII)}$ of the third galaxy (right) is due to an ongoing interaction with another galaxy, which appears in the stellar contours towards the lower left. This highlights the role of mergers and environment in influencing MgII halo properties, which we explore below.

\begin{figure}
\centering
\includegraphics[angle=0,width=3.3in]{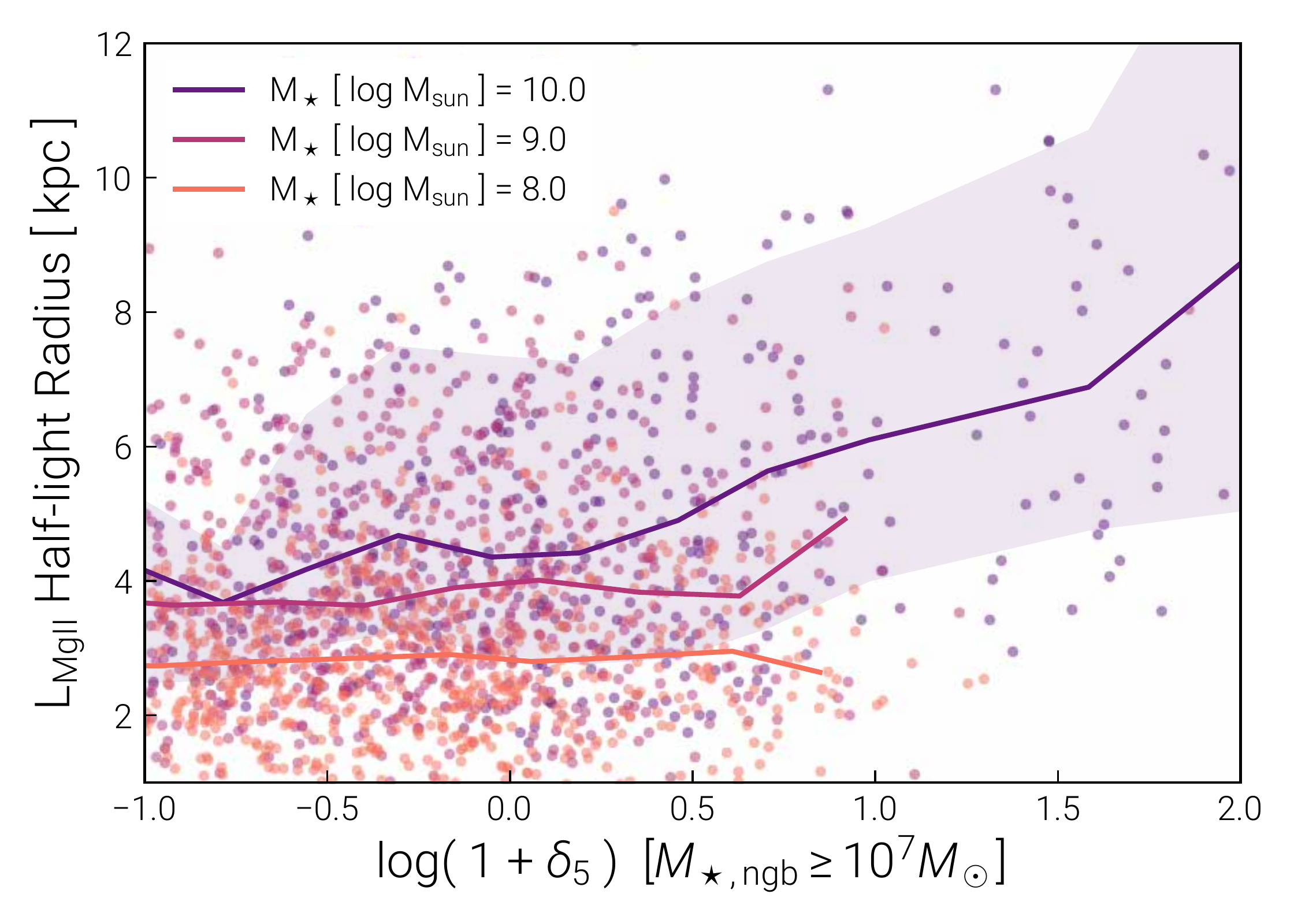}
\caption{ Dependence of MgII halo extent on environment: the half-light radius as a function of the $\log{(1+\delta_5)}$ measure of local galaxy overdensity, based on neighbors with $M_\star \geq 10^7$\msun. We show only centrals galaxies at $z=0.7$, and the three colored lines show three central stellar mass bins, shaded bands indicating $\pm 1\sigma$ scatter. For the lower two mass bins ($M_\star \lesssim 10^9$\msun) we find no strong dependence of $r_{\rm 1/2,MgII}$ on $\delta_5$. In contrast, more massive galaxies ($M_\star \sim 10^{10}$\msun) reach larger overdensities, where a clear trend emerges: MgII halos are more extended for galaxies in overdense environments, i.e. those which are surrounded by relatively many nearby neighboring galaxies.
 \label{fig_env}}
\end{figure}

We examine whether galaxy environment can influence the properties of MgII halos in emission. Namely, their total luminosity, size, shape. We quantify environment by the local galaxy overdensity $(1 + \delta_{\rm N})$, where $\delta_{\rm N} = \rho_{\rm N} / \bar{\rho}_{\rm N} - 1$ and the density $\rho_{\rm N} = N / (4 \pi R_{\rm N}^3/3)$ is the number of neighbors normalized by their enclosing volume. We adopt a definition based on $N=5$, the distance to the fifth nearest neighbor \citep[e.g.][]{kovac10}, above a threshold $M_\star \geq 10^7$\msun, of these three MgII halo properties on $\delta_5$.

Figure \ref{fig_env} shows the relation between MgII half-light radius and the $\log{(1+\delta_5)}$ local galaxy overdensity measure, showing centrals galaxies at $z=0.7$, for three different stellar mass bins of $M_\star = 10^{\{8,9,10\}}$\msun. For the last case the shaded band indicates the $\pm 1\sigma$ scatter, and individual points are included. The x-axis implies that systems to the left of this figure live in underdense environments (are surrounded by relatively few other galaxies), while systems to the right live in overdense environments. The highest stellar mass bin reveals a clear trend: galaxies in overdense environments host preferentially more extended MgII halos in emission. On the other hand, lower mass galaxies do not sample such dense environments, as indicated by the median lines which terminate when bins no longer contain more than ten galaxies, resulting in no clear trend of $r_{\rm 1/2,MgII}$ with $\delta_5$ for $M_\star \leq 10^9$\msun.

For the high stellar mass galaxies the trend in the median is mild, roughly a factor of two larger size at fixed stellar mass across 2 dex in overdensity. At the same time, the scatter also increases significantly, particularly towards large value of $r_{\rm 1/2,MgII}$, suggesting that high outliers are exactly those galaxies lying in the most overdense environments. This corresponds qualitatively to our three examples from Figure \ref{fig_size}, where we saw that a large outlier on the MgII halo size-stellar mass ($R_{\rm MgII}-M_\star$) relation featured an ongoing interaction with a second, nearby satellite. Our visual inspection suggests that this inflates the half-light radii (a) in part due to the contribution of the dense MgII emitting gas within the body of the satellite galaxy itself, but also (b) because of more widespread cold gas in the vicinity, either stripped from the satellites or generally distributed due to interactions in the relatively denser environment \citep{burchett18}. Note that for the galaxies in the high mass bin, a value of $\delta_5 = 10$ corresponds to an enclosing radius of $\sim 200-250$\,kpc, roughly $\sim 2 r_{\rm vir}$.

We do not find any significant trends of $L_{\rm MgII}$ or shape (axis ratio) on $\delta_5$, nor other proxies for local environment including neighbor counts within various different mass ranges and distances. We also measure three non-parametric structural parameters of MgII emission: the Gini, M20, and concentration indices \citep{rodriguezgomez19}, and find no significant trends with $\delta_5$ at $z=0.7$ for MUSE UDF-like maps. Taking advantage of the information content of spatially resolved imaging, beyond one-dimensional profiles and size measures, will clearly benefit from carefully designed statistics and new methods \citep[e.g.][]{chen21}.

We also consider whether recent merger activity, rather than contemporaneous neighbors, influence MgII halos. To do so we look for trends with the number of major (stellar mass ratio $\mu > 1/4$) and minor ($1/10 < \mu < 1/4$) mergers, in the past Gyr, or for all time. At fixed stellar mass, we also find no significant trends of MgII halo shape (axis ratio), size (half light radius), or total luminosity on the number of major nor minor mergers, over either time period. On the other hand, if we do not fix stellar mass, the total $L_{\rm MgII}$, for example, naturally increases with the number of recent mergers, as well as with $\delta_5$, but this predominantly reflects the underlying mass trends.

\subsection{MgII Halo Shape} \label{subsec_shape}

\begin{figure}
\centering
\includegraphics[angle=0,width=3.4in]{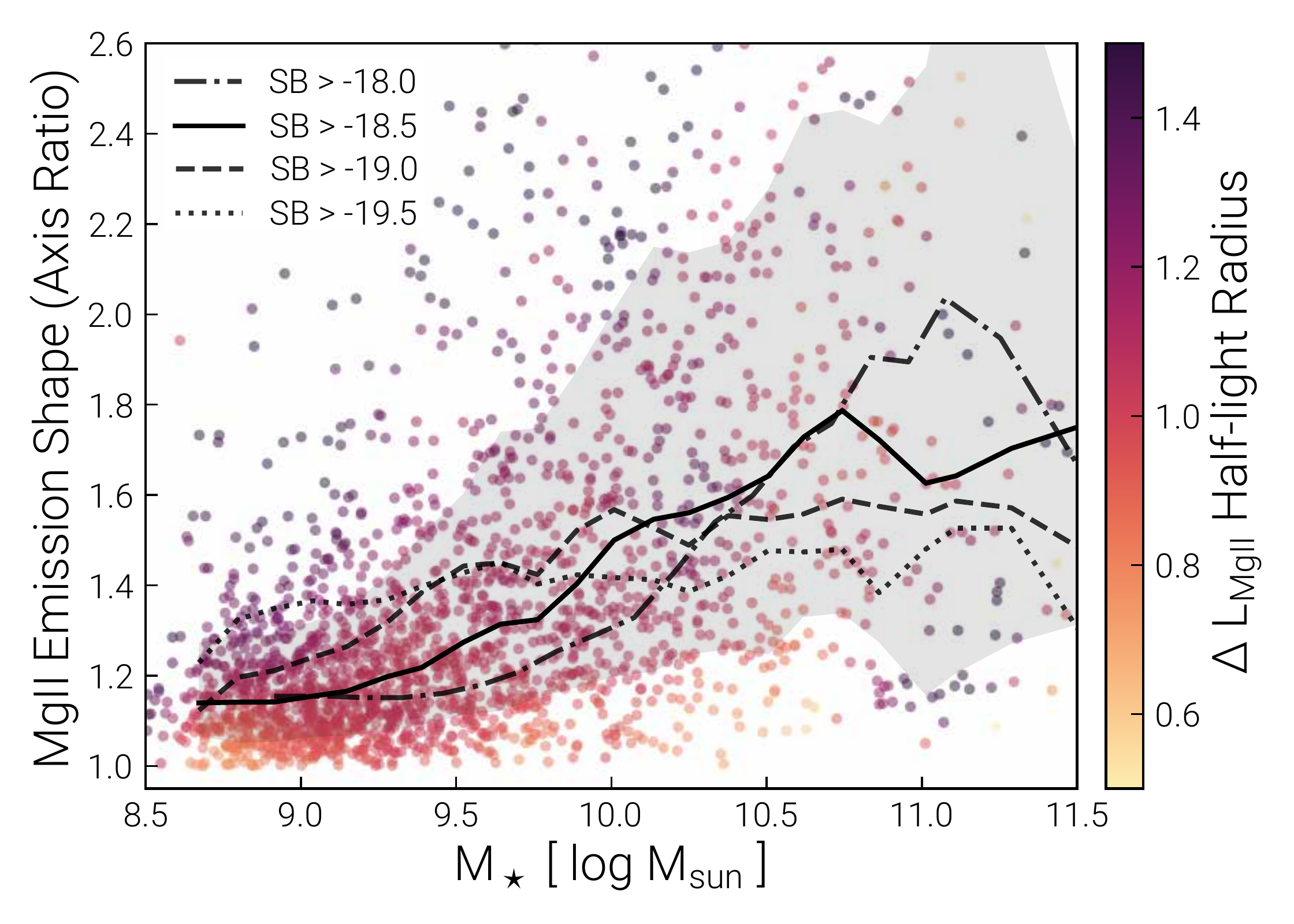}
\caption{ Shape of MgII halos in emission, measured through the axis ratio $(a/b)$ of an ellipse fit to a given isophotal contour of the surface brightness image. Each galaxy/halo is viewed from a random direction. Four different SB thresholds are considered, as indicated in the legend, $\{-18, -18.5, -19, -19.5\}$ in the usual units of log erg s$^{-1}$ cm$^{-2}$ arcsec$^{-2}$. For each the median relation shows that MgII halos tend to become less circular $(a/b \sim 1)$ with increasing galaxy stellar mass, reaching $a/b \sim 1.5$ and above for $M_\star \gtrsim 10^{10}$\msun. The trend of increasing asymmetry with mass is largest for the brightest isophotes, i.e. the inner emission. The scatter is significant, with the gray band indicating 16-84 percentiles for the $10^{-18.5}$ case, where we also show individual galaxies as colored points. The color indicates relative ($\Delta$) MgII half-light radius, with respect to the median at that $M_\star$, which is larger for more asymmetric halos at fixed mass.
 \label{fig_shape}}
\end{figure}

The findings thus far suggest that not only the extent, but also the shape, of the MgII halo in emission could encode physical information. In general, halos tend to be highly asymmetric when observed in projection. To quantify shapes we start with projected surface brightness maps from random viewing angles, apply the representative PSF, and then fit the minimum volume bounding ellipse (MVBE) to the isophotal contour at a given surface brightness value. We adopt the axis ratio $(a/b)$ of the major to minor axes as our shape measure, a value of one indicating a circular halo, and values greater than 1.5 indicating highly elongated emission. This procedure is designed to be as simple as possible and fully compatible with an application to real observational data.

Figure \ref{fig_shape} shows the shape of MgII halos as a function of galaxy stellar mass at $z=0.7$, for four different surface brightness isophotes (different line styles). We find a clear trend that emission from lower mass galaxies is more circular, becoming progressively more asymmetric towards higher mass. The median value increases from $\sim 1.2$ to $\sim 1.8$ at $10^{-18.5}$ erg s$^{-1}$ cm$^{-2}$ arcsec$^{-2}$. The mass trend is strongest for the highest SB values, i.e. for the inner bright portion of a MgII halo, and weaker for lower SB values, i.e. for the extended outskirts. For example, at the lowest surface brightness value explored here ($10^{-19.5}$ erg s$^{-1}$ cm$^{-2}$ arcsec$^{-2}$) the median trend with stellar mass is nearly flat.

Note that the scatter in axis ratio is significant at any given mass; the gray band indicates 16-84 percentiles for the $10^{-18.5}$ case, where we also show individual galaxies as colored points, coloring by the MgII half-light radius. We find that, at fixed galaxy mass, larger MgII halos are also preferentially more asymmetric, whereas smaller $r_{\rm 1/2,L(MgII)}$ implies more circularly symmetric emission. We examined a number of other properties for correlations with MgII halo shape, including halo mass, total MgII luminosity, black hole mass, total number of galaxy-galaxy mergers, stellar size, halo spin, and sSFR, and found no significant trends (not shown).

We also study the relationship between the size, and shape, of MgII halos in emission, and galaxy inclination (not shown). In particular, we select TNG50 galaxies at $z=0.7$ with $M_\star \geq 10^8$\msun and choose five random inclinations for each. For each effective viewing angle we re-project MgII emission, and re-measure both the half-light radius and shape (axis ratio). Overall, we find only minor trends. As galaxies are rotated from $i=0^\circ$ (edge-on) to $i=90^\circ$ (face-on), the median MgII half-light radii increases. There is however significant scatter, and the difference in the median value across the sample is only $\sim 25$\%. Similarly, the axis ratio $(a/b)$ measurement decreases: MgII halos are more elongated for edge-on systems, and more circularly symmetric for face-on systems, in qualitative agreement with the image gallery of Figure \ref{fig_vis_gallery}. Across this range of inclination the median axis ratio also changes by $\sim 20$\%, although the 16-84th percentile at fixed inclination is similar to the magnitude of the trend itself.

\begin{figure}
\centering
\includegraphics[angle=0,width=3.4in]{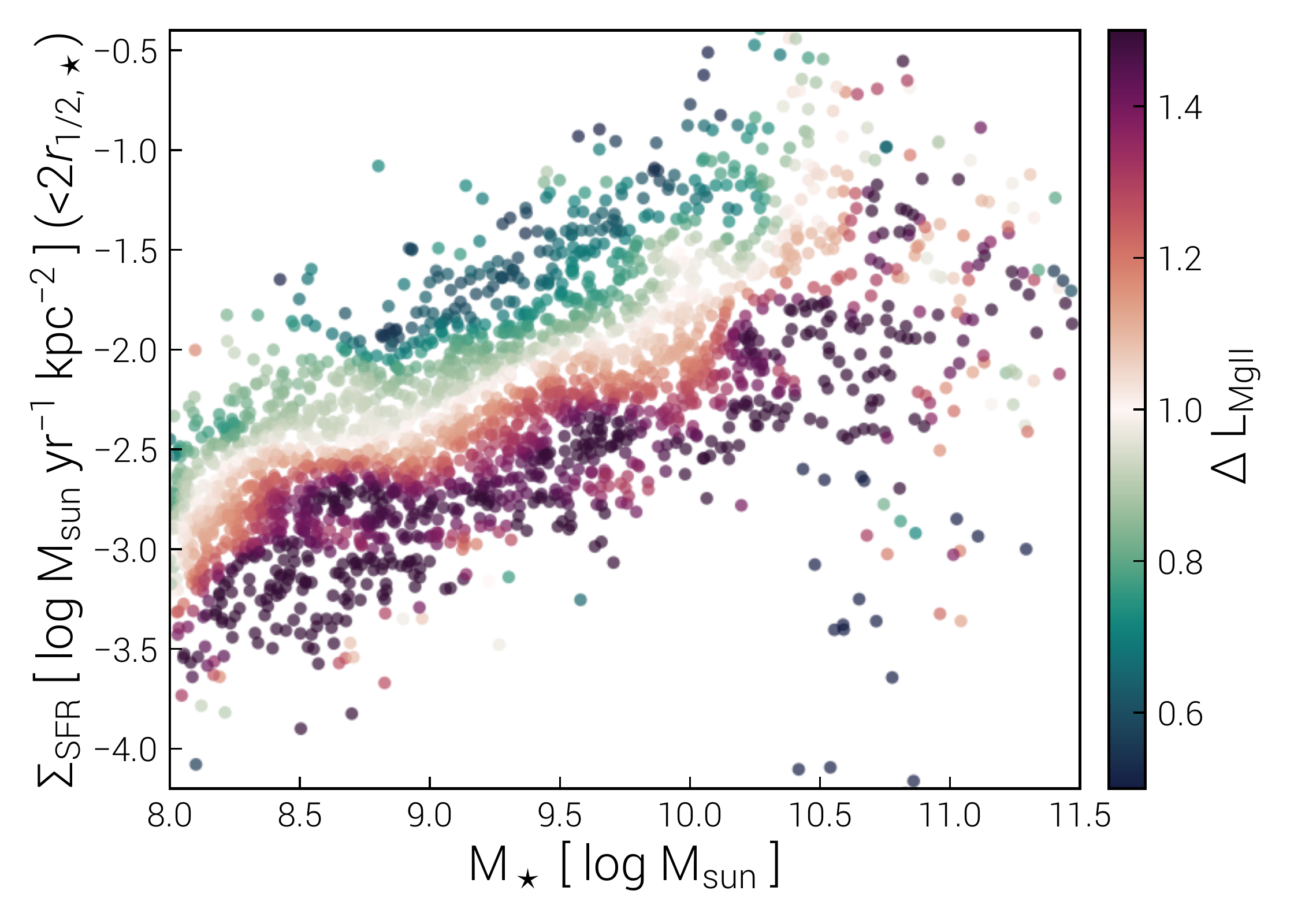}
\includegraphics[angle=0,width=3.4in]{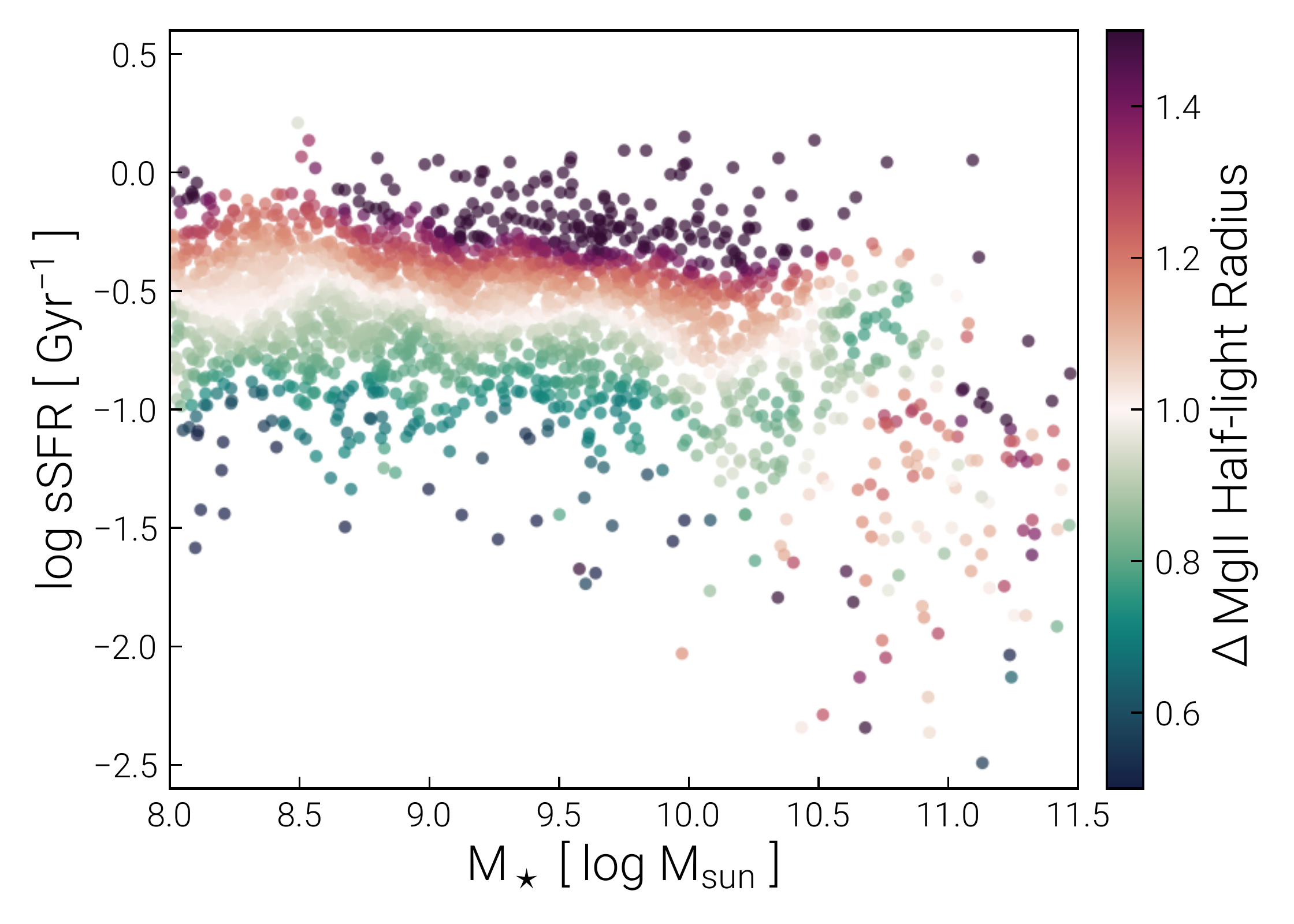}
\caption{ Correlation of MgII emission and the surface density of star formation $\Sigma_{\rm SFR}$ (top) or sSFR (bottom) of galaxies, as a function of stellar mass at $z=0.7$. The top panel colors points by total MgII luminosity $L_{\rm MgII}$, while the bottom panel colors points by the half-light radius i.e. spatial extent of the MgII halo. In both panels the color represents not the actual physical value, but is instead relative $(\Delta$) to the median value of that property at fixed stellar mass, computed in 0.1\,dex bins. We restrict to a maximum of 1000 galaxies per dex for visual clarity, randomly subsampling, and smooth the relative coloring with the LOWESS algorithm \protect\citep[][$l=0.2$]{cleveland79}. We see that, at fixed mass, total MgII luminosity increases for galaxies with less compact star formation. At the same time, galaxies with high sSFRs tend to have more spatially extended MgII halos than their counterparts on, or below, the star-forming main sequence. 
 \label{fig_lum_ssfr}}
\end{figure}

In Figure \ref{fig_lum_ssfr} we examine the relationship between star formation activity in the galaxy and MgII emission. The top panel shows the relationship between the star formation rate surface density $\Sigma_{\rm SFR}$ and galaxy stellar mass $M_\star$ at $z=0.7$ in TNG50. Color indicates $\Delta L_{\rm MgII}$, the luminosity relative to the median value at that stellar mass, red (green) signaling relatively more (less) MgII luminosity. We see a clear correlation that, at fixed mass, galaxies with higher $\Sigma_{\rm SFR}$ actually have less luminous MgII halos. If we color by the MgII halo half-light radius instead of total luminosity, we find a different behavior where higher $\Sigma_{\rm SFR}$ galaxies tend to have more extended emission (not shown). As a result, $L_{\rm MgII}$ decreases for galaxies with more compact star formation, and this emission is more centrally concentrated. The magnitude of this effect is $\sim 0.5$ dex in $L_{\rm MgII}$ between the most extreme outliers in $\Sigma_{\rm SFR}$, at fixed mass.\footnote{We note that this trend is consistent with high outliers in the $L_{\rm MgII}-M_\star$ relation (Figure \ref{fig_lum_vs_mstar}) having lower $\Sigma_{\rm SFR}$ values, due not to decreased total star formation, but instead to lower star formation surface densities when normalized by their (larger) stellar sizes.}

\begin{figure}
\centering
\includegraphics[angle=0,width=3.4in]{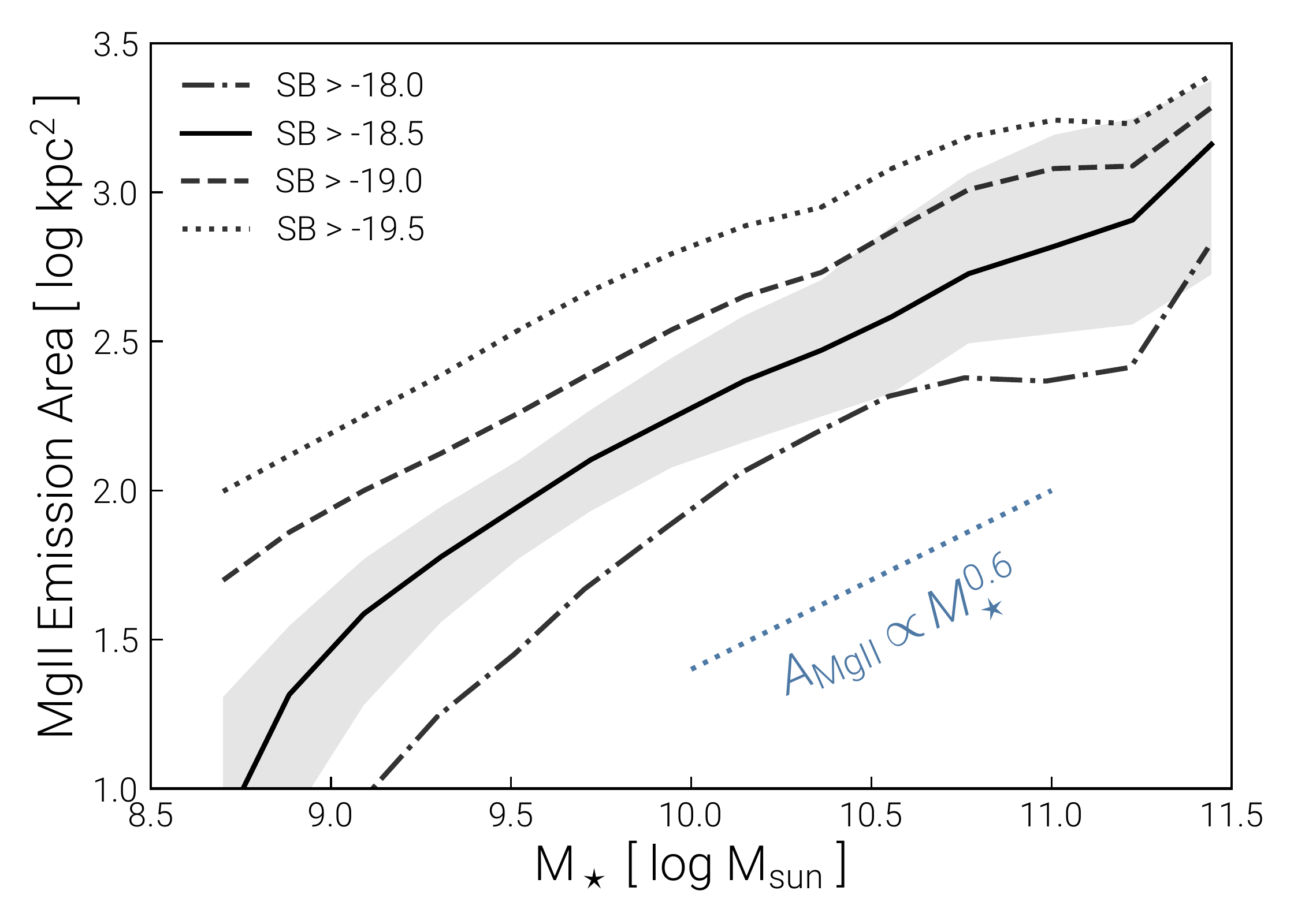}
\caption{ Total area of MgII halos in emission, in square kiloparsecs, above four given thresholds in surface brightness, as indicated in the legend: $\{-18, -18.5, -19, -19.5\}$ log erg s$^{-1}$ cm$^{-2}$ arcsec$^{-2}$. The area increases rapidly with galaxy stellar mass, and with lower surface brightness threshold. In the fiducial case (solid line), the total area increases from $\sim$\,10 kpc$^2$ for $M_\star \simeq 10^{8.5}$\msun to $\sim$\,1000 kpc$^2$ at $M_\star \simeq 10^{11.5}$\msun. The blue dotted line shows the scaling relation $A_{\rm MgII} \propto M_\star^{0.6}$ which approximately reproduces the behavior seen.
 \label{fig_area}}
\end{figure}

The bottom panel of Figure \ref{fig_lum_ssfr} shows the relationship between specific star formation rate and MgII emission, however here we consider the size of the MgII halo, rather than its total luminosity. We see the clear signature that, at fixed stellar mass, galaxies with larger total sSFRs produce larger (more spatially extended) MgII halos. Across $\sim 1$\,dex of scatter in the SFMS, the magnitude of this effect is a factor of a few. This secondary correlation holds true only where the SFMS is well defined, for $M_\star \leq 10^{10.5}$\msun. In contrast, we note that if we instead color by MgII luminosity, as in the top panel, there is no strong trend with offset from the star-forming main sequence (not shown). This implies that $L_{\rm MgII}$ is independent of sSFR to zeroth order, while $r_{\rm 1/2, MgII}$ increases rapidly with sSFR. While several physical mechanisms may contribute to these relations, we speculate that it is related in part to our previous finding that galaxies above the SFMS drive faster winds \citep{nelson19a}, which would then propagate to larger distances. 

\begin{figure*}
\centering
\includegraphics[angle=0,width=3.4in]{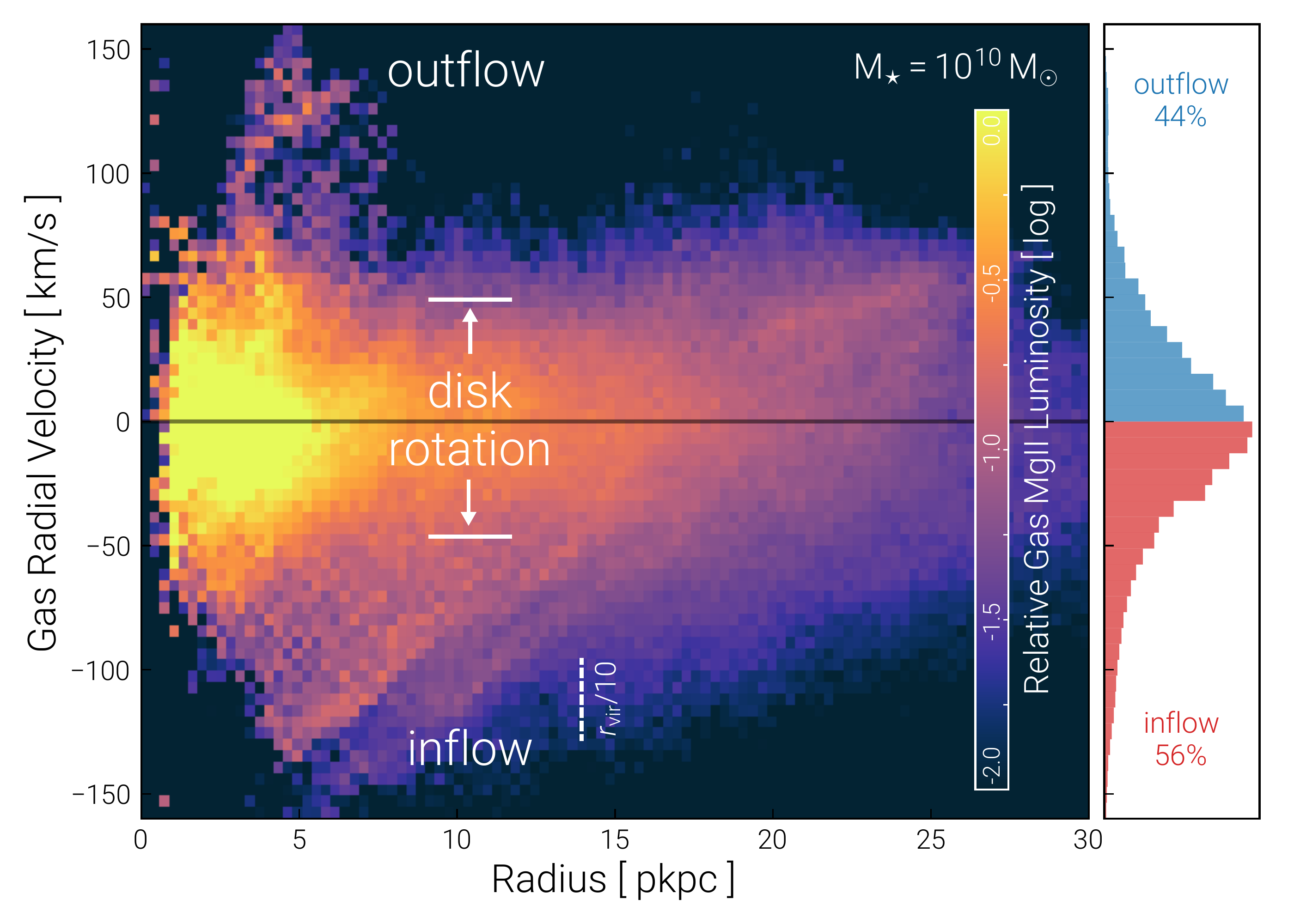}
\includegraphics[angle=0,width=3.35in]{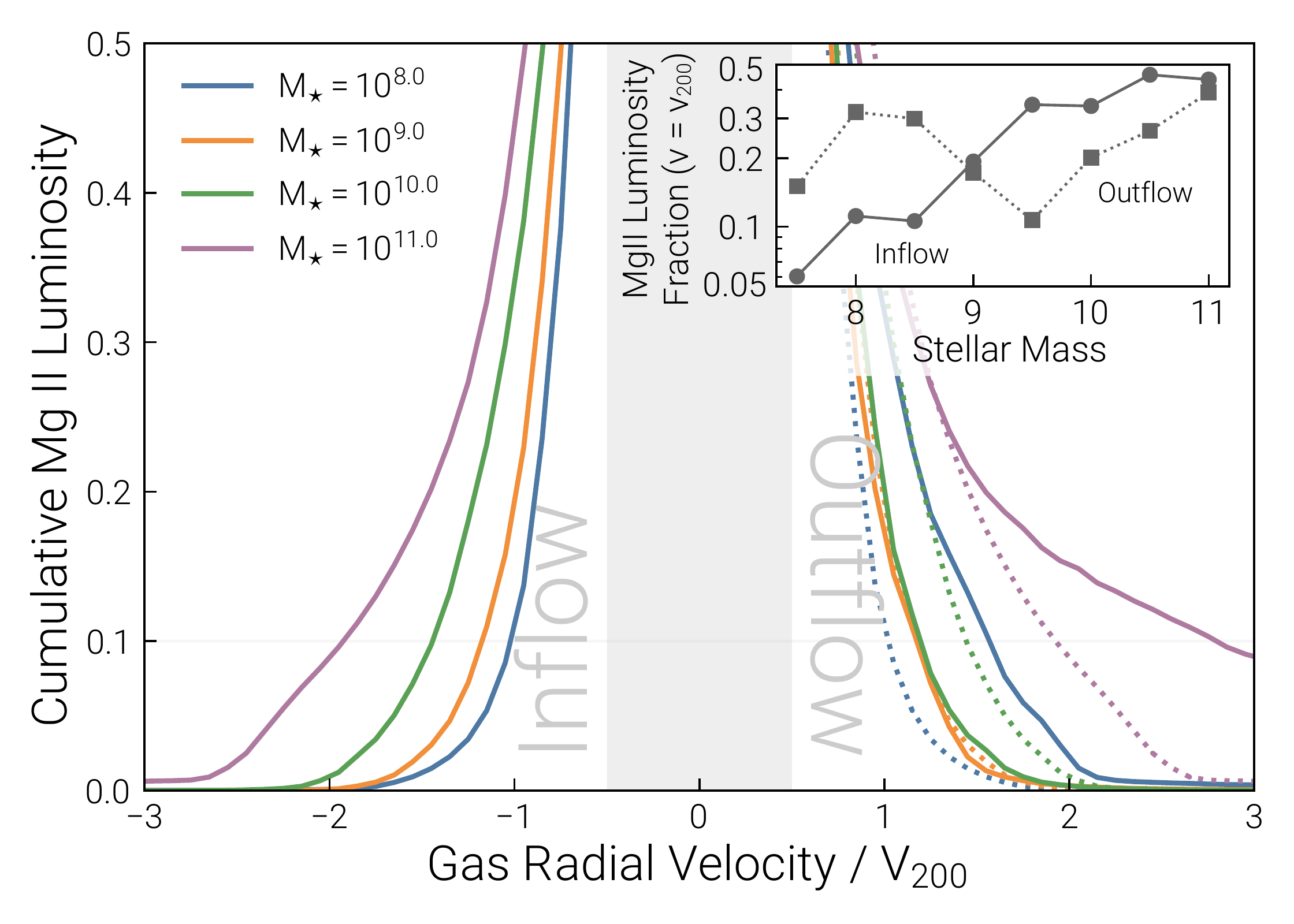}
\caption{ Relationship between MgII luminous gas and inflowing versus outflowing kinematics. \textbf{Left:} We focus on $M_\star = 10^{10}$\msun, stacking galaxies within $\pm 0.01$ dex of this mass at $z=0.7$. The main panel shows radial velocity versus radius, weighted by MgII emission. The right sub-panel shows the marginalized distribution of radial velocity (positive denoting outflow; negative denoting inflow) over the same radial range. The two fractions indicate the amount of MgII flux contributed by inflowing and outflowing gas, respectively. These two components are nearly equal, implying that MgII emission from TNG50 halos predominantly traces halo fountain flows which have, in the MgII emission weighted average, roughly equilibrium kinematics with small $|v_{\rm rad}|$. \textbf{Right:} Cumulative contribution of gas to the total MgII luminosity of a halo, as a function of the radial velocity of that gas. We focus on gas with significant radial motion, i.e. exclude any rotating gaseous disk, by excluding from consideration gas with $|v/v_{\rm 200}| < 1/2$ (gray shaded bands). The lines therefore show the fractional contribution of gas, moving at the given velocity or faster, to the total MgII emission. The left half of the figure shows inflow ($v_{\rm rad} < 0$, solid lines), while the right half shows outflow ($v_{\rm rad} > 0$, again solid lines). For comparison the inflow lines have been mirrored and are shown on the right side of the figure in dotted lines. The inset provides a slice of this data at a particular fixed velocity of $v_{\rm rad} / v_{\rm 200} = 1$, comparing the luminosity fraction as a function of galaxy stellar mass between outflow (dotted) and inflow (solid). The two crossover at $M_\star \sim 10^9$\msun, indicating that galaxies at this mass have MgII halos whose emission arises from a roughly balanced mix of inflows and outflows -- i.e., an equilibrium fountain -- whereas outflows dominate towards lower galaxy masses.
 \label{fig_kinematics}}
\end{figure*}

As a final measure of the extent of MgII halos, we consider their total projected area on the sky, in kpc$^2$, above specific thresholds in surface brightness (SB). Figure \ref{fig_area} shows $A_{\rm MgII}$ as a function of galaxy stellar mass for four thresholds: $\geq 10^{-18}$ (dot dashed), $\geq 10^{-18.5}$ (solid, with shaded band showing $\pm 1 \sigma$ halo-to-halo scatter), $\geq 10^{-19}$ (dashed), and $\geq 10^{-19.5}$ (dotted), all SB values in units of erg s$^{-1}$ cm$^{-2}$ arcsec$^{-2}$. The total area of the MgII halo increases towards fainter SB limits, as expected. It also increases rapidly with stellar mass: we show the approximate scaling found in the simulation of $A_{\rm MgII} \propto M_\star^{0.6}$.

\subsection{Linking MgII Emission to Kinematics} \label{subsec_kinematics}

Broadly speaking, MgII emitting gas could have one of three possible kinematic characteristics: inflowing, outflowing, or approximate centrifugal support (i.e. ISM in a disk). Here we start to decompose these components and address the relationship between MgII emission and gas kinematics.

Figure \ref{fig_kinematics} (left panel) shows the distribution of gas in the two-dimensional plane of radial velocity versus distance, where $v_{\rm rad} > 0$ denotes outflow (upper half) and $v_{\rm rad} < 0$ denotes inflow (lower half). We weight by MgII luminosity, as shown logarithmically by the color, over a factor of one hundred -- i.e., the dimmest pixels contain $\sim$1\% the luminosity of the brightest. We stack twenty TNG50 galaxies at $z=0.7$ with $M_\star \simeq 10^{10}$\msun to obtain a representative view.

At small radii $\lesssim 5$\,kpc MgII emission is dominated by gas with little to no radial motion, i.e. material which is on largely circular orbits in a disk. The plume of emission up to $v_{\rm rad} \sim 100-150$ km/s is a signpost of outflows, while the roughly diagonal stripes with $v_{\rm rad} \sim -100$ km/s at similar radii trace inflows. Both components are clearly subdominant to small $v_{\rm rad}$ gas, if we integrate over all radii. The histogram to the right shows the marginalized distribution of $v_{\rm rad}$ weighted by MgII luminosity, integrating over $0-30$ kpc. As indicated by the two fractions, outflows and inflows (blue and red, respectively), both contribute more or less half to the total MgII luminosity. That is, we do not see any clear indication that MgII emission is specifically tracing outflowing gas. Although we have shown this diagnostic for $M_\star \simeq 10^{10}$\msun, the same qualitative picture holds at lower stellar masses $\sim 10^9$\msun and $\sim 10^8$\msun.

Figure \ref{fig_kinematics} (right panel) quantifies the relationship between MgII emission and gas kinematics. Here we show the cumulative, fractional contribution of gas to the total MgII luminosity of a halo, as a function of the radial velocity of that gas. That is, lines show the fraction of $L_{\rm MgII}$ contributed by gas moving at the given velocity, or faster. In order to focus on gas with non-negligible radial motion, i.e. the non-rotationally supported disk-like component, we exclude all gas with $|v/v_{\rm 200}| < 1/2$ (gray shaded bands) from consideration. 

The left half of the main panel shows inflow ($v_{\rm rad} < 0$). The right half shows outflow ($v_{\rm rad} > 0$), also in solid lines. To ease comparison, we mirror the lines of inflow about zero, and show them on the right half as dotted lines. Different line colors indicate eight different galaxy stellar mass bins, from $M_\star = 10^{8}$\msun to $M_\star = 10^{11}$\msun, always at $z=0.7$. For reference, for $M_\star = \{10^8,10^9,10^{10},10^{11}\}$\msun the mean values are $v_{\rm 200} = \{50, 70, 110, 200\} \,\rm{km\,s^{-1}}$.

First, we find that rapidly flowing gas generally provides only a modest contribution to the total MgII luminosity. For example, all lines at $|v_{\rm rad} / v_{\rm 200}| < 2$ drop below 0.1, indicating that more than 90 percent of MgII emission comes from gas with slower kinematics. Nonetheless, we see that both inflowing and outflowing material do contribute, depending on galaxy mass. Whenever a solid line is rightwards of a dotted line of the same color, outflows contribute more than inflows within that velocity range. This is true of both the lowest stellar mass bin of $\sim 10^{8}$\msun (blue line) and the highest at $\sim 10^{11}$\msun (purple line), but not at intermediate masses of $\sim 10^{9-10}$\msun (orange and green lines).

To quantify this behavior, the inset panel shows the value of the y-axis, i.e. the fractional contribution to the MgII luminosity, as a function of stellar mass, for a fixed velocity threshold of $v_{\rm rad} / v_{\rm 200} > 1$. We see that at $M_\star = 10^8$\msun inflowing gas with $|v/v_{\rm 200}| > 1$ contributes $\sim$10\% to the total MgII emission, while outflowing gas above the same velocity threshold contributes $\sim$30\%. That is, for gas with such characteristically high flow speeds, outflows dominate the contribution to the MgII halo. The situation reverses at higher stellar mass, with a crossover occurring at $M_\star \sim 10^9$\msun, where inflows and outflows both contribute an equal $\sim$20\%. At the highest stellar masses considered, $M_\star = 10^{11}$\msun, rapidly flowing gas contributes significantly to the total luminosity. The crossover at $10^9$\msun is notable, as it indicates a mass regime where MgII halos in emission result from quasi-equilibrium fountain behavior -- balanced inflows and outflows. In the inset shows that, for the velocity threshold of $v_{\rm rad} / v_{\rm 200} > 1$, outflows dominate at the smallest stellar masses, while towards intermediate and high galaxy mass inflows instead tend to contribute more, with the two again converging at the highest masses.

These differences are at most a few tens of percent. We also reiterate that the complement of these fractions, i.e. roughly $50-70$ percent of the MgII luminosity, arises from gas with $v_{\rm rad} < v_{\rm 200}$, much of which is in a central, disk-like morphology. Note that in this analysis we include all gravitationally bound gas in each halo; if we instead include only gas within 30 kpc, for instance, the results are qualitatively unchanged. We have also checked if the situation changes at high redshift ($z=2$), where in general gas fractions, star formation rates, gas accretion rates, and feedback-driven outflows are all stronger. We find a similar picture, where inflows remain the dominant contribution to the MgII emission of halos across the main sequence (not shown). Only for the highest mass galaxies $M_\star \gtrsim 10^{11} \rm{M}_\odot$ do outflows begin to contribute significantly, as they also do at lower redshift.


\section{Discussion} \label{sec_discussion}

\subsection{MgII Emission in Outflows}

In TNG50, there is copious amounts of MgII bearing cold gas aligned with the major axes of galaxies, tracing extended disks and/or inflows. The abundance and kinematics of this gas is in good agreement with data from the MEGAFLOW survey \citep{defelippis21}. In contrast, we do not yet have a quantitative comparison between the TNG simulations for cool gas along minor axis sightlines in such galaxies. In this work we have shown that the bulk of the MgII emission within 10s of kpc of typical galaxies at $z \sim 1$ does not arise in rapidly outflowing gas. Similarly, visual inspection hints that it may be the case that there is relatively less $\sim 10^4$\,K gas oriented along the minor axis than the major axis. Outflows preferentially propagate aligned with the minor axis of well-formed disk galaxies \citep{nelson19b}, where observations also find strong enhancements of MgII in absorption \citep{martin19}. This minor axis excess is typically assumed to be due to galactic winds \citep{bordoloi11,bouche12,kacprzak12}, and is also seen for higher mass galaxies at earlier times \citep{lan18}.

A lack of MgII emitting gas in outflows in TNG50 could plausibly be due to the finite resolution of the simulations. With a baryon mass resolution of $\sim 8 \times 10^4$\msun and a maximum spatial resolution of $\sim 100$\,pc in dense gas, processes occurring below this scale cannot be resolved. This may inhibit the formation, and/or the survival, of cold components within galactic winds \citep{thompson16} similar to those seen in high resolution idealized calculations \citep{scannapieco17,schneider18b} and in our early results with super-Lagrangian circumgalactic medium refinement in cosmological zoom simulations \citep{suresh19}. In this sense, unresolved `clumping' would increase the expected surface brightness values \citep{corlies20}. Although the effective temperature floor of the TNG model at $\sim 10^4$\,K would prevent e.g. the formation of molecular phases \citep[e.g.][]{cicone18}, it is unlikely to inhibit MgII.

Observationally, down the barrel studies show outflowing MgII in absorption in the vast majority of cases, and rarely find clear signatures of inflowing MgII \citep{weiner09,martin12,rubin14}. The observability of the inflowing MgII gas highlighted in Figure \ref{fig_kinematics} should be carefully considered in the future -- together with the regimes where MgII in absorption versus emission trace the same, or different, gaseous flows. What our results show is that, contrary to down the barrel inferences and what is commonly assumed, MgII in emission is a good probe of gas inflowing through the inner CGM. The ultimate balance of inflow versus outflow in the origin of cool gas in the CGM remains an open theoretical question \citep{voit21,afruni21}.

\subsection{MgII Resonant Scattering}

We have investigated the properties of cool circumgalactic gas as visible in the ultraviolet emission line of the MgII ion. Specifically, the doublet at $\lambda \simeq \{2796.352, 2803.532\}$\AA. However, these transitions are resonant, meaning that photons which are emitted can be efficiently reabsorbed, and then rapidly re-emitted. This `resonant scattering' process means that such photons may scatter tens to thousands of times between their origin and the point at which they eventually free stream towards the observer. This process makes the interpretation of MgII observations challenging, in a similar manner as with the Ly$\alpha$ line of hydrogen \citep{prochaska11a}.

Our approach in this study has been to consider the observable signatures of MgII halos under the simplifying assumption that MgII photons, once emitted, do not scatter. That is, we have intentionally neglected the physics of the resonant transition, instead treating the ISM and surrounding CGM as optically thin. Our results therefore provide a benchmark for MgII halo properties under the assumption that resonant scattering is not dominant. 

\begin{figure}
\centering
\includegraphics[angle=0,width=3.3in]{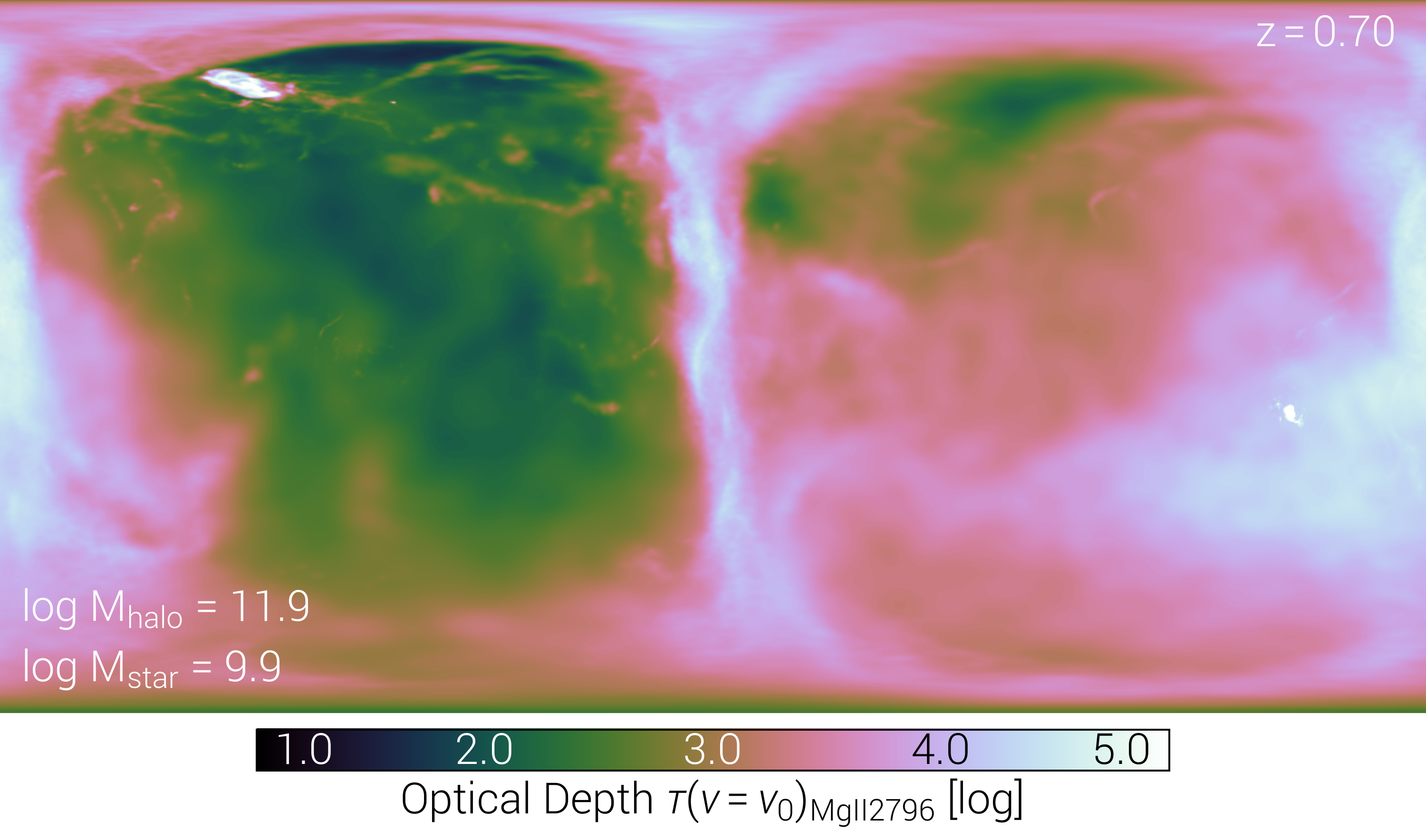}
\caption{ Optical depth $\tau_0 = n \sigma_0 L$ at line center $(\nu = \nu_0)$ for the MgII 2796\AA\ transition. We show an all-sky view, from the exact center of the same galaxy as in Figure \ref{fig_vis_single}, in equirectangular projection. The resulting sightlines therefore escape from the potential well minimum, representing an upper limit on opacity. Optical depths of $\sim$\,100 and below are highlighted in green/blue/black, while the galactic disk itself is prominently visible as the high $\tau \gtrsim 10^4$ stripes in pink/white. Photons traveling into the majority of the solid angle have on average a reasonably low number $\tau < 100$ of scattering paths, which constraints the importance of future resonant scattering. 
 \label{fig_tau}}
\end{figure}

The main reasons for this choice are: (i) such a baseline has not yet been studied within hydrodynamical galaxy formation simulations, and we present it here; (ii) future comparison of the `no-scattering' case with observational data will shed light on the importance of this process for MgII halo emission in the real Universe; (iii) we can not yet treat this physical process. The final point requires the development of a Monte Carlo radiative transfer approach, similar to those employed for Ly$\alpha$, and is an active direction for future work, following the methodology developed in \cite{byrohl20}.

Figure \ref{fig_tau} provides an initial assessment of the validity of this assumption. Here we show an all-sky view, from the center of a single $M_\star \sim 10^{10}$\msun galaxy at $z=0.7$. For each pixel, i.e. along each radially outward line-of-sight, we derive the line-center optical depth $\tau_0 = n \sigma_0 L$ of the MgII 2796\AA\ transition, which ranges from $\mathcal{O}(1)$ along the least dense escape paths, to $\mathcal{O}(10^5)$ when traversing through the entire galactic disk of dense ISM material (vertical pink/white features). Note that by placing an `observer' at the galactic center we assess the highest possible $\tau$ case.

In general, photons escaping into most of the solid angle have a reasonably low number $\tau \lesssim 100$ of interactions. For comparison, the average optical depth per cell at line center, across the galaxy and halo, is $\sim 10^3$ for MgII 2796, as compared to $\sim 10^6$ for Lyman-$\alpha$. Despite the significantly smaller values, the optical depth is far from unity, signifying that photons emitted from the nuclear regions of galaxies are certainly impacted by resonant scattering. 

As in the case of Ly$\alpha$, we anticipate that scattering will redistribute the emergent MgII emission from smaller to larger radii, flattening radial surface brightness profiles \citep{byrohl20}. Similarly, some amount of the flux we observe to originate from the central galaxy itself will be processed in halo-scale flows -- either inflows and/or outflows -- which should boost the fraction of MgII emission appearing to originate from this material. In addition to the difference in optical depths between these two lines, Doppler broadening is also significantly smaller for MgII versus Lyman-$\alpha$, which will lead to smaller frequency diffusion. The overall effect, however, will be similar: scattering will obscure the link between observed emission and the underlying velocity structure of circumgalactic gas, potentially altering the relationships to kinematics we have investigated herein.

The key challenge with the new wealth of observational data on emission from the cool circumgalactic medium will be modeling the complex kinematics, convolved with the impact of resonant line transfer, in order to decipher the underlying physical structure and kinematics of this gas \citep{martin19_kcwi}. Our followup study will quantitatively address both points and thus provide a more refined prediction and point of comparison against observations of extended MgII halo emission around galaxies.

\subsection{Observability with KCWI/MUSE}

We now consider the degree to which current instrumentation could detect the predicted extended MgII emission. We first consider the observational setup used by \cite{burchett20}, a study optimized to detect such emission with KCWI at $z\sim0.7$. That work uses the BL grating with the medium KCWI slicer, yielding individual spaxel sizes of $0.7\arcsec \times0.3\arcsec = 0.21~\rm arcsec^2$. With a total exposure time of 3.5 hours they construct a narrowband image from the IFU datacube centered on the observed-frame wavelength of the $\lambda2796$ transition (alone) with a 5\,\AA\ width. The resulting image has a $1\sigma$ surface brightness sensitivity of \mbox{$4.8\times 10^{-19}$ erg s$^{-1}$ cm$^{-2}$ arcsec$^{-2}$} for individual pixels, corresponding to \mbox{$2.2\times 10^{-19}$ erg s$^{-1}$ cm$^{-2}$ arcsec$^{-2}$} per $1~\rm arcsec^2$ binned aperture.

\begin{figure}
\centering
\includegraphics[angle=0,width=3.3in]{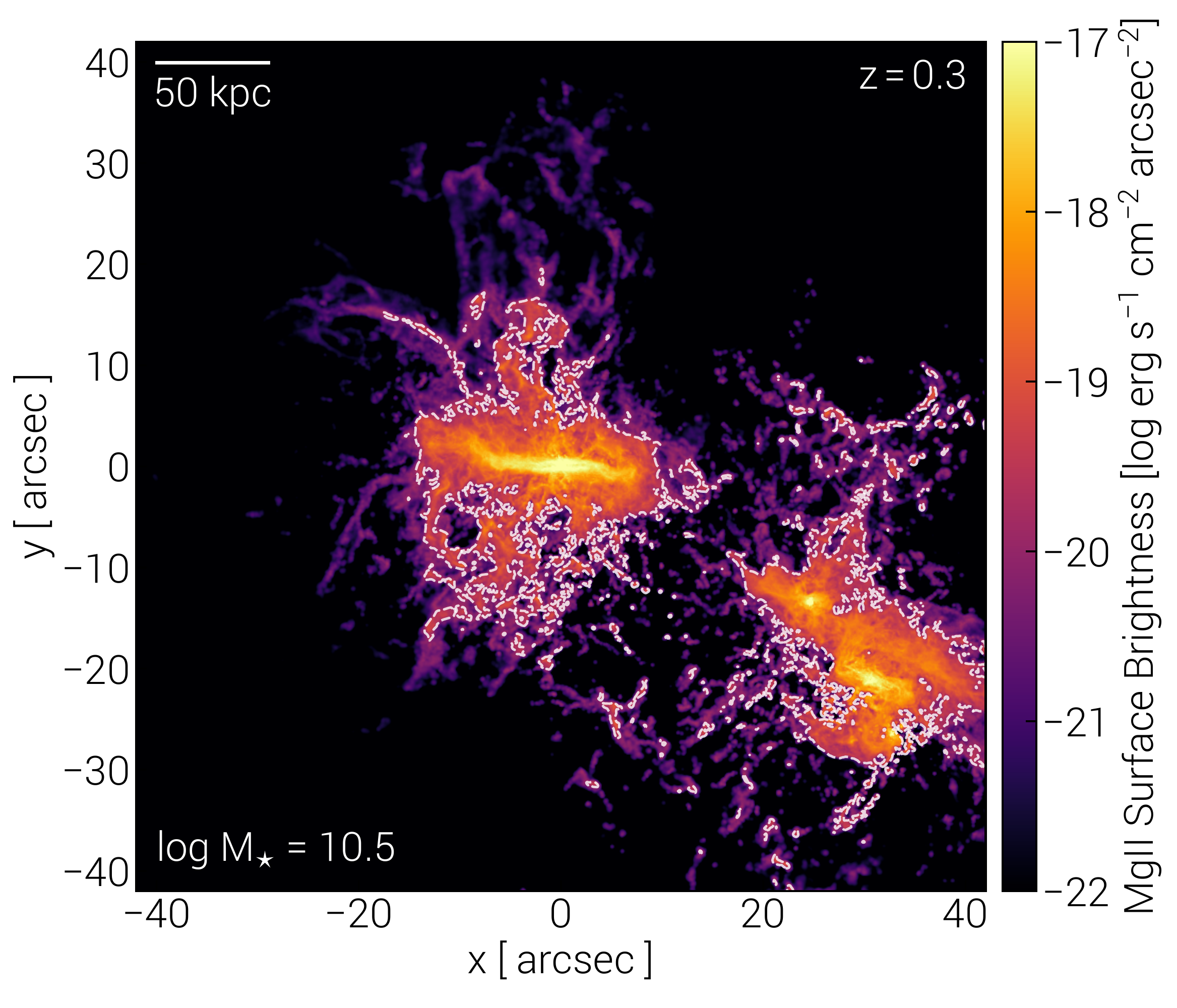}
\includegraphics[angle=0,width=3.3in]{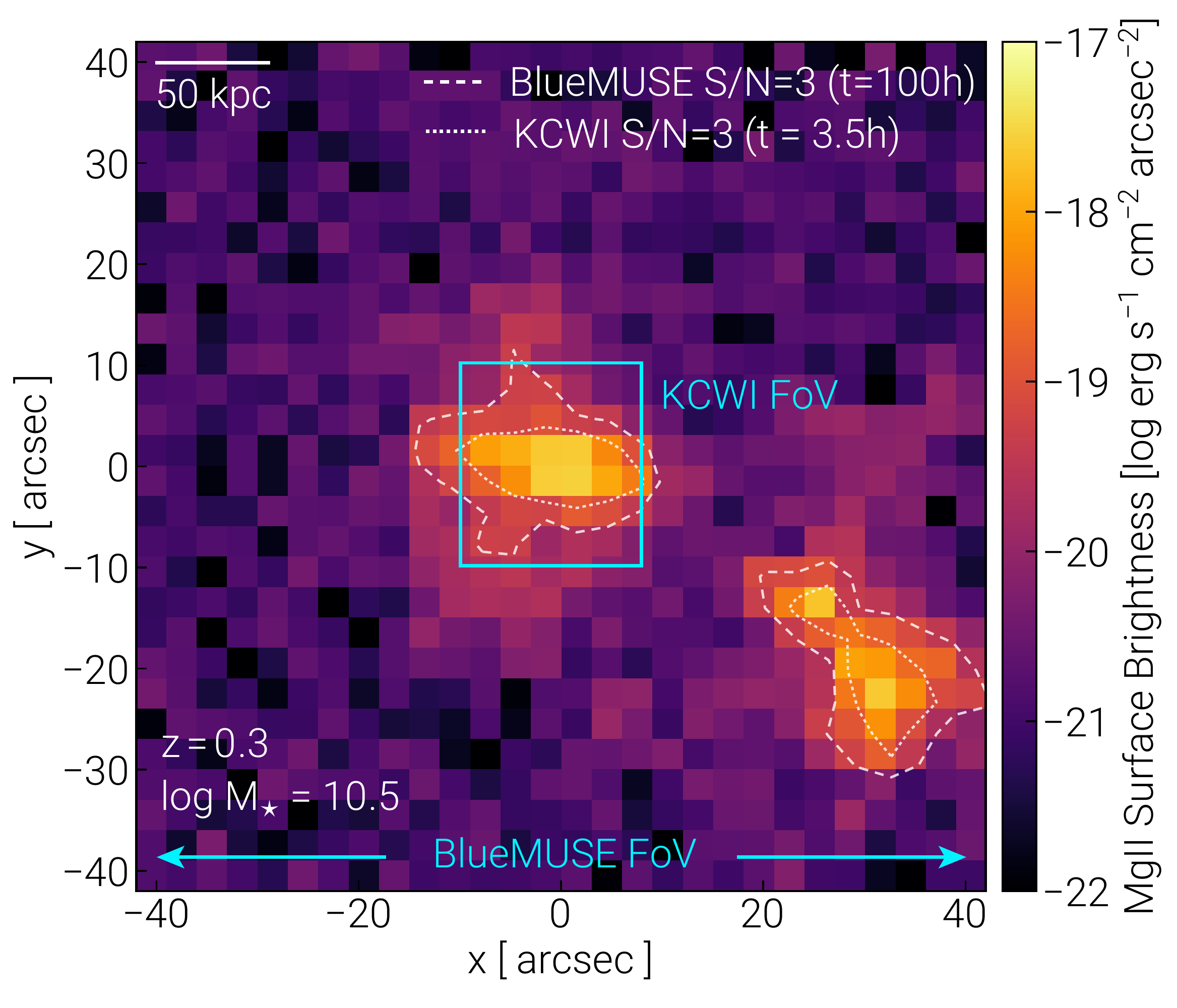}
\caption{ Predicted MgII emission structure around a galaxy observable with KCWI and the future VLT/BlueMUSE instruments. We show one example $M_\star \simeq 10^{10.5}$\msun galaxy at $z=0.3$. The image extent corresponds to the full BlueMUSE field of view of $1.4' \times 1.4'$, while the blue rectangle in the bottom panel shows the smaller $16.5\arcsec \times 20\arcsec$ (M slicer) FoV of KCWI. The top panel depicts the direct output of the simulation, while the bottom panel shows a synthetic BlueMUSE observation for the noise level resulting from a 100 hour exposure time, binning 10x10 pixels spatially, and applying a PSF smoothing of 0.7\arcsec. The primary advantage of BlueMUSE will be its substantially larger field of view enabling blind field observations, and to compare these two instruments we assume a longer integration time of 100 hours for BlueMUSE. In the top panel the white contour highlights an ambitious surface brightness limit of $1.5 \times 10^{-20}$ erg s$^{-1}$ cm$^{-2}$ arcsec$^{-2}$, while in the lower panel we include two empirically motivated surface brightness limits: the $3\sigma$ limit for the 3.5 hour KCWI case at $2.3 \times 10^{-19}$ erg s$^{-1}$ cm$^{-2}$ arcsec$^{-2}$ (see text; inner dotted contour), and the 100 hour BlueMUSE case at $4.3 \times 10^{-20}$ erg s$^{-1}$ cm$^{-2}$ arcsec$^{-2}$ (outer dashed contour).
 \label{fig_bluemuse}}
\end{figure}

At a higher observed wavelength (redshift), an example of data available from the MUSE instrument is the MUSE ultra deep field \citep[UDF;][]{bacon17}. The MUSE UDF mosaic data ranges in exposure time from $10-30$ hours, and achieves a $1\sigma$ limiting surface brightness sensitivity of $2.8-5.5\times10^{-20}~\rm erg~s^{-1}~cm^{-2}$ \AA$\rm ^{-1}~arcsec^{-2}$ per $1~\rm arcsec^2$ binned aperture over the wavelength range $7000-8500$ \AA\ \citep{leclercq20}. Summing these data over a $500~\rm km~s^{-1}$ spectral window yields narrowband images with $1\sigma$ sensitivities of $\approx 0.9-2.4\times10^{-19}$ erg s$^{-1}$ cm$^{-2}$ arcsec$^{-2}$ per $1~\rm arcsec^2$ binned aperture. The new MUSE Extremely Deep Field \citep{bacon21}, with a total exposure time of 140 hours, has a typical $5\sigma$ surface brightness limit of $1.3\times10^{-19}$ erg s$^{-1}$ cm$^{-2}$ arcsec$^{-2}$ per $1~\rm arcsec^2$ binned aperture, summed over a 3.75 \AA-wide spectral window centered at 7000\,\AA. This corresponds to a $1\sigma$ limit of $\approx 4-5 \times 10^{-20}$ erg s$^{-1}$ cm$^{-2}$ arcsec$^{-2}$ when summed over a $500~\rm km~s^{-1}$ spectral window (as above). For comparison, the recent MgII emission observation of \citet{zabl21} achieves a $\{1,2,3\}\sigma$ surface brightness limit of $\{7,14,21\} \times 10^{-19}$ erg s$^{-1}$ cm$^{-2}$ arcsec$^{-2}$ per $1.5~\rm arcsec^2$ binned aperture and total spectral window of $1200$ km s$^{-1}$, with a $11.2$ hour MUSE exposure. The equivalent $1\sigma$ limit for a 1 arcsec$^2$ aperture and a $500~\rm km~s^{-1}$ spectral window is $5.5 \times 10^{-19}$ erg s$^{-1}$ cm$^{-2}$ arcsec$^{-2}$, consistent with \citet{leclercq20}. 

Targeted observations with KCWI can reasonably survey multiple systems with e.g. $\sim\,3-5$ hour exposure times per target halo \citep[e.g.][]{chen21}. The MgII observation of \citet{burchett20} achieves an effective $3\sigma$ surface brightness sensitivity of $\sim 2.3\times10^{-19}$ erg s$^{-1}$ cm$^{-2}$ arcsec$^{-2}$ measured in a 3x3 arcsec$^2$ aperture (represented with the dotted contour in Figure \ref{fig_bluemuse} for an example TNG50 halo), whereas blind searches e.g. within the deep MUSE datacubes described above can achieve $3\sigma$ surface brightness limits in the same aperture from $0.4-5.5\times10^{-19}$ erg s$^{-1}$ cm$^{-2}$ arcsec$^{-2}$. 

Our modeling predicts that surface brightness values $\gtrsim 10^{-19}$ erg s$^{-1}$ cm$^{-2}$ arcsec$^{-2}$ will arise within $\lesssim 2\arcsec$ ($\lesssim 15$ kpc) of the typical central galaxy with $M_{\star} = 10^{10.5}M_{\odot}$ at $z=0.7$, increasing to $\lesssim 3\arcsec$ ($\lesssim 20$ kpc) for more massive $10^{11}$\msun systems. We note that these values combine both lines of the doublet, whereas the observational limits discussed in the prior paragraphs are for $\lambda2796$ alone, but the difference is negligible. For individual systems this suggests that KCWI and MUSE will be sensitive mostly to emission from the inner halo where gas densities are highest, i.e. within $\sim 15$\% of $r_{\rm vir}$ for these halos, and stacking will be required to capture fainter emission.

\subsection{Observability with BlueMUSE}

Here we estimate the observability of MgII emission based on the current design of the future VLT/BlueMUSE instrument \citep{richard19}. To compare BlueMUSE and KCWI, we note that these two instruments will have similar efficiencies, with total throughputs of $\sim 0.3$, and are hosted on telescopes with similar collecting areas. The primary advantage of BlueMUSE will be its substantially larger field of view, i.e. grasp, enabling blind fields and survey type operation. To make a reasonable comparison between these two instruments, we assume a longer integration time of 100 hours for BlueMUSE, similar to the existing MUSE XDF, and simply scale the previous $3\sigma$ surface brightness limit estimated for KCWI of $2.3 \times 10^{-19}$ erg s$^{-1}$ cm$^{-2}$ arcsec$^{-2}$ by $(100/3.5)^{-1/2}$, resulting in a $3\sigma$ surface brightness limit for MgII line emission for BlueMUSE of $4.3 \times 10^{-20}$ erg s$^{-1}$ cm$^{-2}$ arcsec$^{-2}$ per $3x3~\rm arcsec^2$ binned aperture.

Figure \ref{fig_bluemuse} shows a mock image of an intermediate mass $z = 0.3$ galaxy. The top panel shows the direct result for MgII emission from TNG50, fully idealized. The bottom panel includes a 10x10 spatial binning, convolution by a 0.7'' PSF, and an approximate sky background noise level from above. The images are 1.4 arcmin across, corresponding to the BlueMUSE field of view. The more limited field of view of KCWI, indicated with the blue rectangle (for the M slicer), suggests that it will mainly detect MgII halos with targeted observations of specific galaxies, as opposed to the wide field mapping capabilities of BlueMUSE, which is why we present comparative surface brightness limit contours for rather different exposure times. The 100 hour BlueMUSE limit is only a factor of a few improved with respect to the 3.5 hour KCWI limit, and the contours of Figure \ref{fig_bluemuse} suggest that both will mainly be sensitive to the bright, inner emission ($\lesssim 10\arcsec$ or $\lesssim 40$ kpc) of such a $z=0.3$ halo.

We have also made an estimate using the BlueMUSE exposure time calculator, assuming a 10x10 spatial pixel binning (reducing from the native 0.3'' px$^{-1}$), as well as 10x spectral binning (reducing from the native 0.55 \AA\, px$^{-1}$). Requiring a signal-to-noise ratio of three and with an on-source integration time of 100 hours this results in a consistent estimate to the surface brightness limit above for individual halos at $z \sim 0.3$, i.e. without stacking. Our predicted profiles from Figure \ref{fig_sbr_stacked} imply that extended MgII emission will be detected out to $\sim 10$ kpc for low-mass star-forming galaxies with $M_\star = 10^9$\msun, and out to $\sim 50$ kpc for higher-mass $M_\star = 10^{11}$\msun systems, while stacking would enable the detection of even more diffuse emission. These estimates are somewhat conservative with the exclusive of resonant radiative transfer effects, which would boost surface brightness values and make the detection of individual halos, if anything, easier.

The galaxy in the center of this field has a stellar mass of $M_\star \simeq 10^{10.5}$\msun, and is oriented roughly edge-on. Prominent outflows both above and below the disk are evident in MgII emission. Two smaller nearby companion galaxies are visible towards the lower left corner. Their ongoing interaction produces a complex structure of MgII emitting gas which can be clearly captured, at moderate spatial resolution, by BlueMUSE.


\section{Summary of Conclusions} \label{sec_conclusions}

In this paper we use the TNG50 cosmological magnetohydrodynamical simulation to explore the rest-frame ultraviolet emission from the MgII doublet at $\lambda \simeq \{2796.352, 2803.532\}$\AA. This bright transition is visible from the ground at $z \geq 0.3$, and traces cool $\sim 10^4$\,K gas in the dense interstellar medium of galaxies, in feedback driven galactic-scale outflows, and in the diffuse circumgalactic medium of halos. We develop a model for the abundance of gas-phase MgII and its emission by post-processing the TNG50 simulation with an ionization model, assuming that MgII is optically thin and neglecting the impact of resonant scattering. We characterize MgII halos in emission surrounding a population of thousands of simulated galaxies which span $7.5 < \log{(M_\star / \rm{M}_\odot)} < 11.0$ over the redshift range $0.3 < z < 2$. Our main results are:

\begin{itemize}
    \item Independent of mass and redshift, extended halos of MgII emission are an ubiquitous feature of normal star-forming galaxies.
    \item Resolved emission maps show that MgII halos extend out to tens of kpc, i.e. far beyond the stellar bodies of galaxies themselves. MgII halos exhibit a diversity of morphologies, with emission tracing: extended disk-like structures, fountain flows with signatures of infalling gas clouds and wind-driven outflows, filamentary features, X-shaped central components, and in general, a complex morphology of small-scale ($\sim$ kpc), inhomogenously distributed, clumpy gas.
    \item Stacked, median radial surface brightness profiles of MgII decline smoothly and rapidly with projected distance, from values exceeding $10^{-18}$ erg s$^{-1}$ cm$^{-2}$ arcsec$^{-2}$ in the centers of halos to an order of magnitude fainter by $\sim 10$ kpc ($z=0.7$). Total MgII halo luminosity increases nearly linearly with galaxy stellar mass, and is largely independent of redshift, reaching \mbox{$10^{40.5}$ erg s$^{-1}$} for $M_\star \sim 10^{10}$\msun. At fixed mass, $L_{\rm MgII}$ is larger for galaxies with smaller star formation rate surface densities $\Sigma_{\rm SFR}(<2r_{\rm 1/2,\star})$.
    \item The size of MgII halos, measured in terms of their half-light radii, increases from $\sim$ a few kpc for low-mass $M_\star \sim 10^{8-9}$\msun galaxies to $\sim$ tens of kpc for $M_\star \gtrsim 10^{10.5}$\msun. At fixed mass, galaxies with higher sSFR have larger MgII halos. The extent of MgII emission correlates with environment: $r_{\rm 1/2,MgII}$ increases by a factor of two across the local galaxy overdensities sampled by $M_\star = 10^{10}$\msun galaxies, due to the contributions of additional cool gas within, as well as stripped from, nearby satellites.
    \item MgII halo shapes in projection are far from circular. Parameterizing shape in terms of the axis ratio of the bounding ellipse for a given MgII isophotal contour, we find that MgII halos become progressively more asymmetric towards higher galaxy mass, from a median axis ratio of $\sim 1.1$ (nearly circularly symmetric) at $M_\star \simeq 10^{8}$\msun to $\sim 1.6$ (decidedly elongated) for $M_\star \simeq 10^{10.5}$\msun. MgII halos are more asymmetric in their centers, i.e. at higher surface brightness thresholds, and more circular in their outskirts.
    \item MgII emission arises from gas with different kinematics. Overall, in the central tens of kpc, it is dominated by gas with little radial velocity -- near-equilibrium fountain flows and material with non-negligible rotational support. Rapidly flowing gas, either inflows or outflows, are a subdominant contribution. Considering such gas with $v_{\rm rad} / v_{\rm 200} > 1/2$, we find that outflows dominate the contribution to MgII halos for low mass galaxies, whereas inflows are more prevalent at high mass. A crossover occurs at $M_\star \simeq 10^{9}$\msun, where outflows and inflows provide roughly the same fractional contribution to the total MgII luminosity of the halo.
\end{itemize}

The properties, correlations, and origins of MgII halos in emission explored herein provide a significant number of empirically testable outcomes of the TNG50 simulation, while simultaneously illuminating the physics of the cool CGM. Observationally, several recent instruments have come online which offer the distinct possibility to characterize the cool circumgalactic medium of galaxies via \textit{direct imaging}, rather than in absorption, as has been more accessible in the past. In particular, the VLT/MUSE and Keck/KCWI integral field unit  spectrographs, as well as the Dragonfly and Fireball-2 instruments, have already begun to demonstrate the power of this technique.

Whereas Lyman-alpha halos have begun to be statistically characterized across the galaxy population, no such breadth of empirical measurements yet exists for MgII. Together with FeII, these three transitions all offer complementary views into the abundance, physical properties, and kinematics of cool circumgalactic gas -- a significant reservoir of baryons in the Universe. Interpretation of this data is, however, challenging. In this work we have presented a first look at predictions for MgII halos from TNG50, demonstrating how cosmological simulations offer a powerful framework from which to decipher the gas flows which make up the cosmic baryon cycle.


\section*{Data Availability}

Data directly related to this publication and its figures is available on request from the corresponding author. The IllustrisTNG simulations, including TNG50, are publicly available and accessible at \url{www.tng-project.org/data} as described in \cite{nelson19a}.

\section*{Acknowledgements}
DN acknowledges funding from the Deutsche Forschungsgemeinschaft (DFG) through an Emmy Noether Research Group (grant number NE 2441/1-1).
The primary TNG simulations were realized with compute time granted by the Gauss Centre for Supercomputing (GCS) under GCS Large-Scale Projects GCS-ILLU (2014) and GCS-DWAR (2016) on the GCS share of the supercomputer Hazel Hen at the High Performance Computing Center Stuttgart (HLRS). GCS is the alliance of the three national supercomputing centres HLRS (Universit{\"a}t Stuttgart), JSC (Forschungszentrum J{\"u}lich), and LRZ (Bayerische Akademie der Wissenschaften), funded by the German Federal Ministry of Education and Research (BMBF) and the German State Ministries for Research of Baden-W{\"u}rttemberg (MWK), Bayern (StMWFK) and Nordrhein-Westfalen (MIWF). Additional simulations and analysis were carried out on the Hydra and Draco supercomputers at the Max Planck Computing and Data Facility (MPCDF). 

\bibliographystyle{mnras}
\bibliography{refs}

\begin{thebibliography}{}
\makeatletter
\relax
\def\mn@urlcharsother{\let\do\@makeother \do\$\do\&\do\#\do\^\do\_\do\%\do\~}
\def\mn@doi{\begingroup\mn@urlcharsother \@ifnextchar [ {\mn@doi@}
  {\mn@doi@[]}}
\def\mn@doi@[#1]#2{\def\@tempa{#1}\ifx\@tempa\@empty \href
  {http://dx.doi.org/#2} {doi:#2}\else \href {http://dx.doi.org/#2} {#1}\fi
  \endgroup}
\def\mn@eprint#1#2{\mn@eprint@#1:#2::\@nil}
\def\mn@eprint@arXiv#1{\href {http://arxiv.org/abs/#1} {{\tt arXiv:#1}}}
\def\mn@eprint@dblp#1{\href {http://dblp.uni-trier.de/rec/bibtex/#1.xml}
  {dblp:#1}}
\def\mn@eprint@#1:#2:#3:#4\@nil{\def\@tempa {#1}\def\@tempb {#2}\def\@tempc
  {#3}\ifx \@tempc \@empty \let \@tempc \@tempb \let \@tempb \@tempa \fi \ifx
  \@tempb \@empty \def\@tempb {arXiv}\fi \@ifundefined
  {mn@eprint@\@tempb}{\@tempb:\@tempc}{\expandafter \expandafter \csname
  mn@eprint@\@tempb\endcsname \expandafter{\@tempc}}}

\bibitem[\protect\citeauthoryear{{Afruni}, {Fraternali}  \&
  {Pezzulli}}{{Afruni} et~al.}{2021}]{afruni21}
{Afruni} A.,  {Fraternali} F.,   {Pezzulli} G.,  2021, \mn@doi [\mnras]
  {10.1093/mnras/staa3759}, \href
  {https://ui.adsabs.harvard.edu/abs/2021MNRAS.501.5575A} {501, 5575}

\bibitem[\protect\citeauthoryear{{Anand}, {Nelson}  \& {Kauffmann}}{{Anand}
  et~al.}{2021}]{anand21}
{Anand} A.,  {Nelson} D.,   {Kauffmann} G.,  2021, \mn@doi [\mnras]
  {10.1093/mnras/stab871}, \href
  {https://ui.adsabs.harvard.edu/abs/2021MNRAS.504...65A} {504, 65}

\bibitem[\protect\citeauthoryear{{Anderson}, {Gaspari}, {White}, {Wang}  \&
  {Dai}}{{Anderson} et~al.}{2015}]{anderson15}
{Anderson} M.~E.,  {Gaspari} M.,  {White} S.~D.~M.,  {Wang} W.,   {Dai} X.,
  2015, \mn@doi [\mnras] {10.1093/mnras/stv437}, \href
  {http://adsabs.harvard.edu/abs/2015MNRAS.449.3806A} {449, 3806}

\bibitem[\protect\citeauthoryear{{Arrigoni Battaia}, {Hennawi}, {Prochaska},
  {O{\~n}orbe}, {Farina}, {Cantalupo}  \& {Lusso}}{{Arrigoni Battaia}
  et~al.}{2019}]{battaia19}
{Arrigoni Battaia} F.,  {Hennawi} J.~F.,  {Prochaska} J.~X.,  {O{\~n}orbe} J.,
  {Farina} E.~P.,  {Cantalupo} S.,   {Lusso} E.,  2019, \mn@doi [\mnras]
  {10.1093/mnras/sty2827}, \href
  {https://ui.adsabs.harvard.edu/abs/2019MNRAS.482.3162A} {482, 3162}

\bibitem[\protect\citeauthoryear{{Augustin} et~al.,}{{Augustin}
  et~al.}{2019}]{augustin19}
{Augustin} R.,  et~al., 2019, \mn@doi [\mnras] {10.1093/mnras/stz2238}, \href
  {https://ui.adsabs.harvard.edu/abs/2019MNRAS.489.2417A} {489, 2417}

\bibitem[\protect\citeauthoryear{{Bacon} et~al.,}{{Bacon}
  et~al.}{2017}]{bacon17}
{Bacon} R.,  et~al., 2017, \mn@doi [\aap] {10.1051/0004-6361/201730833}, \href
  {https://ui.adsabs.harvard.edu/abs/2017A&A...608A...1B} {608, A1}

\bibitem[\protect\citeauthoryear{{Bacon} et~al.,}{{Bacon}
  et~al.}{2021}]{bacon21}
{Bacon} R.,  et~al., 2021, \mn@doi [\aap] {10.1051/0004-6361/202039887}, \href
  {https://ui.adsabs.harvard.edu/abs/2021A&A...647A.107B} {647, A107}

\bibitem[\protect\citeauthoryear{{Barnes} et~al.,}{{Barnes}
  et~al.}{2018}]{barnes18}
{Barnes} D.~J.,  et~al., 2018, \mn@doi [\mnras] {10.1093/mnras/sty2078}, \href
  {https://ui.adsabs.harvard.edu/abs/2018MNRAS.481.1809B} {481, 1809}

\bibitem[\protect\citeauthoryear{{Bertone} \& {Schaye}}{{Bertone} \&
  {Schaye}}{2012}]{bertone12}
{Bertone} S.,  {Schaye} J.,  2012, \mn@doi [\mnras]
  {10.1111/j.1365-2966.2011.19742.x}, \href
  {http://adsabs.harvard.edu/abs/2012MNRAS.419..780B} {419, 780}

\bibitem[\protect\citeauthoryear{{Bertone}, {Schaye}, {Dalla Vecchia}, {Booth},
  {Theuns}  \& {Wiersma}}{{Bertone} et~al.}{2010a}]{bertone10a}
{Bertone} S.,  {Schaye} J.,  {Dalla Vecchia} C.,  {Booth} C.~M.,  {Theuns} T.,
   {Wiersma} R.~P.~C.,  2010a, \mn@doi [\mnras]
  {10.1111/j.1365-2966.2010.16932.x}, \href
  {http://adsabs.harvard.edu/abs/2010MNRAS.407..544B} {407, 544}

\bibitem[\protect\citeauthoryear{{Bertone}, {Schaye}, {Booth}, {Dalla Vecchia},
  {Theuns}  \& {Wiersma}}{{Bertone} et~al.}{2010b}]{bertone10b}
{Bertone} S.,  {Schaye} J.,  {Booth} C.~M.,  {Dalla Vecchia} C.,  {Theuns} T.,
   {Wiersma} R.~P.~C.,  2010b, \mn@doi [\mnras]
  {10.1111/j.1365-2966.2010.17188.x}, \href
  {http://adsabs.harvard.edu/abs/2010MNRAS.408.1120B} {408, 1120}

\bibitem[\protect\citeauthoryear{{Bird}, {Vogelsberger}, {Haehnelt}, {Sijacki},
  {Genel}, {Torrey}, {Springel}  \& {Hernquist}}{{Bird} et~al.}{2014}]{bird14}
{Bird} S.,  {Vogelsberger} M.,  {Haehnelt} M.,  {Sijacki} D.,  {Genel} S.,
  {Torrey} P.,  {Springel} V.,   {Hernquist} L.,  2014, \mn@doi [\mnras]
  {10.1093/mnras/stu1923}, \href
  {http://adsabs.harvard.edu/abs/2014MNRAS.445.2313B} {445, 2313}

\bibitem[\protect\citeauthoryear{{Boksenberg} \& {Sargent}}{{Boksenberg} \&
  {Sargent}}{1978}]{boksenberg78}
{Boksenberg} A.,  {Sargent} W.~L.~W.,  1978, \mn@doi [\apj] {10.1086/155880},
  \href {https://ui.adsabs.harvard.edu/abs/1978ApJ...220...42B} {220, 42}

\bibitem[\protect\citeauthoryear{{Bordoloi} et~al.,}{{Bordoloi}
  et~al.}{2011}]{bordoloi11}
{Bordoloi} R.,  et~al., 2011, \mn@doi [\apj] {10.1088/0004-637X/743/1/10},
  \href {https://ui.adsabs.harvard.edu/abs/2011ApJ...743...10B} {743, 10}

\bibitem[\protect\citeauthoryear{{Borisova} et~al.,}{{Borisova}
  et~al.}{2016}]{borisova16}
{Borisova} E.,  et~al., 2016, \mn@doi [\apj] {10.3847/0004-637X/831/1/39},
  \href {https://ui.adsabs.harvard.edu/abs/2016ApJ...831...39B} {831, 39}

\bibitem[\protect\citeauthoryear{{Bouch{\'e}}, {Murphy}, {P{\'e}roux}, {Csabai}
   \& {Wild}}{{Bouch{\'e}} et~al.}{2006}]{bouche06}
{Bouch{\'e}} N.,  {Murphy} M.~T.,  {P{\'e}roux} C.,  {Csabai} I.,   {Wild} V.,
  2006, \mn@doi [\mnras] {10.1111/j.1365-2966.2006.10685.x}, \href
  {https://ui.adsabs.harvard.edu/abs/2006MNRAS.371..495B} {371, 495}

\bibitem[\protect\citeauthoryear{{Bouch{\'e}}, {Hohensee}, {Vargas},
  {Kacprzak}, {Martin}, {Cooke}  \& {Churchill}}{{Bouch{\'e}}
  et~al.}{2012}]{bouche12}
{Bouch{\'e}} N.,  {Hohensee} W.,  {Vargas} R.,  {Kacprzak} G.~G.,  {Martin}
  C.~L.,  {Cooke} J.,   {Churchill} C.~W.,  2012, \mn@doi [\mnras]
  {10.1111/j.1365-2966.2012.21114.x}, \href
  {https://ui.adsabs.harvard.edu/abs/2012MNRAS.426..801B} {426, 801}

\bibitem[\protect\citeauthoryear{{Bouch{\'e}} et~al.,}{{Bouch{\'e}}
  et~al.}{2021}]{bouche21}
{Bouch{\'e}} N.~F.,  et~al., 2021, arXiv e-prints, \href
  {https://ui.adsabs.harvard.edu/abs/2021arXiv210112250B} {p. arXiv:2101.12250}

\bibitem[\protect\citeauthoryear{{Burchett}, {Tripp}, {Wang}, {Willmer},
  {Bowen}  \& {Jenkins}}{{Burchett} et~al.}{2018}]{burchett18}
{Burchett} J.~N.,  {Tripp} T.~M.,  {Wang} Q.~D.,  {Willmer} C. N.~A.,  {Bowen}
  D.~V.,   {Jenkins} E.~B.,  2018, \mn@doi [\mnras] {10.1093/mnras/stx3170},
  \href {https://ui.adsabs.harvard.edu/abs/2018MNRAS.475.2067B} {475, 2067}

\bibitem[\protect\citeauthoryear{{Burchett}, {Rubin}, {Prochaska}, {Coil},
  {Rickards Vaught}  \& {Hennawi}}{{Burchett} et~al.}{2020}]{burchett20}
{Burchett} J.~N.,  {Rubin} K. H.~R.,  {Prochaska} J.~X.,  {Coil} A.~L.,
  {Rickards Vaught} R.,   {Hennawi} J.~F.,  2020, arXiv e-prints, \href
  {https://ui.adsabs.harvard.edu/abs/2020arXiv200503017B} {p. arXiv:2005.03017}

\bibitem[\protect\citeauthoryear{{Byler}, {Dalcanton}, {Conroy}  \&
  {Johnson}}{{Byler} et~al.}{2017}]{byler17}
{Byler} N.,  {Dalcanton} J.~J.,  {Conroy} C.,   {Johnson} B.~D.,  2017, \mn@doi
  [\apj] {10.3847/1538-4357/aa6c66}, \href
  {http://adsabs.harvard.edu/abs/2017ApJ...840...44B} {840, 44}

\bibitem[\protect\citeauthoryear{{Byrohl}, {Nelson}, {Behrens}, {Pillepich},
  {Hernquist}, {Marinacci}  \& {Vogelsberger}}{{Byrohl}
  et~al.}{2020}]{byrohl20}
{Byrohl} C.,  {Nelson} D.,  {Behrens} C.,  {Pillepich} A.,  {Hernquist} L.,
  {Marinacci} F.,   {Vogelsberger} M.,  2020, arXiv e-prints, \href
  {https://ui.adsabs.harvard.edu/abs/2020arXiv200907283B} {p. arXiv:2009.07283}

\bibitem[\protect\citeauthoryear{{Cai} et~al.,}{{Cai} et~al.}{2019}]{cai19}
{Cai} Z.,  et~al., 2019, \mn@doi [\apjs] {10.3847/1538-4365/ab4796}, \href
  {https://ui.adsabs.harvard.edu/abs/2019ApJS..245...23C} {245, 23}

\bibitem[\protect\citeauthoryear{{Cantalupo} et~al.,}{{Cantalupo}
  et~al.}{2019}]{cantalupo19}
{Cantalupo} S.,  et~al., 2019, \mn@doi [\mnras] {10.1093/mnras/sty3481}, \href
  {https://ui.adsabs.harvard.edu/abs/2019MNRAS.483.5188C} {483, 5188}

\bibitem[\protect\citeauthoryear{{Chen}, {Lanzetta}  \& {Webb}}{{Chen}
  et~al.}{2001}]{chen01}
{Chen} H.-W.,  {Lanzetta} K.~M.,   {Webb} J.~K.,  2001, \mn@doi [\apj]
  {10.1086/321537}, \href
  {https://ui.adsabs.harvard.edu/abs/2001ApJ...556..158C} {556, 158}

\bibitem[\protect\citeauthoryear{{Chen} et~al.,}{{Chen} et~al.}{2021}]{chen21}
{Chen} Y.,  et~al., 2021, arXiv e-prints, \href
  {https://ui.adsabs.harvard.edu/abs/2021arXiv210410173C} {p. arXiv:2104.10173}

\bibitem[\protect\citeauthoryear{{Chisholm}, {Prochaska}, {Schaerer},
  {Gazagnes}  \& {Henry}}{{Chisholm} et~al.}{2020}]{chisholm20}
{Chisholm} J.,  {Prochaska} J.~X.,  {Schaerer} D.,  {Gazagnes} S.,   {Henry}
  A.,  2020, \mn@doi [\mnras] {10.1093/mnras/staa2470}, \href
  {https://ui.adsabs.harvard.edu/abs/2020MNRAS.498.2554C} {498, 2554}

\bibitem[\protect\citeauthoryear{{Cicone}, {Brusa}, {Ramos Almeida}, {Cresci},
  {Husemann}  \& {Mainieri}}{{Cicone} et~al.}{2018}]{cicone18}
{Cicone} C.,  {Brusa} M.,  {Ramos Almeida} C.,  {Cresci} G.,  {Husemann} B.,
  {Mainieri} V.,  2018, \mn@doi [Nature Astronomy] {10.1038/s41550-018-0406-3},
  \href {http://adsabs.harvard.edu/abs/2018NatAs...2..176C} {2, 176}

\bibitem[\protect\citeauthoryear{Cleveland}{Cleveland}{1979}]{cleveland79}
Cleveland W.~S.,  1979, Journal of the American Statistical Association, 74,
  829

\bibitem[\protect\citeauthoryear{{Corlies} \& {Schiminovich}}{{Corlies} \&
  {Schiminovich}}{2016}]{corlies16}
{Corlies} L.,  {Schiminovich} D.,  2016, \mn@doi [\apj]
  {10.3847/0004-637X/827/2/148}, \href
  {http://adsabs.harvard.edu/abs/2016ApJ...827..148C} {827, 148}

\bibitem[\protect\citeauthoryear{{Corlies}, {Peeples}, {Tumlinson}, {O'Shea},
  {Lehner}, {Howk}, {O'Meara}  \& {Smith}}{{Corlies} et~al.}{2020}]{corlies20}
{Corlies} L.,  {Peeples} M.~S.,  {Tumlinson} J.,  {O'Shea} B.~W.,  {Lehner} N.,
   {Howk} J.~C.,  {O'Meara} J.~M.,   {Smith} B.~D.,  2020, \mn@doi [\apj]
  {10.3847/1538-4357/ab9310}, \href
  {https://ui.adsabs.harvard.edu/abs/2020ApJ...896..125C} {896, 125}

\bibitem[\protect\citeauthoryear{{De Cia}, {Ledoux}, {Mattsson}, {Petitjean},
  {Srianand}, {Gavignaud}  \& {Jenkins}}{{De Cia} et~al.}{2016}]{decia16}
{De Cia} A.,  {Ledoux} C.,  {Mattsson} L.,  {Petitjean} P.,  {Srianand} R.,
  {Gavignaud} I.,   {Jenkins} E.~B.,  2016, \mn@doi [\aap]
  {10.1051/0004-6361/201527895}, \href
  {https://ui.adsabs.harvard.edu/abs/2016A&A...596A..97D} {596, A97}

\bibitem[\protect\citeauthoryear{{De Cia}, {Ledoux}, {Petitjean}  \&
  {Savaglio}}{{De Cia} et~al.}{2018}]{decia18}
{De Cia} A.,  {Ledoux} C.,  {Petitjean} P.,   {Savaglio} S.,  2018, \mn@doi
  [\aap] {10.1051/0004-6361/201731970}, \href
  {https://ui.adsabs.harvard.edu/abs/2018A&A...611A..76D} {611, A76}

\bibitem[\protect\citeauthoryear{{DeFelippis}, {Bouch{\'e}}, {Genel}, {Bryan},
  {Nelson}, {Marinacci}  \& {Hernquist}}{{DeFelippis}
  et~al.}{2021}]{defelippis21}
{DeFelippis} D.,  {Bouch{\'e}} N.~F.,  {Genel} S.,  {Bryan} G.~L.,  {Nelson}
  D.,  {Marinacci} F.,   {Hernquist} L.,  2021, arXiv e-prints, \href
  {https://ui.adsabs.harvard.edu/abs/2021arXiv210208383D} {p. arXiv:2102.08383}

\bibitem[\protect\citeauthoryear{{Donnari} et~al.,}{{Donnari}
  et~al.}{2020}]{donnari20a}
{Donnari} M.,  et~al., 2020, arXiv e-prints, \href
  {https://ui.adsabs.harvard.edu/abs/2020arXiv200800005D} {p. arXiv:2008.00005}

\bibitem[\protect\citeauthoryear{{Dutta} et~al.,}{{Dutta}
  et~al.}{2020}]{dutta20}
{Dutta} R.,  et~al., 2020, \mn@doi [\mnras] {10.1093/mnras/staa3147}, \href
  {https://ui.adsabs.harvard.edu/abs/2020MNRAS.499.5022D} {499, 5022}

\bibitem[\protect\citeauthoryear{{Erb}, {Quider}, {Henry}  \& {Martin}}{{Erb}
  et~al.}{2012}]{erb12}
{Erb} D.~K.,  {Quider} A.~M.,  {Henry} A.~L.,   {Martin} C.~L.,  2012, \mn@doi
  [\apj] {10.1088/0004-637X/759/1/26}, \href
  {http://adsabs.harvard.edu/abs/2012ApJ...759...26E} {759, 26}

\bibitem[\protect\citeauthoryear{{Fang}, {Croft}, {Sanders}, {Houck},
  {Dav{\'e}}, {Katz}, {Weinberg}  \& {Hernquist}}{{Fang} et~al.}{2005}]{fang05}
{Fang} T.,  {Croft} R.~A.~C.,  {Sanders} W.~T.,  {Houck} J.,  {Dav{\'e}} R.,
  {Katz} N.,  {Weinberg} D.~H.,   {Hernquist} L.,  2005, \mn@doi [\apj]
  {10.1086/428656}, \href {http://adsabs.harvard.edu/abs/2005ApJ...623..612F}
  {623, 612}

\bibitem[\protect\citeauthoryear{{Farina} et~al.,}{{Farina}
  et~al.}{2019}]{farina19}
{Farina} E.~P.,  et~al., 2019, \mn@doi [\apj] {10.3847/1538-4357/ab5847}, \href
  {https://ui.adsabs.harvard.edu/abs/2019ApJ...887..196F} {887, 196}

\bibitem[\protect\citeauthoryear{{Faucher-Gigu{\`e}re}, {Lidz}, {Zaldarriaga}
  \& {Hernquist}}{{Faucher-Gigu{\`e}re} et~al.}{2009}]{fg09}
{Faucher-Gigu{\`e}re} C.-A.,  {Lidz} A.,  {Zaldarriaga} M.,   {Hernquist} L.,
  2009, \mn@doi [\apj] {10.1088/0004-637X/703/2/1416}, \href
  {http://adsabs.harvard.edu/abs/2009ApJ...703.1416F} {703, 1416}

\bibitem[\protect\citeauthoryear{{Feltre} et~al.,}{{Feltre}
  et~al.}{2018}]{feltre18}
{Feltre} A.,  et~al., 2018, preprint, \href
  {http://adsabs.harvard.edu/abs/2018arXiv180601864F} {} (\mn@eprint {arXiv}
  {1806.01864})

\bibitem[\protect\citeauthoryear{{Ferland} et~al.,}{{Ferland}
  et~al.}{2017}]{ferland17}
{Ferland} G.~J.,  et~al., 2017, \rmxaa, \href
  {https://ui.adsabs.harvard.edu/abs/2017RMxAA..53..385F} {53, 385}

\bibitem[\protect\citeauthoryear{{Finley} et~al.,}{{Finley}
  et~al.}{2017a}]{finley17a}
{Finley} H.,  et~al., 2017a, \mn@doi [\aap] {10.1051/0004-6361/201730428},
  \href {http://adsabs.harvard.edu/abs/2017A%26A...605A.118F} {605, A118}

\bibitem[\protect\citeauthoryear{{Finley} et~al.,}{{Finley}
  et~al.}{2017b}]{finley17b}
{Finley} H.,  et~al., 2017b, \mn@doi [\aap] {10.1051/0004-6361/201731499},
  \href {http://adsabs.harvard.edu/abs/2017A%26A...608A...7F} {608, A7}

\bibitem[\protect\citeauthoryear{{Frank} et~al.,}{{Frank}
  et~al.}{2012}]{frank12}
{Frank} S.,  et~al., 2012, \mn@doi [\mnras] {10.1111/j.1365-2966.2011.20172.x},
  \href {http://adsabs.harvard.edu/abs/2012MNRAS.420.1731F} {420, 1731}

\bibitem[\protect\citeauthoryear{{Fraternali} \& {Binney}}{{Fraternali} \&
  {Binney}}{2006}]{fraternali06}
{Fraternali} F.,  {Binney} J.~J.,  2006, \mn@doi [\mnras]
  {10.1111/j.1365-2966.2005.09816.x}, \href
  {https://ui.adsabs.harvard.edu/abs/2006MNRAS.366..449F} {366, 449}

\bibitem[\protect\citeauthoryear{{Fraternali} \& {Binney}}{{Fraternali} \&
  {Binney}}{2008}]{fraternali08}
{Fraternali} F.,  {Binney} J.~J.,  2008, \mn@doi [\mnras]
  {10.1111/j.1365-2966.2008.13071.x}, \href
  {https://ui.adsabs.harvard.edu/abs/2008MNRAS.386..935F} {386, 935}

\bibitem[\protect\citeauthoryear{{Ginolfi} et~al.,}{{Ginolfi}
  et~al.}{2020}]{ginolfi20}
{Ginolfi} M.,  et~al., 2020, \mn@doi [\aap] {10.1051/0004-6361/201936872},
  \href {https://ui.adsabs.harvard.edu/abs/2020A&A...633A..90G} {633, A90}

\bibitem[\protect\citeauthoryear{{Grevesse}, {Asplund}, {Sauval}  \&
  {Scott}}{{Grevesse} et~al.}{2010}]{grevesse10}
{Grevesse} N.,  {Asplund} M.,  {Sauval} A.~J.,   {Scott} P.,  2010, \mn@doi
  [\apss] {10.1007/s10509-010-0288-z}, \href
  {http://adsabs.harvard.edu/abs/2010Ap%26SS.328..179G} {328, 179}

\bibitem[\protect\citeauthoryear{{Hayes}, {Melinder}, {{\"O}stlin}, {Scarlata},
  {Lehnert}  \& {Mannerstr{\"o}m-Jansson}}{{Hayes} et~al.}{2016}]{hayes16}
{Hayes} M.,  {Melinder} J.,  {{\"O}stlin} G.,  {Scarlata} C.,  {Lehnert} M.~D.,
    {Mannerstr{\"o}m-Jansson} G.,  2016, \mn@doi [\apj]
  {10.3847/0004-637X/828/1/49}, \href
  {http://adsabs.harvard.edu/abs/2016ApJ...828...49H} {828, 49}

\bibitem[\protect\citeauthoryear{{Henry}, {Berg}, {Scarlata}, {Verhamme}  \&
  {Erb}}{{Henry} et~al.}{2018}]{henry18}
{Henry} A.,  {Berg} D.~A.,  {Scarlata} C.,  {Verhamme} A.,   {Erb} D.,  2018,
  \mn@doi [\apj] {10.3847/1538-4357/aab099}, \href
  {https://ui.adsabs.harvard.edu/abs/2018ApJ...855...96H} {855, 96}

\bibitem[\protect\citeauthoryear{{Kacprzak}, {Churchill}  \&
  {Nielsen}}{{Kacprzak} et~al.}{2012}]{kacprzak12}
{Kacprzak} G.~G.,  {Churchill} C.~W.,   {Nielsen} N.~M.,  2012, \mn@doi [\apjl]
  {10.1088/2041-8205/760/1/L7}, \href
  {https://ui.adsabs.harvard.edu/abs/2012ApJ...760L...7K} {760, L7}

\bibitem[\protect\citeauthoryear{{Katz} et~al.,}{{Katz} et~al.}{2019}]{katz19}
{Katz} H.,  et~al., 2019, \mn@doi [\mnras] {10.1093/mnras/stz1672}, \href
  {https://ui.adsabs.harvard.edu/abs/2019MNRAS.487.5902K} {487, 5902}

\bibitem[\protect\citeauthoryear{{Kova{\v{c}}} et~al.,}{{Kova{\v{c}}}
  et~al.}{2010}]{kovac10}
{Kova{\v{c}}} K.,  et~al., 2010, \mn@doi [\apj] {10.1088/0004-637X/708/1/505},
  \href {https://ui.adsabs.harvard.edu/abs/2010ApJ...708..505K} {708, 505}

\bibitem[\protect\citeauthoryear{{Kravtsov}, {Klypin}  \& {Hoffman}}{{Kravtsov}
  et~al.}{2002}]{kravtsov02}
{Kravtsov} A.~V.,  {Klypin} A.,   {Hoffman} Y.,  2002, \mn@doi [\apj]
  {10.1086/340046}, \href {http://adsabs.harvard.edu/abs/2002ApJ...571..563K}
  {571, 563}

\bibitem[\protect\citeauthoryear{{Lan} \& {Mo}}{{Lan} \& {Mo}}{2018}]{lan18}
{Lan} T.-W.,  {Mo} H.,  2018, \mn@doi [\apj] {10.3847/1538-4357/aadc08}, \href
  {https://ui.adsabs.harvard.edu/abs/2018ApJ...866...36L} {866, 36}

\bibitem[\protect\citeauthoryear{{Lan}, {M{\'e}nard}  \& {Zhu}}{{Lan}
  et~al.}{2014}]{lan14}
{Lan} T.-W.,  {M{\'e}nard} B.,   {Zhu} G.,  2014, \mn@doi [\apj]
  {10.1088/0004-637X/795/1/31}, \href
  {https://ui.adsabs.harvard.edu/abs/2014ApJ...795...31L} {795, 31}

\bibitem[\protect\citeauthoryear{{Leclercq} et~al.,}{{Leclercq}
  et~al.}{2017}]{leclercq17}
{Leclercq} F.,  et~al., 2017, \mn@doi [\aap] {10.1051/0004-6361/201731480},
  \href {https://ui.adsabs.harvard.edu/abs/2017A&A...608A...8L} {608, A8}

\bibitem[\protect\citeauthoryear{{Leclercq} et~al.,}{{Leclercq}
  et~al.}{2020}]{leclercq20}
{Leclercq} F.,  et~al., 2020, \mn@doi [\aap] {10.1051/0004-6361/201937339},
  \href {https://ui.adsabs.harvard.edu/abs/2020A&A...635A..82L} {635, A82}

\bibitem[\protect\citeauthoryear{{Li}, {Bregman}, {Wang}, {Crain}, {Anderson}
  \& {Zhang}}{{Li} et~al.}{2017}]{lijt17}
{Li} J.-T.,  {Bregman} J.~N.,  {Wang} Q.~D.,  {Crain} R.~A.,  {Anderson} M.~E.,
    {Zhang} S.,  2017, \mn@doi [\apjs] {10.3847/1538-4365/aa96fc}, \href
  {https://ui.adsabs.harvard.edu/abs/2017ApJS..233...20L} {233, 20}

\bibitem[\protect\citeauthoryear{{Lokhorst}, {Abraham}, {van Dokkum}, {Wijers}
  \& {Schaye}}{{Lokhorst} et~al.}{2019}]{lokhorst19}
{Lokhorst} D.,  {Abraham} R.,  {van Dokkum} P.,  {Wijers} N.,   {Schaye} J.,
  2019, \mn@doi [\apj] {10.3847/1538-4357/ab184e}, \href
  {https://ui.adsabs.harvard.edu/abs/2019ApJ...877....4L} {877, 4}

\bibitem[\protect\citeauthoryear{{Lopez} et~al.,}{{Lopez}
  et~al.}{2018}]{lopez18}
{Lopez} S.,  et~al., 2018, \mn@doi [\nat] {10.1038/nature25436}, \href
  {https://ui.adsabs.harvard.edu/abs/2018Natur.554..493L} {554, 493}

\bibitem[\protect\citeauthoryear{{Mackenzie} et~al.,}{{Mackenzie}
  et~al.}{2021}]{mackenzie21}
{Mackenzie} R.,  et~al., 2021, \mn@doi [\mnras] {10.1093/mnras/staa3277}, \href
  {https://ui.adsabs.harvard.edu/abs/2021MNRAS.502..494M} {502, 494}

\bibitem[\protect\citeauthoryear{{Marinacci} et~al.,}{{Marinacci}
  et~al.}{2018}]{marinacci18}
{Marinacci} F.,  et~al., 2018, \mn@doi [\mnras] {10.1093/mnras/sty2206}, \href
  {http://adsabs.harvard.edu/abs/2018MNRAS.480.5113M} {480, 5113}

\bibitem[\protect\citeauthoryear{{Martin}, {Shapley}, {Coil}, {Kornei},
  {Bundy}, {Weiner}, {Noeske}  \& {Schiminovich}}{{Martin}
  et~al.}{2012}]{martin12}
{Martin} C.~L.,  {Shapley} A.~E.,  {Coil} A.~L.,  {Kornei} K.~A.,  {Bundy} K.,
  {Weiner} B.~J.,  {Noeske} K.~G.,   {Schiminovich} D.,  2012, \mn@doi [\apj]
  {10.1088/0004-637X/760/2/127}, \href
  {http://adsabs.harvard.edu/abs/2012ApJ...760..127M} {760, 127}

\bibitem[\protect\citeauthoryear{{Martin}, {Shapley}, {Coil}, {Kornei},
  {Murray}  \& {Pancoast}}{{Martin} et~al.}{2013}]{martin13}
{Martin} C.~L.,  {Shapley} A.~E.,  {Coil} A.~L.,  {Kornei} K.~A.,  {Murray} N.,
    {Pancoast} A.,  2013, \mn@doi [\apj] {10.1088/0004-637X/770/1/41}, \href
  {http://adsabs.harvard.edu/abs/2013ApJ...770...41M} {770, 41}

\bibitem[\protect\citeauthoryear{{Martin} et~al.,}{{Martin}
  et~al.}{2019a}]{martin19_kcwi}
{Martin} D.~C.,  et~al., 2019a, \mn@doi [Nature Astronomy]
  {10.1038/s41550-019-0791-2}, \href
  {https://ui.adsabs.harvard.edu/abs/2019NatAs...3..822M} {3, 822}

\bibitem[\protect\citeauthoryear{{Martin}, {Ho}, {Kacprzak}  \&
  {Churchill}}{{Martin} et~al.}{2019b}]{martin19}
{Martin} C.~L.,  {Ho} S.~H.,  {Kacprzak} G.~G.,   {Churchill} C.~W.,  2019b,
  \mn@doi [\apj] {10.3847/1538-4357/ab18ac}, \href
  {https://ui.adsabs.harvard.edu/abs/2019ApJ...878...84M} {878, 84}

\bibitem[\protect\citeauthoryear{{Mitchell}, {Blaizot}, {Cadiou}, {Dubois},
  {Garel}  \& {Rosdahl}}{{Mitchell} et~al.}{2021}]{mitchell21}
{Mitchell} P.~D.,  {Blaizot} J.,  {Cadiou} C.,  {Dubois} Y.,  {Garel} T.,
  {Rosdahl} J.,  2021, \mn@doi [\mnras] {10.1093/mnras/stab035}, \href
  {https://ui.adsabs.harvard.edu/abs/2021MNRAS.501.5757M} {501, 5757}

\bibitem[\protect\citeauthoryear{{Naiman} et~al.,}{{Naiman}
  et~al.}{2018}]{naiman18}
{Naiman} J.~P.,  et~al., 2018, \mn@doi [\mnras] {10.1093/mnras/sty618}, \href
  {http://adsabs.harvard.edu/abs/2018MNRAS.477.1206N} {477, 1206}

\bibitem[\protect\citeauthoryear{{Nelson} et~al.,}{{Nelson}
  et~al.}{2018a}]{nelson18a}
{Nelson} D.,  et~al., 2018a, \mn@doi [\mnras] {10.1093/mnras/stx3040}, \href
  {http://adsabs.harvard.edu/abs/2018MNRAS.475..624N} {475, 624}

\bibitem[\protect\citeauthoryear{{Nelson} et~al.,}{{Nelson}
  et~al.}{2018b}]{nelson18b}
{Nelson} D.,  et~al., 2018b, \mn@doi [\mnras] {10.1093/mnras/sty656}, \href
  {http://adsabs.harvard.edu/abs/2018MNRAS.477..450N} {477, 450}

\bibitem[\protect\citeauthoryear{{Nelson} et~al.,}{{Nelson}
  et~al.}{2019a}]{nelson19a}
{Nelson} D.,  et~al., 2019a, \mn@doi [Computational Astrophysics and Cosmology]
  {10.1186/s40668-019-0028-x}, \href
  {https://ui.adsabs.harvard.edu/abs/2019ComAC...6....2N} {6, 2}

\bibitem[\protect\citeauthoryear{{Nelson} et~al.,}{{Nelson}
  et~al.}{2019b}]{nelson19b}
{Nelson} D.,  et~al., 2019b, \mn@doi [\mnras] {10.1093/mnras/stz2306}, \href
  {https://ui.adsabs.harvard.edu/abs/2019MNRAS.490.3234N} {490, 3234}

\bibitem[\protect\citeauthoryear{{Nelson} et~al.,}{{Nelson}
  et~al.}{2020}]{nelson20}
{Nelson} D.,  et~al., 2020, \mn@doi [\mnras] {10.1093/mnras/staa2419}, \href
  {https://ui.adsabs.harvard.edu/abs/2020MNRAS.498.2391N} {498, 2391}

\bibitem[\protect\citeauthoryear{{O'Sullivan}, {Martin}, {Matuszewski},
  {Hoadley}, {Hamden}, {Neill}, {Lin}  \& {Parihar}}{{O'Sullivan}
  et~al.}{2020}]{osullivan20}
{O'Sullivan} D.~B.,  {Martin} C.,  {Matuszewski} M.,  {Hoadley} K.,  {Hamden}
  E.,  {Neill} J.~D.,  {Lin} Z.,   {Parihar} P.,  2020, \mn@doi [\apj]
  {10.3847/1538-4357/ab838c}, \href
  {https://ui.adsabs.harvard.edu/abs/2020ApJ...894....3O} {894, 3}

\bibitem[\protect\citeauthoryear{{Pakmor} \& {Springel}}{{Pakmor} \&
  {Springel}}{2013}]{pakmor13}
{Pakmor} R.,  {Springel} V.,  2013, \mn@doi [\mnras] {10.1093/mnras/stt428},
  \href {http://adsabs.harvard.edu/abs/2013MNRAS.432..176P} {432, 176}

\bibitem[\protect\citeauthoryear{{Pakmor}, {Bauer}  \& {Springel}}{{Pakmor}
  et~al.}{2011}]{pakmor11}
{Pakmor} R.,  {Bauer} A.,   {Springel} V.,  2011, \mn@doi [\mnras]
  {10.1111/j.1365-2966.2011.19591.x}, \href
  {http://adsabs.harvard.edu/abs/2011MNRAS.418.1392P} {418, 1392}

\bibitem[\protect\citeauthoryear{{P{\'e}roux} \& {Howk}}{{P{\'e}roux} \&
  {Howk}}{2020}]{peroux20a}
{P{\'e}roux} C.,  {Howk} J.~C.,  2020, \mn@doi [\araa]
  {10.1146/annurev-astro-021820-120014}, \href
  {https://ui.adsabs.harvard.edu/abs/2020ARA&A..58..363P} {58, 363}

\bibitem[\protect\citeauthoryear{{P{\'e}roux}, {Rahmani}, {Arrigoni Battaia}
  \& {Augustin}}{{P{\'e}roux} et~al.}{2018}]{peroux18}
{P{\'e}roux} C.,  {Rahmani} H.,  {Arrigoni Battaia} F.,   {Augustin} R.,  2018,
  \mn@doi [\mnras] {10.1093/mnrasl/sly090}, \href
  {https://ui.adsabs.harvard.edu/abs/2018MNRAS.479L..50P} {479, L50}

\bibitem[\protect\citeauthoryear{{Pillepich} et~al.,}{{Pillepich}
  et~al.}{2018a}]{pillepich18a}
{Pillepich} A.,  et~al., 2018a, \mn@doi [\mnras] {10.1093/mnras/stx2656}, \href
  {http://adsabs.harvard.edu/abs/2018MNRAS.473.4077P} {473, 4077}

\bibitem[\protect\citeauthoryear{{Pillepich} et~al.,}{{Pillepich}
  et~al.}{2018b}]{pillepich18b}
{Pillepich} A.,  et~al., 2018b, \mn@doi [\mnras] {10.1093/mnras/stx3112}, \href
  {http://adsabs.harvard.edu/abs/2018MNRAS.475..648P} {475, 648}

\bibitem[\protect\citeauthoryear{{Pillepich} et~al.,}{{Pillepich}
  et~al.}{2019}]{pillepich19}
{Pillepich} A.,  et~al., 2019, \mn@doi [\mnras] {10.1093/mnras/stz2338}, \href
  {https://ui.adsabs.harvard.edu/abs/2019MNRAS.490.3196P} {490, 3196}

\bibitem[\protect\citeauthoryear{{Planck Collaboration}}{{Planck
  Collaboration}}{2016}]{planck2015_xiii}
{Planck Collaboration} 2016, \mn@doi [\aap] {10.1051/0004-6361/201525830},
  \href {http://adsabs.harvard.edu/abs/2016A%26A...594A..13P} {594, A13}

\bibitem[\protect\citeauthoryear{{Predehl} et~al.,}{{Predehl}
  et~al.}{2020}]{predehl20}
{Predehl} P.,  et~al., 2020, \mn@doi [\nat] {10.1038/s41586-020-2979-0}, \href
  {https://ui.adsabs.harvard.edu/abs/2020Natur.588..227P} {588, 227}

\bibitem[\protect\citeauthoryear{{Prochaska}, {Kasen}  \& {Rubin}}{{Prochaska}
  et~al.}{2011}]{prochaska11a}
{Prochaska} J.~X.,  {Kasen} D.,   {Rubin} K.,  2011, \mn@doi [\apj]
  {10.1088/0004-637X/734/1/24}, \href
  {https://ui.adsabs.harvard.edu/abs/2011ApJ...734...24P} {734, 24}

\bibitem[\protect\citeauthoryear{{Rahmati}, {Pawlik}, {Rai{\v c}evic}  \&
  {Schaye}}{{Rahmati} et~al.}{2013}]{rahmati13}
{Rahmati} A.,  {Pawlik} A.~H.,  {Rai{\v c}evic} M.,   {Schaye} J.,  2013,
  \mn@doi [\mnras] {10.1093/mnras/stt066}, \href
  {http://adsabs.harvard.edu/abs/2013MNRAS.430.2427R} {430, 2427}

\bibitem[\protect\citeauthoryear{{Rauch}, {Sargent}, {Barlow}  \&
  {Carswell}}{{Rauch} et~al.}{2001}]{rauch01}
{Rauch} M.,  {Sargent} W.~L.~W.,  {Barlow} T.~A.,   {Carswell} R.~F.,  2001,
  \mn@doi [\apj] {10.1086/323523}, \href
  {http://adsabs.harvard.edu/abs/2001ApJ...562...76R} {562, 76}

\bibitem[\protect\citeauthoryear{{Richard} et~al.,}{{Richard}
  et~al.}{2019}]{richard19}
{Richard} J.,  et~al., 2019, arXiv e-prints, \href
  {https://ui.adsabs.harvard.edu/abs/2019arXiv190601657R} {p. arXiv:1906.01657}

\bibitem[\protect\citeauthoryear{{Rickards Vaught}, {Rubin}, {Arrigoni
  Battaia}, {Prochaska}  \& {Hennawi}}{{Rickards Vaught}
  et~al.}{2019}]{rickards19}
{Rickards Vaught} R.~J.,  {Rubin} K. H.~R.,  {Arrigoni Battaia} F.,
  {Prochaska} J.~X.,   {Hennawi} J.~F.,  2019, \mn@doi [\apj]
  {10.3847/1538-4357/ab211f}, \href
  {https://ui.adsabs.harvard.edu/abs/2019ApJ...879....7R} {879, 7}

\bibitem[\protect\citeauthoryear{{Rodriguez-Gomez} et~al.,}{{Rodriguez-Gomez}
  et~al.}{2019}]{rodriguezgomez19}
{Rodriguez-Gomez} V.,  et~al., 2019, \mn@doi [\mnras] {10.1093/mnras/sty3345},
  \href {https://ui.adsabs.harvard.edu/abs/2019MNRAS.483.4140R} {483, 4140}

\bibitem[\protect\citeauthoryear{{Rubin}, {Weiner}, {Koo}, {Martin},
  {Prochaska}, {Coil}  \& {Newman}}{{Rubin} et~al.}{2010}]{rubin10}
{Rubin} K.~H.~R.,  {Weiner} B.~J.,  {Koo} D.~C.,  {Martin} C.~L.,  {Prochaska}
  J.~X.,  {Coil} A.~L.,   {Newman} J.~A.,  2010, \mn@doi [\apj]
  {10.1088/0004-637X/719/2/1503}, \href
  {http://adsabs.harvard.edu/abs/2010ApJ...719.1503R} {719, 1503}

\bibitem[\protect\citeauthoryear{{Rubin}, {Prochaska}, {M{\'e}nard}, {Murray},
  {Kasen}, {Koo}  \& {Phillips}}{{Rubin} et~al.}{2011}]{rubin11}
{Rubin} K.~H.~R.,  {Prochaska} J.~X.,  {M{\'e}nard} B.,  {Murray} N.,  {Kasen}
  D.,  {Koo} D.~C.,   {Phillips} A.~C.,  2011, \mn@doi [\apj]
  {10.1088/0004-637X/728/1/55}, \href
  {http://adsabs.harvard.edu/abs/2011ApJ...728...55R} {728, 55}

\bibitem[\protect\citeauthoryear{{Rubin}, {Prochaska}, {Koo}, {Phillips},
  {Martin}  \& {Winstrom}}{{Rubin} et~al.}{2014}]{rubin14}
{Rubin} K.~H.~R.,  {Prochaska} J.~X.,  {Koo} D.~C.,  {Phillips} A.~C.,
  {Martin} C.~L.,   {Winstrom} L.~O.,  2014, \mn@doi [\apj]
  {10.1088/0004-637X/794/2/156}, \href
  {http://adsabs.harvard.edu/abs/2014ApJ...794..156R} {794, 156}

\bibitem[\protect\citeauthoryear{{Rupke} et~al.,}{{Rupke}
  et~al.}{2019}]{rupke19}
{Rupke} D. S.~N.,  et~al., 2019, \mn@doi [\nat] {10.1038/s41586-019-1686-1},
  \href {https://ui.adsabs.harvard.edu/abs/2019Natur.574..643R} {574, 643}

\bibitem[\protect\citeauthoryear{{Scannapieco}}{{Scannapieco}}{2017}]{scannapieco17}
{Scannapieco} E.,  2017, \mn@doi [\apj] {10.3847/1538-4357/aa5d0d}, \href
  {http://adsabs.harvard.edu/abs/2017ApJ...837...28S} {837, 28}

\bibitem[\protect\citeauthoryear{{Schneider}, {Robertson}  \&
  {Thompson}}{{Schneider} et~al.}{2018}]{schneider18b}
{Schneider} E.~E.,  {Robertson} B.~E.,   {Thompson} T.~A.,  2018, \mn@doi
  [\apj] {10.3847/1538-4357/aacce1}, \href
  {http://adsabs.harvard.edu/abs/2018ApJ...862...56S} {862, 56}

\bibitem[\protect\citeauthoryear{{Springel}}{{Springel}}{2010}]{spr10}
{Springel} V.,  2010, \mn@doi [\mnras] {10.1111/j.1365-2966.2009.15715.x}, 401,
  791

\bibitem[\protect\citeauthoryear{{Springel} \& {Hernquist}}{{Springel} \&
  {Hernquist}}{2003}]{spr03}
{Springel} V.,  {Hernquist} L.,  2003, \mn@doi [\mnras]
  {10.1046/j.1365-8711.2003.06206.x}, 339, 289

\bibitem[\protect\citeauthoryear{{Springel}, {White}, {Tormen}  \&
  {Kauffmann}}{{Springel} et~al.}{2001}]{spr01}
{Springel} V.,  {White} S.~D.~M.,  {Tormen} G.,   {Kauffmann} G.,  2001,
  \mn@doi [\mnras] {10.1046/j.1365-8711.2001.04912.x}, \href
  {http://adsabs.harvard.edu/abs/2001MNRAS.328..726S} {328, 726}

\bibitem[\protect\citeauthoryear{{Springel} et~al.,}{{Springel}
  et~al.}{2018}]{springel18}
{Springel} V.,  et~al., 2018, \mn@doi [\mnras] {10.1093/mnras/stx3304}, \href
  {http://adsabs.harvard.edu/abs/2018MNRAS.475..676S} {475, 676}

\bibitem[\protect\citeauthoryear{{Sravan} et~al.,}{{Sravan}
  et~al.}{2016}]{sravan16}
{Sravan} N.,  et~al., 2016, \mn@doi [\mnras] {10.1093/mnras/stw1962}, \href
  {https://ui.adsabs.harvard.edu/abs/2016MNRAS.463..120S} {463, 120}

\bibitem[\protect\citeauthoryear{{Suresh}, {Nelson}, {Genel}, {Rubin}  \&
  {Hernquist}}{{Suresh} et~al.}{2019}]{suresh19}
{Suresh} J.,  {Nelson} D.,  {Genel} S.,  {Rubin} K. H.~R.,   {Hernquist} L.,
  2019, \mn@doi [\mnras] {10.1093/mnras/sty3402}, \href
  {https://ui.adsabs.harvard.edu/abs/2019MNRAS.483.4040S} {483, 4040}

\bibitem[\protect\citeauthoryear{{Terrazas} et~al.,}{{Terrazas}
  et~al.}{2020}]{terrazas20}
{Terrazas} B.~A.,  et~al., 2020, \mn@doi [\mnras] {10.1093/mnras/staa374},
  \href {https://ui.adsabs.harvard.edu/abs/2020MNRAS.493.1888T} {493, 1888}

\bibitem[\protect\citeauthoryear{{Thompson}, {Quataert}, {Zhang}  \&
  {Weinberg}}{{Thompson} et~al.}{2016}]{thompson16}
{Thompson} T.~A.,  {Quataert} E.,  {Zhang} D.,   {Weinberg} D.~H.,  2016,
  \mn@doi [\mnras] {10.1093/mnras/stv2428}, \href
  {http://adsabs.harvard.edu/abs/2016MNRAS.455.1830T} {455, 1830}

\bibitem[\protect\citeauthoryear{{Tripp}, {Savage}  \& {Jenkins}}{{Tripp}
  et~al.}{2000}]{tripp00}
{Tripp} T.~M.,  {Savage} B.~D.,   {Jenkins} E.~B.,  2000, \mn@doi [\apjl]
  {10.1086/312644}, \href
  {https://ui.adsabs.harvard.edu/abs/2000ApJ...534L...1T} {534, L1}

\bibitem[\protect\citeauthoryear{{Truong} et~al.,}{{Truong}
  et~al.}{2020}]{truong20}
{Truong} N.,  et~al., 2020, \mn@doi [\mnras] {10.1093/mnras/staa685}, \href
  {https://ui.adsabs.harvard.edu/abs/2020MNRAS.tmp..642T} {}

\bibitem[\protect\citeauthoryear{{Tumlinson}, {Peeples}  \& {Werk}}{{Tumlinson}
  et~al.}{2017}]{tumlinson17}
{Tumlinson} J.,  {Peeples} M.~S.,   {Werk} J.~K.,  2017, \mn@doi [\araa]
  {10.1146/annurev-astro-091916-055240}, \href
  {http://adsabs.harvard.edu/abs/2017ARA%26A..55..389T} {55, 389}

\bibitem[\protect\citeauthoryear{{Voit}}{{Voit}}{2021}]{voit21}
{Voit} G.~M.,  2021, \mn@doi [\apjl] {10.3847/2041-8213/abe11f}, \href
  {https://ui.adsabs.harvard.edu/abs/2021ApJ...908L..16V} {908, L16}

\bibitem[\protect\citeauthoryear{{Weinberger} et~al.,}{{Weinberger}
  et~al.}{2017}]{weinberger17}
{Weinberger} R.,  et~al., 2017, \mn@doi [\mnras] {10.1093/mnras/stw2944}, \href
  {http://adsabs.harvard.edu/abs/2017MNRAS.465.3291W} {465, 3291}

\bibitem[\protect\citeauthoryear{{Weiner} et~al.,}{{Weiner}
  et~al.}{2009}]{weiner09}
{Weiner} B.~J.,  et~al., 2009, \mn@doi [\apj] {10.1088/0004-637X/692/1/187},
  \href {http://adsabs.harvard.edu/abs/2009ApJ...692..187W} {692, 187}

\bibitem[\protect\citeauthoryear{{Wiersma}, {Schaye}  \& {Smith}}{{Wiersma}
  et~al.}{2009}]{wiersma09}
{Wiersma} R.~P.~C.,  {Schaye} J.,   {Smith} B.~D.,  2009, \mn@doi [\mnras]
  {10.1111/j.1365-2966.2008.14191.x}, \href
  {http://adsabs.harvard.edu/abs/2009MNRAS.393...99W} {393, 99}

\bibitem[\protect\citeauthoryear{{Witstok}, {Puchwein}, {Kulkarni}, {Smit}  \&
  {Haehnelt}}{{Witstok} et~al.}{2019}]{witstok19}
{Witstok} J.,  {Puchwein} E.,  {Kulkarni} G.,  {Smit} R.,   {Haehnelt} M.~G.,
  2019, arXiv e-prints, \href
  {https://ui.adsabs.harvard.edu/abs/2019arXiv190506954W} {p. arXiv:1905.06954}

\bibitem[\protect\citeauthoryear{{Yoshikawa}, {Yamasaki}, {Suto}, {Ohashi},
  {Mitsuda}, {Tawara}  \& {Furuzawa}}{{Yoshikawa} et~al.}{2003}]{yoshikawa03}
{Yoshikawa} K.,  {Yamasaki} N.~Y.,  {Suto} Y.,  {Ohashi} T.,  {Mitsuda} K.,
  {Tawara} Y.,   {Furuzawa} A.,  2003, \mn@doi [\pasj] {10.1093/pasj/55.5.879},
  \href {http://adsabs.harvard.edu/abs/2003PASJ...55..879Y} {55, 879}

\bibitem[\protect\citeauthoryear{{Zabl} et~al.,}{{Zabl} et~al.}{2021}]{zabl21}
{Zabl} J.,  et~al., 2021, arXiv e-prints, \href
  {https://ui.adsabs.harvard.edu/abs/2021arXiv210514090Z} {p. arXiv:2105.14090}

\bibitem[\protect\citeauthoryear{{Zhang}, {Zaritsky}  \& {Behroozi}}{{Zhang}
  et~al.}{2018}]{zhang18}
{Zhang} H.,  {Zaritsky} D.,   {Behroozi} P.,  2018, \mn@doi [\apj]
  {10.3847/1538-4357/aac6b7}, \href
  {https://ui.adsabs.harvard.edu/abs/2018ApJ...861...34Z} {861, 34}

\bibitem[\protect\citeauthoryear{{van de Voort} \& {Schaye}}{{van de Voort} \&
  {Schaye}}{2013}]{vdv13}
{van de Voort} F.,  {Schaye} J.,  2013, \mn@doi [\mnras]
  {10.1093/mnras/stt115}, \href
  {http://adsabs.harvard.edu/abs/2013MNRAS.430.2688V} {430, 2688}

\makeatother
\end{thebibliography}

\appendix
\section{Impact of Modeling Choices}

In this appendix we explore the impact of a number of model choices.

\begin{figure}
\centering
\includegraphics[angle=0,width=3.4in]{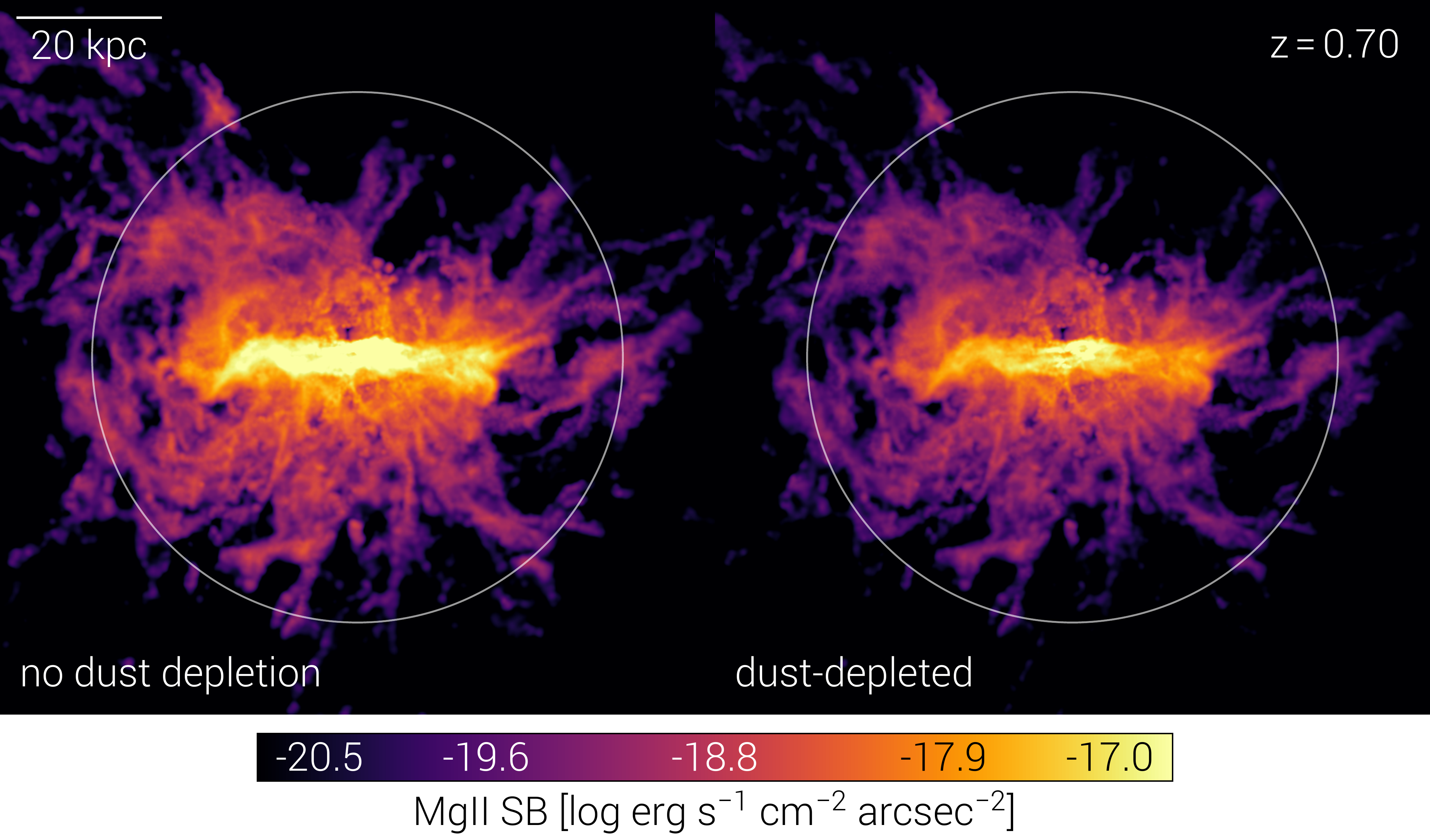}
\caption{ Impact of the MgII dust depletion model on the surface brightness of MgII around a typical $z=0.7$ ($M_\star \simeq 10^{10}$\msun) galaxy, shown edge-on. Magnesium is locked into solid dust grains in greater fractions with higher gas-phase metallicity. As a result, emission from the dense star-forming ISM of the galactic disk is reduced, while extended emission in the CGM is only slightly affected.
 \label{fig_appendix_dustdepletion}}
\end{figure}

Figure \ref{fig_appendix_dustdepletion} shows the role our dust depletion model plays in setting the abundance of MgII. It compares two surface brightness maps for the same galaxy, including dust depletion (right; our fiducial model in this work) and neglecting dust depletion (left), where all Mg is assumed to be gas-phase. Because of the strong metallicity of the dust depletion model MgII emission is preferentially suppressed in the central galaxy itself, while the outskirts are largely unchanged. We believe the dust depleted result to be more faithful to reality.

\begin{figure*}
\centering
\includegraphics[angle=0,width=2.25in]{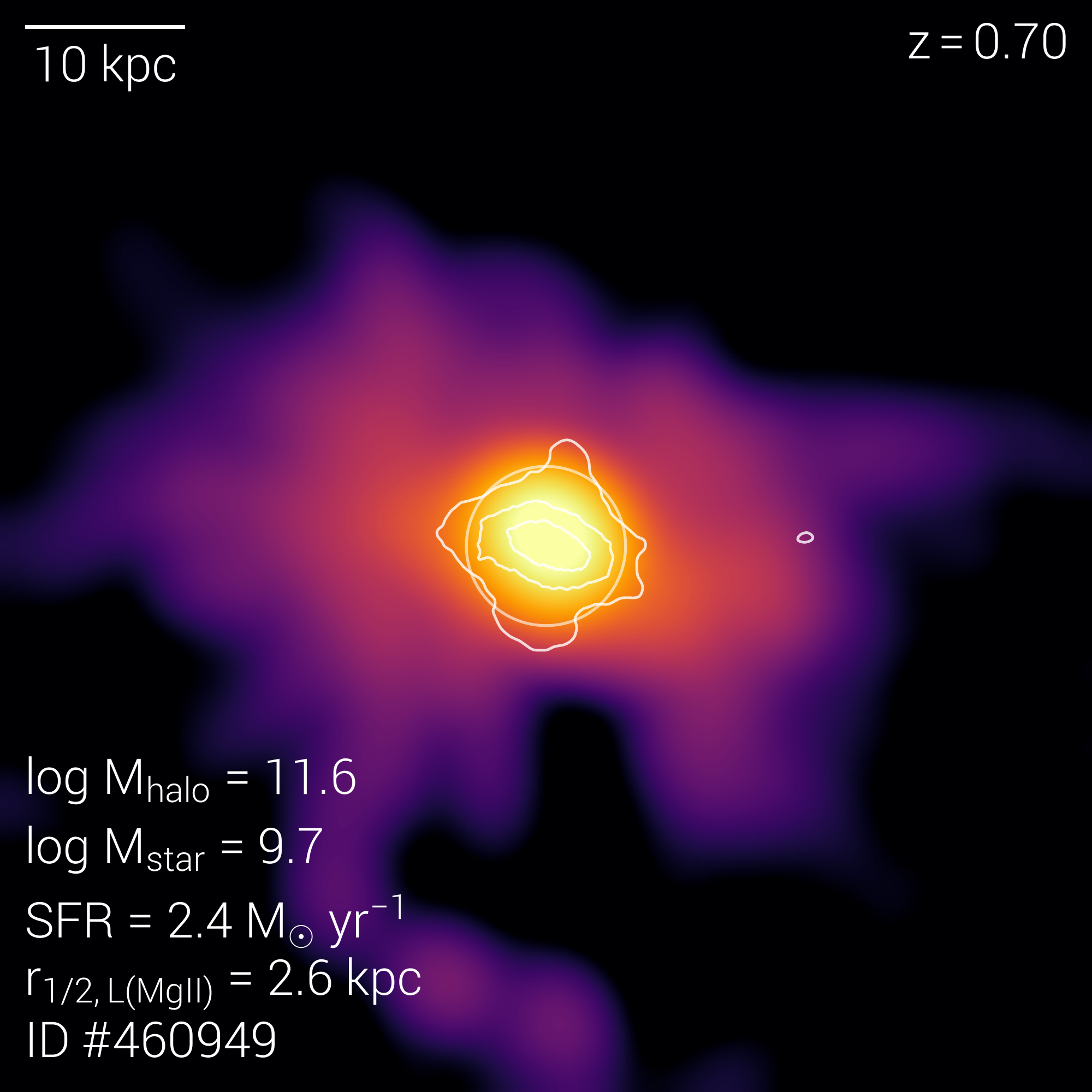}
\includegraphics[angle=0,width=2.25in]{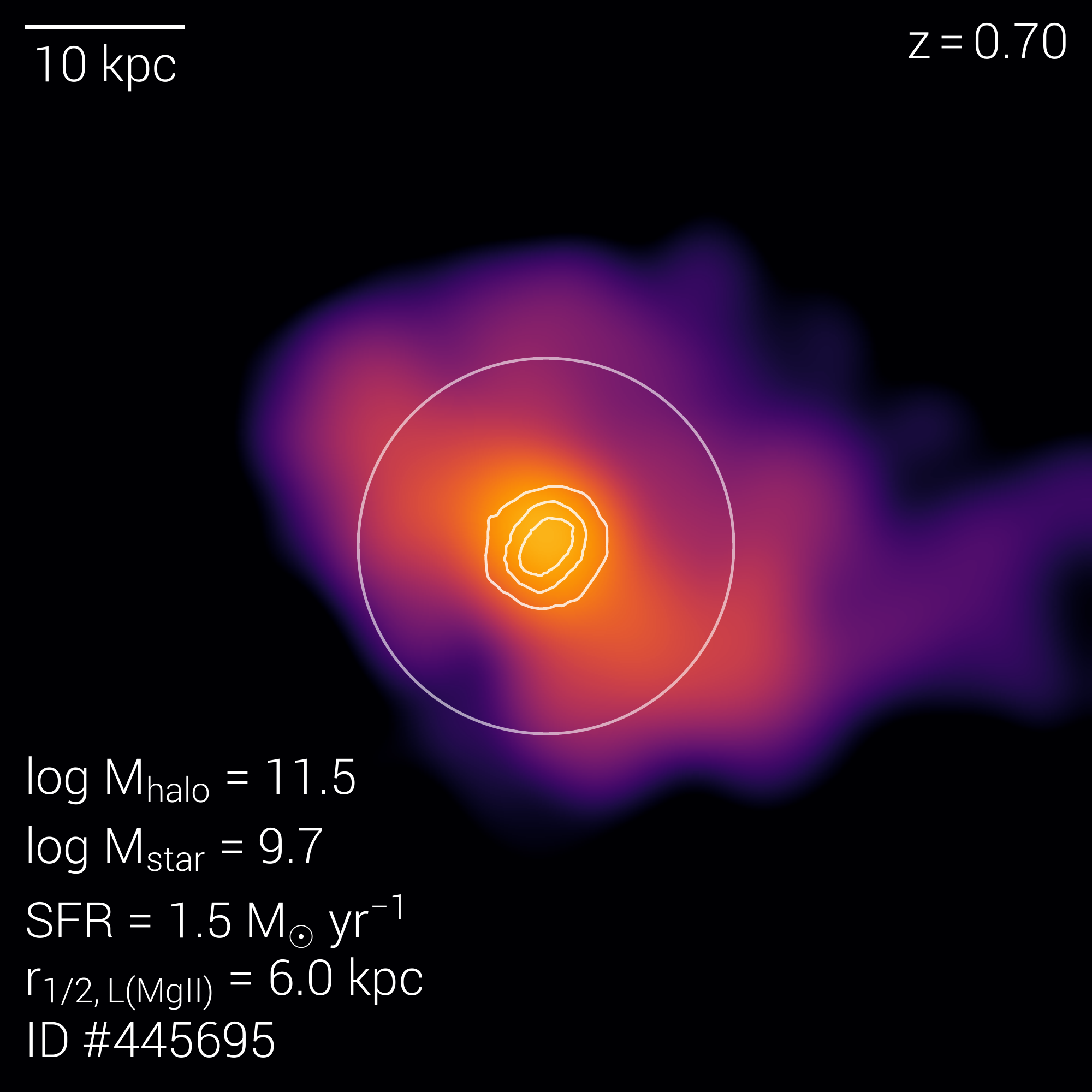}
\includegraphics[angle=0,width=2.25in]{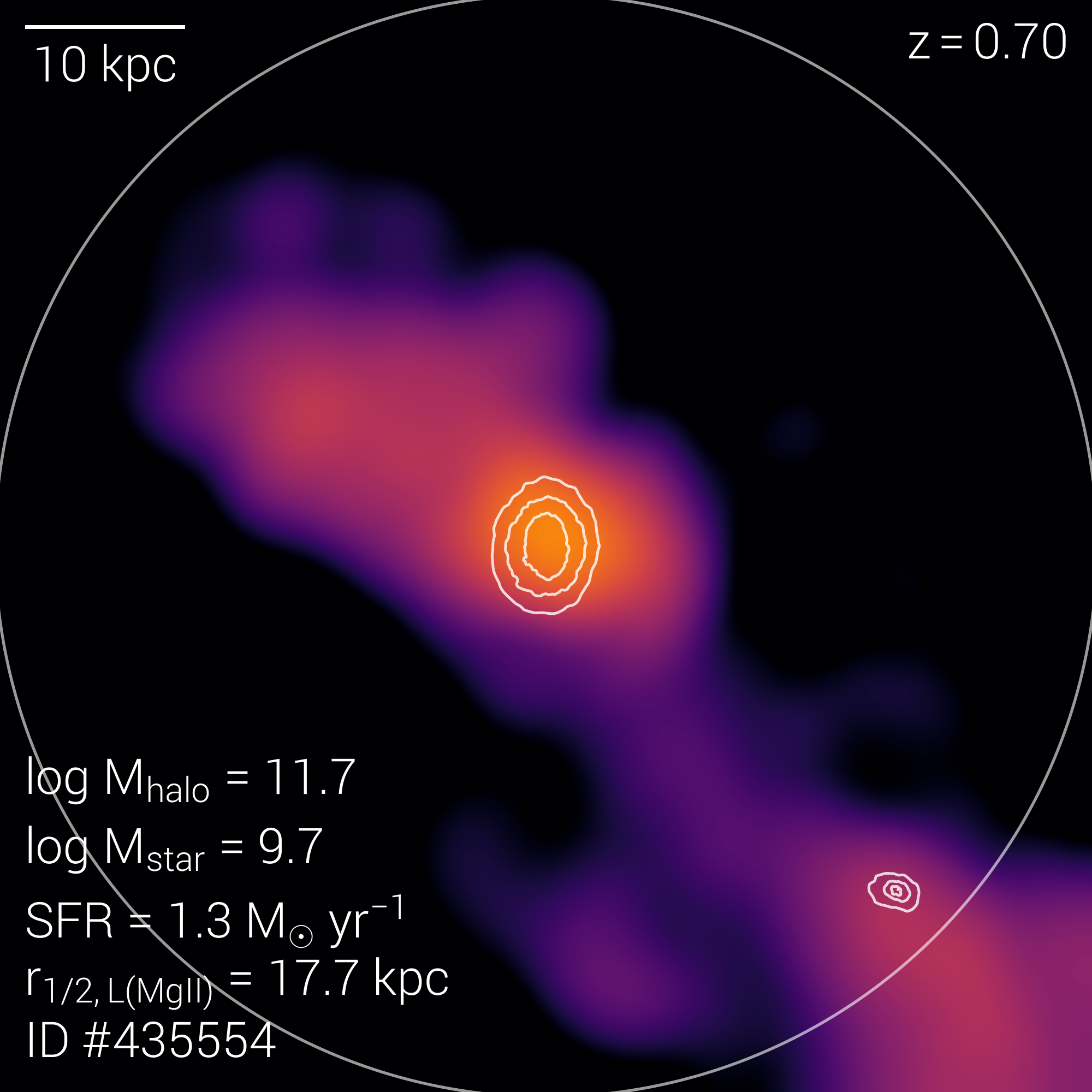}
\caption{ The same three projections from Figure \ref{fig_size} (bottom row), here including the effect of PSF smoothing.
 \label{fig_appendix_psf_images}}
\end{figure*}

\begin{figure}
\centering
\includegraphics[angle=0,width=3.4in]{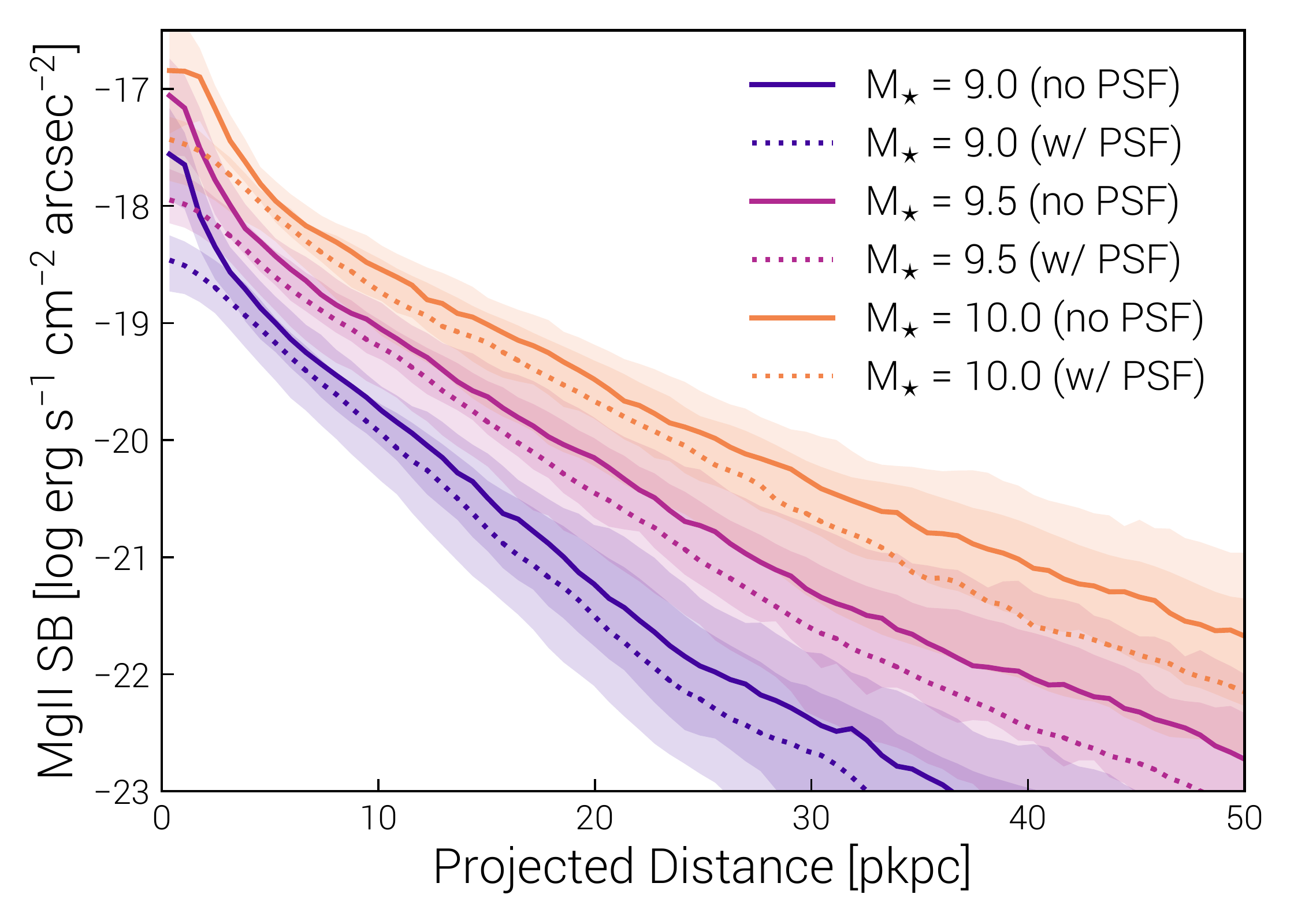}
\caption{ Impact of the PSF convolution, where we adopt a 0.7'' Gaussian PSF as appropriate for the MUSE (non-AO) UDF. Three stellar mass bins are shown at $z=0.7$, as in Figure \ref{fig_sbr_stacked}, solid lines showing intrinsic surface brightness profiles, whereas dotted lines include smoothing by the PSF.
 \label{fig_appendix_psf}}
\end{figure}

Figure \ref{fig_appendix_psf_images} shows the same images as in Figure \ref{fig_size}, where we discussed the origin of MgII halos with substantially different spatial extent (half-light radii) at fixed stellar mass. Here we include the effect of PSF smoothing, for comparison. Note that throughout the entire paper all images have been shown without convolving by the PSF, in order to highlight the intrinsic structure and complexity of MgII emitting gas on small scales as resolved by TNG50. At the same time, all quantitative analysis of the paper -- namely, the profiles, sizes, areas, and shapes derived for MgII halos -- have always included the PSF in order to present more realistic physical predictions for comparison with data.

Figure \ref{fig_appendix_psf} shows the impact of the PSF convolution (i.e. smoothing) on the radial MgII surface brightness profiles. Three stellar mass bins are shown for reference, where the solid lines neglect the PSF, and the dotted lines include the fiducial smoothing of a 0.7'' FWHM Gaussian PSF, as in Figure \ref{fig_sbr_stacked} and throughout this paper. At this resolution the impact is moderate: the intrinsic emission has a strong central peak which is smoothed out.

\begin{figure}
\centering
\includegraphics[angle=0,width=3.4in]{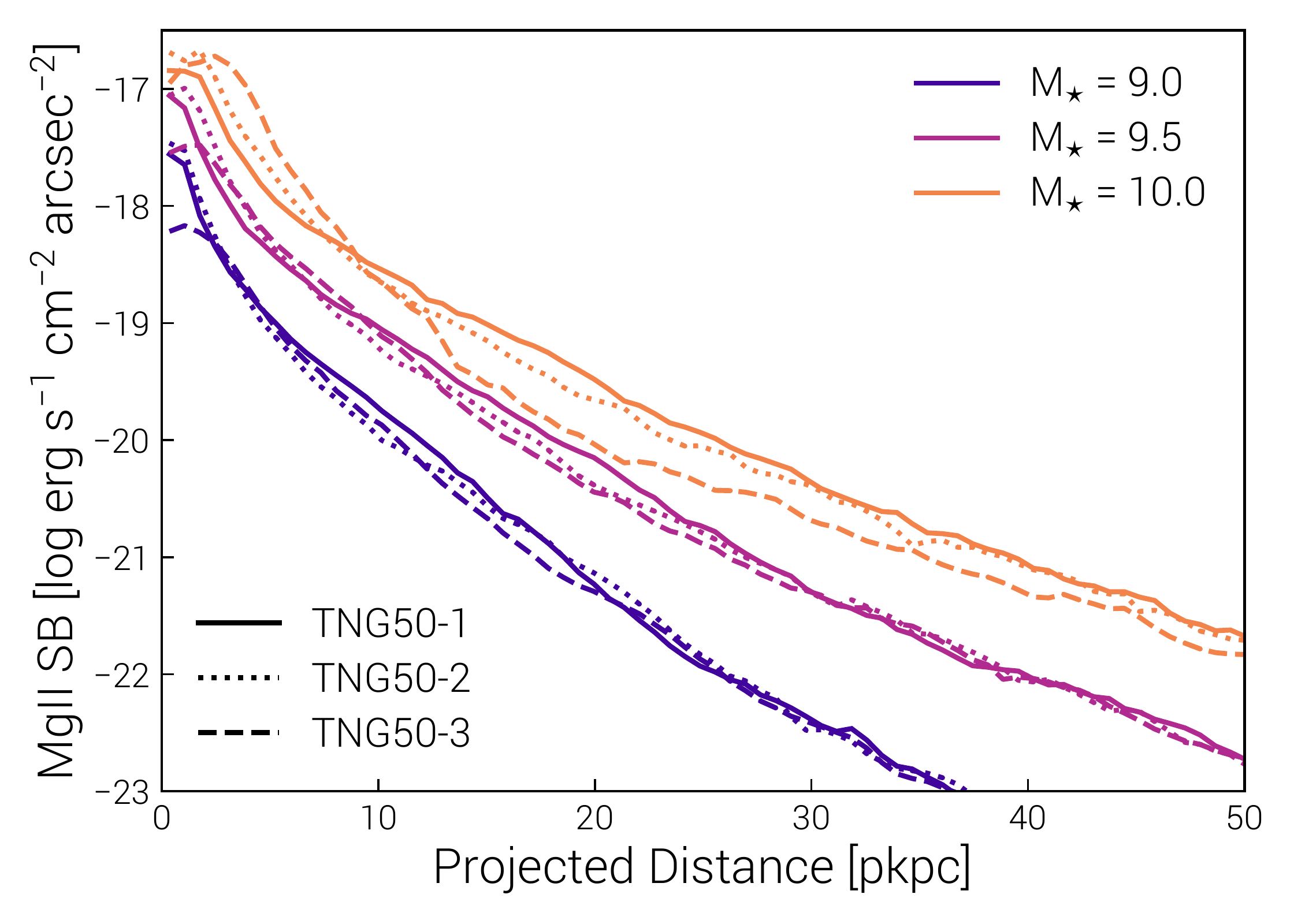}
\caption{ Impact of numerical resolution of the underlying hydrodynamical simulation. We plot stacked radial surface brightness profiles at $z=0.7$ in three stellar mass bins as in Figure \ref{fig_sbr_stacked}. Solid lines show the highest resolution TNG50-1 run, while dotted lines show TNG50-2 (two times lower spatial resolution, eight times lower mass resolution), and dashed lines show TNG50-3 (one further resolution level worse). Compared to the SB variation over the inner 10s of kpc in the CGM, variation due to numerical resolution is generally negligible.
 \label{fig_appendix_res}}
\end{figure}

Finally, Figure \ref{fig_appendix_res} shows a numerical resolution convergence study in terms of the TNG50 simulation itself. We again show three stacked surface brightness profiles, contrasting the highest-resolution main run of TNG50-1 (solid lines) to the lower resolution realizations of the same cosmological volumes: TNG50-2 (dotted lines, eight times lower mass resolution) and TNG50-3 (dashed lines, sixty-four times lower mass resolution). Overall we conclude that the surface brightness profiles of MgII halos are remarkably converged with resolution of the parent hydrodynamical simulation, and this if anything is a subdominant effect.

\end{document}